\documentclass[a4paper,twocolumn,superscriptaddress,nofootinbib,longbibliography,accepted=2019-12-17]{quantumarticle}
\pdfoutput=1

\usepackage[numbers,sort&compress]{natbib}
\usepackage[usenames, dvipsnames]{color}
\usepackage{tikz}
\usetikzlibrary{shapes,snakes,backgrounds,fit,decorations.pathreplacing}
\usepackage{bbm, bm, amsbsy}
\usepackage{amsthm}
\usepackage{amssymb}
\usepackage{amsmath}
\usepackage{dsfont} % for symbols like the identity: \mathds{1}
\usepackage{graphicx} % for graphics
\usepackage{epsfig}
\usepackage{epstopdf}
\usepackage{relsize}

\usepackage[colorlinks,breaklinks]{hyperref}
\makeatletter
\newcommand\org@hypertarget{}
\let\org@hypertarget\hypertarget
\renewcommand\hypertarget[2]{%
  \Hy@raisedlink{\org@hypertarget{#1}{}}#2%
  }
\makeatother
\usepackage{enumerate}%allows different styles of enumerate environment
\usepackage{float}
\hypersetup{
	bookmarksnumbered,
	pdfstartview={FitH},
	citecolor={darkgreen},
	linkcolor={darkred},
	urlcolor={darkblue},
	pdfpagemode={UseOutlines}}
\definecolor{darkgreen}{RGB}{50,190,50}
\definecolor{darkblue}{RGB}{0,0,190}
\definecolor{darkred}{RGB}{238,0,0}
\usepackage{soul}
\newcommand{\pr}{^{\prime}}

\newcommand{\ket}[1]{\ensuremath{\left|\right.\!{#1}\!\left.\right\rangle}}

\newcommand{\bra}[1]{\ensuremath{\left\langle\right.\!{#1}\!\left.\right|}}

\newcommand{\scpr}[2]{\ensuremath{\left\langle\right.\hspace*{-1pt} #1 \hspace*{-1pt}\left|\right.\hspace*{-1pt} #2 \hspace*{-1pt}\left.\right\rangle}}

\newcommand{\subtiny}[3]{\ensuremath{_{\hspace{#1 pt}\protect\raisebox{#2 pt}{\tiny{$ #3$}}}}}
\newcommand{\suptiny}[3]{\ensuremath{^{\hspace{#1 pt}\protect\raisebox{#2 pt}{\tiny{$ #3$}}}}}

\newcommand{\Sys}{\ensuremath{_{\hspace{-0.5pt}\protect\raisebox{0pt}{\tiny{$S$}}}}}
\newcommand{\Poi}{\ensuremath{_{\hspace{-0.5pt}\protect\raisebox{0pt}{\tiny{$P$}}}}}

\newcommand{\Fr}{\ensuremath{_{\hspace{-0.5pt}\protect\raisebox{0pt}{\tiny{$F$}}}}}
\newcommand{\SP}{\ensuremath{_{\hspace{-0.5pt}\protect\raisebox{0pt}{\tiny{$S\hspace*{-1pt}P$}}}}}
\newcommand{\SPE}{\ensuremath{_{\hspace{-0.5pt}\protect\raisebox{0pt}{\tiny{$S\hspace*{-1pt}PE$}}}}}
\newcommand{\Aux}{\ensuremath{_{\hspace{-0.5pt}\protect\raisebox{0pt}{\tiny{$A$}}}}}

\def\phi{\varphi}

\def\be{\begin{equation}}
\def\ee{\end{equation}}

\usetikzlibrary{shapes, arrows}

\newcommand{\tr}{\mathrm{tr}}

\DeclareMathOperator{\diag}{diag}

\newcommand{\djj}{d\kern-0.4em\char"16\kern-0.1em}

\renewcommand{\thesection}{\arabic{section}}
\renewcommand{\thesubsection}{\arabic{section}.\Alph{subsection}}
\makeatletter
\renewcommand{\p@subsection}{}
\renewcommand{\p@subsubsection}{}
\makeatother

\usepackage[customcolors]{hf-tikz}
\tikzset{style green/.style={
    set fill color=green!50!lime!60,
    set border color=white,
  },
  style cyan/.style={
    set fill color=cyan!90!blue!60,
    set border color=white,
  },
  style orange/.style={
    set fill color=orange!80!red!60,
    set border color=white,
  },
  style hordash/.style={
    set fill color=white,
    set border color=black,
  },
  hor/.style={
    above left offset={-0.09,0.25},
    below right offset={0.09,-0.05},
    #1
  },
  ver/.style={
    above left offset={-0.09,0.35},
    below right offset={0.09,-0.1},
    #1
  }
}

\usepackage[framemethod=tikz]{mdframed}
\definecolor{mycolor2}{RGB}{112, 48, 160}
\definecolor{mycolor}{rgb}{0.122, 0.435, 0.698}
\newmdenv[innerlinewidth=0.5pt, roundcorner=4pt,linecolor=mycolor,innerleftmargin=6pt,
innerrightmargin=6pt,innertopmargin=6pt,innerbottommargin=6pt]{mybox}
\usepackage{paracol}
\usepackage{tcolorbox}
\tcbuselibrary{theorems}

\newtcbtheorem{Definitions}{Definition}%
{colback=mycolor!5,colframe=mycolor,fonttitle=\bfseries,
left=4pt,
right=4pt,
top=5pt,
bottom=5pt}{defi}

\newtcbtheorem{Lemmas}{Lemma}%
{colback=green!5,colframe=green!50!blue,fonttitle=\bfseries,
left=4pt,
right=4pt,
top=5pt,
bottom=5pt
}{lemma}

\newtcolorbox[blend into=figures]{boxdefi}[3][]
{ float*=ht,width=\textwidth,lower separated=false, center upper,
title={#2},label= def:#3,#1}

\begin{document}

\title{Ideal Projective Measurements Have Infinite Resource Costs}
\author{Yelena Guryanova}
\email{yelena.guryanova@oeaw.ac.at}
\affiliation{Institute for Quantum Optics and Quantum Information -- IQOQI Vienna, Austrian Academy of Sciences,\\ Boltzmanngasse 3, 1090 Vienna, Austria}
\orcid{0000-0001-9943-9424}
\author{Nicolai Friis}
\email{nicolai.friis@univie.ac.at}
\affiliation{Institute for Quantum Optics and Quantum Information -- IQOQI Vienna, Austrian Academy of Sciences,\\ Boltzmanngasse 3, 1090 Vienna, Austria}
\orcid{0000-0003-1950-8640}
\author{Marcus Huber}
\email{marcus.huber@univie.ac.at}
\affiliation{Institute for Quantum Optics and Quantum Information -- IQOQI Vienna, Austrian Academy of Sciences,\\ Boltzmanngasse 3, 1090 Vienna, Austria}
\orcid{0000-0003-1985-4623}
%%
%%%%
\date{December 18, 2019}

\begin{abstract}
We show that it is impossible to perform ideal projective measurements on quantum systems using finite resources. We identify three fundamental features of ideal projective measurements and show that when limited by finite resources only one of these features can be salvaged. Our framework is general enough to accommodate any system and measuring device (pointer) models, but for illustration we use an explicit model of an $N$-particle pointer. For a pointer that perfectly reproduces the statistics of the system, we provide tight analytic expressions for the energy cost of performing the measurement. This cost may be broken down into two parts. First, the cost of preparing the pointer in a suitable state, and second, the cost of a global interaction between the system and pointer in order to correlate them. Our results show that, even under the assumption that the interaction can be controlled perfectly, achieving perfect correlation is infinitely expensive. We provide protocols for achieving optimal correlation given finite resources for the most general system and pointer Hamiltonians, phrasing our results as fundamental bounds in terms of the dimensions of these systems.
\end{abstract}

\maketitle

%%%%%%%%%%%%%%%%%%%%%%%%%%%%%%%%%%%%%%%%%%%%%%%%%%%%%%%%%%%%%%%%%%%%%%%%%%%%%%%%%%%%%%%%%%%%%%%%%%%%%%%%%%%%%%%%%%%%%

\vspace*{-1mm}
\section{Introduction}
\vspace*{-1mm}

The foundations of any physical theory are laid by its axioms, postulates and laws. In quantum theory, the projection postulate presents one of these central pillars. It says that upon measuring a quantum system, its post-measurement state is given by an eigenstate of the measured observable and the corresponding probability for obtaining this state is given by the Born rule. In this way, an ideal projective measurement leaves the system in a pure state that is perfectly correlated with the measurement outcome.

Similarly, the key tenets of thermodynamics are formed by its three fundamental laws. Intense efforts in quantum thermodynamics \cite{VinjanampathyAnders2016, MillenXuereb2016, GooldHuberRieraDelRioSkrzypczyk2016} have placed these laws on rigorous mathematical footing~\cite{EspositoVanDenBroeck2011, Jacobs2012, BrandaoHorodeckiNgOppenheimWehner2015, LostaglioJenningsRudolph2015, CwiklinskiStudzinskiHorodeckiOppenheim2015, AlhambraMasanesOppenheimPerry2016, WilmingGallegoEisert2016, ScharlauMueller2018, MasanesOppenheim2017, BeraRieraLewensteinWinter2017}. Of particular interest is the third law of thermodynamics in the quantum regime, which tells us that no quantum system can be cooled to the ground state (which, in non-degenerate cases, is a \textit{pure} state) in finite time and with finite resources~\cite{SchulmanMorWeinstein2005, WilmingGallego2017, MasanesOppenheim2017, ClivazSilvaHaackBohrBraskBrunnerHuber2019a, ClivazSilvaHaackBohrBraskBrunnerHuber2019b, ScharlauMueller2018}. This is in apparent contradiction with the projection postulate \cite{Kieu2019} --- how is it that an ideal, error-free, measurement leaves the system in a state forbidden by the laws of thermodynamics?

In reality, we know that measurements in the lab are performed in finite time and with finite resources. These measurements are prone to small errors,  %(e.g., dark counts in photon detectors),
implying that the post-measurement state of the system is never truly pure. However, with technological advances making errors ever smaller, one would assume rising thermodynamic costs as the post-measurement state of the system approaches %a pure state.
purity.

Here, we resolve this apparent contradiction. We show that the resource cost of an ideal quantum measurement in a finite temperature environment is indeed infinite.
Our operational approach is based on correlations between a system and a pointer, allowing us to make quantifiable statements about the cost. Within this framework we identify that an ideal projective measurement has three model-independent properties; it is: \textit{unbiased}, \textit{faithful}, and \textit{non-invasive} --- properties that cannot hold simultaneously for measurements with finite resources (energy and time).
Our framework is general enough to accommodate any measurement model for which we provide quantitative results for an example case.
In doing so, we refrain from making statements about what is commonly perceived as the `measurement problem' (how or why the system is left in a particular state and what it means to obtain a `result'~\cite{KorbiczAguilarCwiklinskiHorodecki2017, Zurek2009}).

Past approaches to quantifying the cost of a quantum measurement have typically assumed that projective measurements can be carried out \textit{perfectly} and that their cost can be attributed to the work value of the measurement outcome~\cite{SagawaUeda2009, Jacobs2012, ElouardHerreraMartiHuardAuffeves2017, LipkaBartosikDemkowiczDobrzanski2018, ElouardJordan2018}. Others adopt the stance that Landauer's erasure bound represents the cost of resetting devices to pure states~\cite{EspositoVanDenBroeck2011, ReebWolf2014, AbdelkhalekNakataReeb2016}, without providing conclusive evidence that the bound is achievable.
These works assume an unlimited supply of pure states, circumventing the third law of thermodynamics and resulting in finite energy costs.
% even for ideal measurements carried out in finite time.
%Indeed, access to pure states implies the ability to perform ideal measurements, in turn allowing the creation of the required pure states.
%In other words, ideal measurements produce pure states, i.e., states at temperature zero.
However, when limited to thermal environments, measurements produce errors, which
% --- the pointer states are not perfectly correlated with those of the system.
% Such errors
can be mitigated by either reducing the temperature of the environment, or by using larger measuring devices. Both
%of these
 strategies can be quantified in terms of their thermodynamic cost, for which we provide exact analytic results. Our results demonstrate that even the simplest quantum measurements on qubits are never for free.\\

%%%%%%%%%%%%%%%%%%%%%%%%%%%%%%%%%%%%%%%%%%%%%%%%%%%%%%%%%%%%%%%%%%%%%%%%%%%%%%%%%%%%%%%%%%%%%%%%%%%%%%%%%%%%%%%%%%%%%

\vspace*{-3mm}
\section{Ideal measurements}
\vspace*{-1mm}

Consider an unknown quantum system $\rho\Sys$ and a measuring device (\textit{pointer}) $\rho\Poi$.  To measure the system, one must couple it to the pointer and effect a joint transformation that correlates them:
%\begin{align}
    $\rho\Sys\otimes \rho\Poi
	\longrightarrow \tilde{\rho}\SP$.
	%\,.
    %\label{eq:interact}
%\end{align}
In an \emph{ideal} measurement, the system and pointer become perfectly correlated, such that upon ``observing" the pointer, one infers which pure state the system is in with probability $1$. More precisely, each eigenstate $\ket{i}\Sys$ of the measured observable of the system is assigned a set $\{\ket{\tilde{\psi}_{n}\suptiny{0}{0}{(i)}}\Poi\}_{n}$ of orthogonal states of the pointer corresponding to a projector $\Pi_{i}=\sum_{n} \ket{\tilde{\psi}_{n}\suptiny{0}{0}{(i)}}\!\!\bra{\tilde{\psi}_{n}\suptiny{0}{0}{(i)}}$. The projectors are orthogonal, forming a complete set, $\Pi_{i}\Pi_{j}=\delta_{ij}\Pi_{i}$ and $\sum_{i}\Pi_{i}=\mathds{1}\Poi$. %Thus, each possible measurement outcome is represented and can be unambiguously identified.
Upon finding the pointer in a state $\ket{\tilde{\psi}_{n}\suptiny{0}{0}{(i)}}\Poi$ (chosen to reflect $\ket{i}\Sys$), one concludes that the measurement outcome is ``$i$", and that the system is left in the state $\ket{i}\Sys$. Up to arbitrary off-diagonal elements w.r.t. the basis $\{\ket{i}\otimes\ket{\tilde{\psi}_{n}\suptiny{0}{0}{(j)}}\}_{i,j,n}$, the ideal post-interaction state with perfect correlation has the form
\begin{align}
    \tilde{\rho}\SP &=\,
    \sum_{i}\rho_{ii}\,\ket{i}\!\!\bra{i}\otimes\rho\suptiny{0}{0}{(i)}
    +\,\text{off-diag.}\,,
    \label{eq:postin}
\end{align}
where $\rho_{ii}=\bra{i}\rho\Sys\ket{i}$ are diagonal elements w.r.t. the basis $\{\ket{i}\}\Sys$ and $\rho\suptiny{0}{0}{(i)}$ is a pointer state, associated to one and only one of the outcomes $i$, i.e., $\Pi_{i}\rho\suptiny{0}{0}{(j)}=\delta_{ij}\rho\suptiny{0}{0}{(j)}$.
% (if all off-diagonal elements vanish).
The form of $\tilde{\rho}\SP$ in Eq.~\eqref{eq:postin} is the result of an \textit{ideal measurement} and can be entangled or simply classically correlated.
%Such an
This ideal measurement satisfies three fundamental properties:
\begin{enumerate}[(i)]
\item{\label{def:unbiased}
\textbf{Unbiased}. The probability of finding the pointer in a state associated with outcome $i$ after the interaction is the same as the probability of finding the system in the state $\ket{i}\Sys$ before the interaction,
\begin{align}
    \tr\bigl[\mathbb{I}\otimes\,\Pi_{i}\tilde{\rho}\SP\bigr]
	   &=\,\tr\bigl[\,\ket{i}\!\!\bra{i}\Sys\rho\Sys]\,=\,\rho_{ii}\quad \forall i\ \forall \rho\Sys.
    \label{eq:unbiased}
\end{align}
A measurement is unbiased if the pointer reproduces the measurement statistics of the system.}
\item{\label{def:faithfulness}
\textbf{Faithful}. There is a one-to-one correspondence between the pointer outcome and the post-measurement system state
\begin{align}
    C(\tilde{\rho}\SP)  &:=\sum_{i} \tr\bigl[\, \ket{i}\!\!\bra{i}\otimes \Pi_{i} \;\tilde{\rho}\SP\bigr]\,=\,1\quad \forall \rho\Sys,
    \label{eq:faithfulness}
\end{align}
%The post-measurement state
i.e., $\tilde{\rho}\SP$ has perfect correlation: on observing the pointer outcome $i$ (associated to $\Pi_{i}$), one concludes that the system is left in the state $\ket{i}\Sys$ with certainty.}
\item{\label{def:noninv}
\textbf{Non-invasive}. The probability of finding the system in the state $\ket{i}\Sys$ is the same before and after the interaction with the pointer,
\begin{align}\label{eq:invasive}
    \tr\bigl[\,\ket{i}\!\!\bra{i}\Sys\tilde{\rho}\SP\bigr]
    &=\,\tr\bigl[\,\ket{i}\!\!\bra{i}\Sys\rho\Sys\bigr]\,=\,\rho_{ii}\ \
\forall \; i\ \forall \rho\Sys.
\end{align}
This property only holds for the basis $\ket{i}\Sys$ and coherences appearing on the off-diagonal can, in general, be destroyed.}
% even for ideal measurements.}
\end{enumerate}
These three properties, stated here without particular hierarchy, capture the pairwise relation between (\ref{def:unbiased}) the pre-measurement system state and the measurement outcome, (\ref{eq:faithfulness}) between the measurement outcome and the post-measurement system state, and (\ref{eq:invasive}) between the pre- and post-measurement system states, respectively.

All quantitative statements we make about the faithfulness of a measurement [property (\ref{eq:faithfulness})] depend on the function $C(\tilde{\rho}\SP)$ in Eq.~\eqref{eq:faithfulness}. This function's value represents the average probability of correctly inferring the post-measurement state upon observing the pointer, which is~$1$ for any unbiased measurement if and only if the post-interaction state is of the form of $\tilde{\rho}\SP$ in Eq.~\eqref{eq:postin}. One could choose more complicated functions or even measures of correlation. However, in our paradigm it is sufficient to be classically correlated to have perfect `correlation' in the sense that $C=1$. Note that quantum correlations are not strictly necessary, since $C(\ket{0}\Sys\ket{0}\Poi)=1$. The advantage of the expression in Eq.~\eqref{eq:faithfulness} is that it quantifies the probability that the pointer indicates an outcome which is \textit{correct} and yields the maximal value $1$ if and only if the post-interaction state is of the form of Eq.~\eqref{eq:postin}.\\

%%%%%%%%%%%%%%%%%%%%%%%%%%%%%%%%%%%%%%%%%%%%%%%%%%%%%%%%%%%%%%%%%%%%%%%%%%%%%%%%%%%%%%%%%%%%%%%

\noindent\textit{Example.}\ Consider a measurement of a qubit system using a single qubit pointer in the ground state. We model the measurement with a controlled NOT operation $U\subtiny{0}{0}{\mathrm{CNOT}}=\ket{0}\!\!\bra{0}\Sys\otimes\mathds{1}\Poi+\ket{1}\!\!\bra{1}\Sys\otimes X\Poi$, where $X=\ket{0}\!\!\bra{1}+\ket{1}\!\!\bra{0}$. The post-measurement state
$\tilde{\rho}\SP = U\subtiny{0}{0}{\mathrm{CNOT}} \;
    (\rho\Sys \otimes \ket{0}\!\!\bra{0}\Poi )\;
	U\subtiny{0}{0}{\mathrm{CNOT}}^\dagger$
%\begin{align}\label{eq:idealmeas}
%	\tilde{\rho}\SP = U\subtiny{0}{0}{\mathrm{CNOT}} \;
%    (\rho\Sys \otimes \ket{0}\!\!\bra{0}\Poi )\;
%	U\subtiny{0}{0}{\mathrm{CNOT}}^\dagger
%\end{align}
is of the form of Eq.~\eqref{eq:postin}, meaning the measurement is unbiased, faithful, and non-invasive. Indeed, whenever both system and pointer have dimension $d\Sys=d\Poi=d$, and the pointer is initially in a pure state (w.l.o.g. the ground state), we can define a unitary $U_{d}:=\ket{0}\!\!\bra{0}\Sys\otimes\mathds{1}\Poi+\sum_{i\neq0}\ket{i}\!\!\bra{i}\Sys\otimes X\suptiny{0}{0}{(i)}\Poi$, where $X\suptiny{0}{0}{(i)}\Poi=\ket{0}\!\!\bra{i}+\ket{i}\!\!\bra{0}+\sum_{j\neq0,i}\ket{j}\!\!\bra{j}$, realizing an ideal measurement. A more detailed discussion of this example is presented in Appendix~\ref{sec:Examples Two-Qubit Measurement Procedures}.\\

%%%%%%%%%%%%%%%%%%%%%%%%%%%%%%%%%%%%%%%%%%%%%%%%%%%%%%%%%%%%%%%%%%%%%%%%%%%%%%%%%%%%%%%%%%%%%%%%%%%%%%%%%%%%%%%%%%%%%

\section{Non-ideal measurements}

\begin{figure}[ht]
\centering
    %%%trim={<left> <lower> <right> <upper>}
     \includegraphics[width=0.47\textwidth,trim={0mm 0mm 0mm 0mm}]{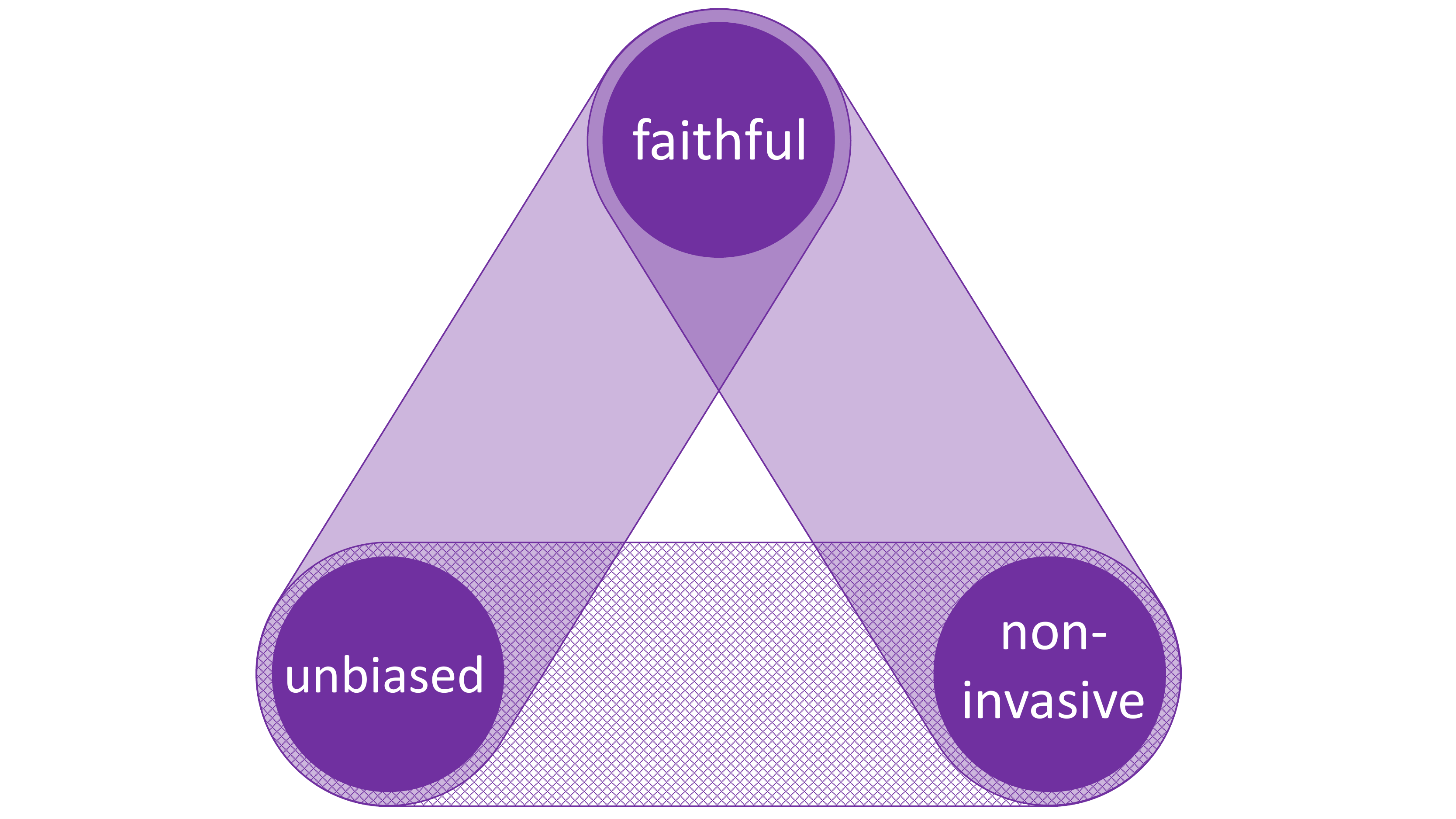}
     %Fig_1_triRelation.pdf}
     \vspace*{-2mm}
     \caption{The properties attributed to an ideal measurement. In a non-ideal measurement these three properties do not hold simultaneously, and satisfying one of them does not imply any of the other two. When $\rho\Sys$ is fixed, in two out of three cases, satisfying a pair of properties implies that the third property also holds. A measurement which is faithful and unbiased implies that it is also non-invasive, and a measurement that is faithful and non-invasive implies that it is unbiased. When $\rho\Sys$ is relaxed to \textit{all} initial system states, then satisfying any pair of properties implies the third, and one recovers the last relation, namely that a measurement which is unbiased and non-invasive (\textit{for all} $\rho\Sys$) implies that it is also faithful.}
\label{fig:triangle}
\end{figure}

We call a measurement in which any of the properties \eqref{def:unbiased} -- \eqref{def:noninv} fails to hold \textit{non-ideal}. This is due to the fact that, in general, the properties do not imply one another, i.e., satisfying a single property does not imply any of the other two, as illustrated in Fig.~\ref{fig:triangle}. Things become more subtle when a \textit{pair} of properties is satisfied. In two cases, satisfying a pair of properties implies the third. As we show in detail in Appendix~\ref{sec:nonid meas procedures}, a measurement which is faithful and unbiased is also non-invasive, and a measurement that is faithful and non-invasive is unbiased. However, in general, a measurement being unbiased and non-invasive for a particular input state \textit{does not} imply it is faithful, unless it is unbiased and non-invasive \emph{for all} input states~$\rho\Sys$, as illustrated in Fig.~\ref{fig:triangle}.

In what follows we prove that faithful measurements (perfect correlations) are possible if and only if one can prepare states with sufficiently many vanishing eigenvalues. Since, by the third law of thermodynamics, one cannot prepare states of non-full rank with finite resources, property~\eqref{def:faithfulness} fails to hold and therefore ideal measurements are not physically feasible.
To see this, consider the most general interaction between a system and pointer ---  a completely positive and trace-preserving (CPTP) map, which can be understood as a unitary on the system and an extended pointer. In order for such a unitary to realize a faithful measurement according to Eq.~\eqref{eq:faithfulness}, the rank of the final state $\tilde{\rho}\SP$ must be bounded from above by the dimension of the pointer $d\Poi$ (with $d\Poi\geq d\Sys$), since $d\Sys\leq\sum_{i}\operatorname{rank}(\rho\suptiny{0}{0}{(i)})\leq d\Poi$. When $d\Sys=d\Poi=d$, this implies that the initial rank of the pointer $\rho\Poi$ must be $1$, i.e., a pure state. For larger pointers, their initial state need not be pure, but it cannot have full rank --- one must have $\operatorname{rank}(\rho\Poi)\leq d\Poi/d\Sys$. Practically, this requires pure state preparation for some non-trivial  pointer subspaces. Thus, faithful and therefore ideal measurements are not possible without a supply of pure states (states at absolute zero temperature). States with non-full rank require infinite time, energy or complexity (interaction range) to be prepared and are prohibited by the third law of thermodynamics~\cite{SchulmanMorWeinstein2005,WilmingGallego2017,MasanesOppenheim2017,ClivazSilvaHaackBohrBraskBrunnerHuber2019a, ClivazSilvaHaackBohrBraskBrunnerHuber2019b,ScharlauMueller2018}. Conversely, whenever the pointer state does not start with full rank, operations such as $U_{d}$ allow one to achieve perfect correlation.

Since faithful (and hence also ideal) measurements are not possible, we want to determine how closely they can be approximated. Since laboratory experiments take place at non-zero temperature, the natural state of a pointer is in thermodynamic equilibrium with its environment, i.e., the state $\tau\Poi(\beta)$, with inverse temperature $\beta=1/k_{\mathrm{B}}T$. At any finite temperature, a thermal state has full rank, and any deviation from it requires an input of work.

%\noindent
%%

While faithfulness is not necessarily the most important property, it is the one that certainly cannot be upheld in practice, whereas one of the other two properties can in principle be maintained also for practical measurements. Here, we take the point of view that the crucial property to demand of \emph{any} measurement is to be \emph{unbiased}. This guarantees that with sufficient repetitions, one obtains a mean value for the measured observable that accurately reflects the mean value of the underlying system $\rho\Sys$ and the degree of trust in this outcome can be quantified using standard statistical methods. Conversely, if one imposed that the measurement were non-invasive, one would be able to perform repeated measurements on the system without changing the statistics of the measured observable. However, without properties~(\ref{def:unbiased}) or~(\ref{def:faithfulness}) it would not be possible to reliably relate the measurement data to statements about $\rho\Sys$. We therefore consider non-ideal measurements between a system $\rho\Sys$ and a thermal pointer $\tau\Poi(\beta)$ with the property that the measurements are \textit{unbiased}. In Appendix~\ref{sec:Unbiased Measurement Procedures}, we derive the general structure of maps realizing unbiased measurements in full detail.\\

%\noindent\textit{Universal bounds.}
\section{Maximally faithful measurements}

After imposing the measurements to be unbiased, we are then interested in the ones that produce the best correlations, in other words those which are as close to faithful as possible and thus approximate ideal measurements in a meaningful way. For such unbiased maximally correlating measurements, one may then determine and minimize the finite energy cost. To provide a self-contained description of this cost, we consider the joint system of $S$ and $P$ to be closed, implying that we restrict to unitary maps. In Appendix~\ref{sec:extremal meas procedures} we further discuss this restriction and show that such unitarily correlating unbiased measurements are indeed possible at any temperature. Moreover, we find that unitarily correlating unbiased measurements have a maximal achievable correlation $C_{\text{max}}$ that can only be reached if sufficient energy is supplied. To see this, let us denote the system and pointer Hamiltonians by $H\Sys$ and $H\Poi$ and the corresponding dimensions by $d\Sys$ and $d\Poi$. For unbiased measurements, the restriction to unitary maps $\rho\Sys\otimes\tau\Poi\rightarrow\tilde{\rho}\SP$ implies a truncation of $d\Poi$ at an integer multiple of $d\Sys$. We then order the spectrum of the pointer Hamiltonian and divide it into $d\Sys$ disjoint sets $\{E\suptiny{0}{0}{(k)}_{i}\}_{i=0,\ldots,d\Poi/d\Sys-1}$ for $k=0,\ldots,d\Sys-1$, such that $H\Poi=\sum_{k=0}^{d\Sys-1}\sum_{i=0}^{d\Poi/d\Sys-1}E\suptiny{0}{0}{(k)}_{i}\ket{E\suptiny{0}{0}{(k)}_{i}}\!\!\bra{E\suptiny{0}{0}{(k)}_{i}}$ and $E\suptiny{0}{0}{(k)}_{i}\geq E\suptiny{0}{0}{(k\pr)}_{j}$ whenever $k\geq k\pr$ or when $k=k\pr$ and $i\geq j$. Up to swaps between degenerate energies, the set $\{E\suptiny{0}{0}{(0)}_{i}\}_{i}$ contains the $d\Poi/d\Sys$ smallest energies and consequently $\{\tfrac{1}{\mathcal{Z}}e^{-\beta E\suptiny{0}{0}{(0)}_{i}\}}\}_{i}$ are the largest populations. These populations are assigned to the `correlated subspace'. For unbiased unitaries, there is an algebraic maximum to the achievable correlations between any system and thermal pointer, given by
\begin{align}
 C_{\mathrm{max}}(\beta) =\sum_{i=0}^{d\Poi/d\Sys-1}e^{-\beta E\suptiny{0}{0}{(0)}_{i}}/\mathcal{Z}
 \,,
    \label{eq:universal bound}
\end{align}
which is independent of $\rho_{ii}$ precisely due to unbiasedness. A more detailed derivation can be found in Appendix~\ref{app:max corr unbiased}.

Note that, because we assigned the largest populations to the `correlated subspace', $ C_{\mathrm{max}}$ can be interpreted as the maximum probability of the post-measurement system being in the same state as the pointer. For an arbitrary unbiased measurement generally $ C(\tilde{\rho}\SP)  \leq  C_{\mathrm{max}}(\beta)$. For an arbitrary unbiased measurement achieving $C_{\mathrm{max}}$ one can select a pointer basis $\{\ket{\tilde{\psi}\suptiny{0}{0}{(k)}_{i}}\}_{i,k}$ such that the resulting state can be written
\begin{align}
   	&\tilde{\rho}\SP =  \sum_{k=0}^{d\Sys-1}\frac{\rho_{kk}}{\mathcal{Z}}
   	\Bigl(\sum_{i=0}^{d\Poi/d\Sys-1}e^{-\beta E\suptiny{0}{0}{(0)}_{i}}\ket{k}\!\!\bra{k}\otimes
   	    \ket{\tilde{\psi}_{i}\suptiny{0}{0}{(k)}}\!\!\bra{\tilde{\psi}_{i}\suptiny{0}{0}{(k)}}
   	    \nonumber\\
  & +\!%\sum_{k=1}^{d_S}\frac{\rho_{kk}}{\mathcal{Z}}
  \sum_{m\neq k}\!\!\!\sum_{i=0}^{d\Poi/d\Sys-1}\!\!\!\!e^{-\beta \tilde{E}_{i,m}}\, U\suptiny{0}{0}{(k)}_{\mathrm{nc}}\ket{m}\!\!\bra{m} \otimes \ket{\tilde{\psi}_{i}\suptiny{0}{0}{(k)}}\!\!\bra{\tilde{\psi}_{i}\suptiny{0}{0}{(k)}}U\suptiny{0}{0}{(k)\dagger}_{\mathrm{nc}}\Bigr),\label{eq:generalform}
\end{align}
where the $U\suptiny{0}{0}{(k)}_{\mathrm{nc}}$ for $k=0,\ldots,d\Sys$ are unitaries on the non-correlated subspaces spanned by the vectors $\ket{m}\ket{\tilde{\psi}_{i}\suptiny{0}{0}{(k)}}$ for $i=0,\ldots, d\Poi/d\Sys-1$ and $m\neq k$, $\{\tilde{E}_{i,m\neq k}\}_{i,m}=\{E\suptiny{0}{0}{(n>0)}_{i}\}_{n,i}$. From this form, we see that perfect correlation $C=1$ is only possible if the pointer temperature reaches absolute zero, or, more generally, if the rank of the pointer is bounded by $\text{rank}(\rho\Poi)\leq d\Poi/d\Sys$. Note that the way the system state is altered through measurement is not completely fixed by Eq.~(\ref{eq:generalform}). The relation between the bases $\{\tilde{\psi}_{i}\suptiny{0}{0}{(k)}\}_{i,k}$ and $\{\ket{E\suptiny{0}{0}{(k)}_{i}}\}_{i,k}$, as well as the choices of $U\suptiny{0}{0}{(k)}_{\mathrm{nc}}$ and the ordering of the energies $\tilde{E}_{i,m}$ leave room for adjusting the final energy cost.

%At the same time, unbiased measurements with non-perfect correlations are generally invasive, but the way in which system states are altered through measurement is not entirely fixed by Eq.~(\ref{eq:generalform}). Depending on $\pi[m,k]$, the disturbance can either induce a unital map on the system (leaving the maximally mixed state invariant) if $\pi[m,k]$ is a Latin square, or otherwise create a disturbance with a specific bias. In terms of energy cost, such a bias of invasiveness can even be beneficial, i.e., system changes can be emphasized to favour states with lower energy w.r.t. $H\Sys$. Independently of this, the basis $\{\ket{\tilde{\psi}_{i}\suptiny{0}{0}{(k)}}\}_{i,k}$ of pointer-outcome states in $\mathcal{H}\Poi$
%, with $\Pi_{k}=\sum_{i}\ket{\tilde{\psi}_{i}\suptiny{0}{0}{(k)}}\!\!\bra{\tilde{\psi}_{i}\suptiny{0}{0}{(k)}}$, is related to the energy eigenbasis $\{\ket{E\suptiny{0}{0}{(k)}_{i}}\}_{i,k}$ of $H\Poi$ by an (in principle) arbitrary unitary, allowing for further freedom to adjust the energy of $\tilde{\rho}\SP$.
The exact form of the unbiased and maximally (but not perfectly) correlated lowest energy states depends both on $\rho\Sys$ and requires diagonalization of $H\Poi$. We now present a solution for an example case and refer to Appendix~\ref{app:optimal} for a detailed step-by-step instruction.\\

%%%%%%%%%%%%%%%%%%%%%%%%%%%%%%%%%%%%%%%%%%%%%%%%%%%%%%%%%%%%%%%%%%%%%%%%%%%%%%%%%%%%%%%%%%%%%%%%%%%%%%%%%%%%%%%%%%%%%

\section{Energy cost of %unbiased
measurements}

We now investigate the relation between the energy cost $\Delta E$ of an unbiased measurement achieving the maximum correlation $C_{\mathrm{max}}$ between a system and the pointer.
Here, we wish to showcase different ways of increasing $C_{\mathrm{max}}$, which depends on temperature and dimension. It is readily seen from Eq.~\eqref{eq:universal bound} that $C_{\mathrm{max}}$ increases when the pointer size is increased at fixed temperature or when the initial temperature is lowered at fixed pointer size. While Eq.~(\ref{eq:generalform}) provides a general form for $\tilde{\rho}\SP$, quantitative insight about $\Delta E$ cannot be gained without fixing the pointer Hamiltonian. An exception is when the pointer dimension is infinite. There, the third law of thermodynamics can be circumvented by using (a part of) the pointer as a `fridge' and creating asymptotically pure subspaces (see, e.g.,~\cite{AllahverdyanHovhannisyanJanzingMahler2011}), a scenario which we include as a limiting case in our analysis. For the general case we refer to Appendix~\ref{app:optimal}, but here, as a concrete example (described in more detail in Appendix~\ref{app:qubits}), we will consider a single-qubit system and a pointer consisting of $N$ initially non-interacting qubits with identical Hamiltonians $H\Poi$.

We have %(otherwise one would have to diagonalise the Hamiltonian, which is known to be hard for generic Hamiltonians).
%Furthermore, we assume the correlations between the system and pointer are generated unitarily such that the corresponding energy cost can be expressed without reference to
%external,
%auxiliary systems,
%In this paradigm, the post-measurement state is
%\begin{align}
    $\tilde{\rho}\SP =\,	
    U_{\text{corr}}\,(\rho\Sys\otimes \tau\Poi(\beta)^{\otimes N})\, U_{\text{corr}}^\dagger$,
    %\label{eq:Ucorr}
%\end{align}
where from now on we take $\tau\Poi(\beta) = {1}/{\mathcal{Z}\Poi}(\ket{0}\!\!\bra{0} +e^{-\beta E\Poi} \ket{1}\!\!\bra{1})$ with the partition function $\mathcal{Z}\Poi = \tr[e^{-\beta H\Poi}]$.
Since we would like to increase $C_{\mathrm{max}}$ as much as possible and we know that $C_{\mathrm{max}}$ depends on the initial temperature of the pointer,
%Since we would like to bring $\tilde{\rho}\SP$ as close as possible to the form in~\eqref{eq:postin},
we also consider cooling the pointer prior to the correlating interaction in order to get closer to a faithful result.

\begin{figure}[t]
\begin{center}
\includegraphics[scale=0.4]{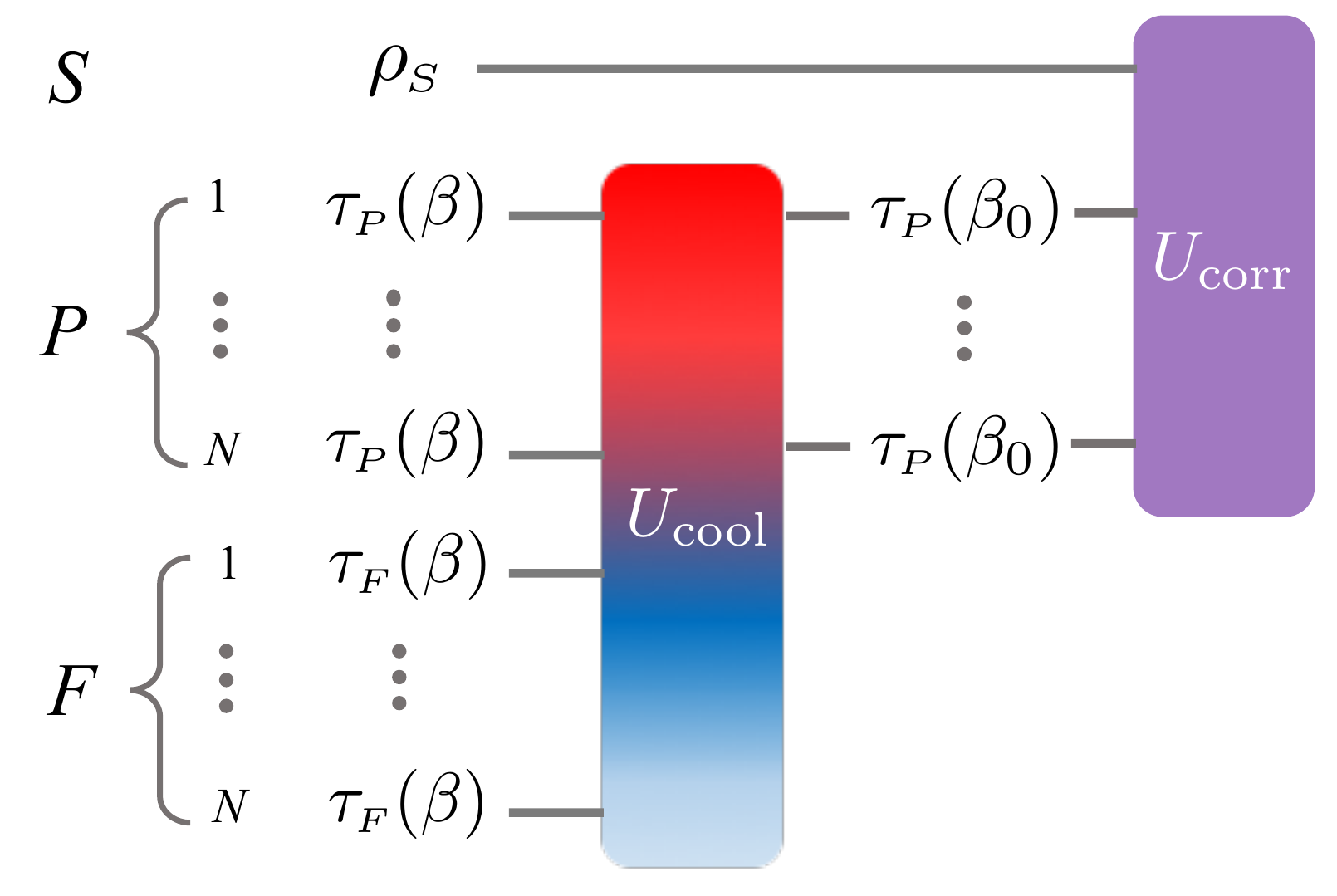}
\vspace*{-2mm}
\caption{The measurement procedure. In step ${\mathrm{I}}$ an $N$-qubit pointer is coupled to an $N$-qubit fridge and cooled from $\beta$ to $\beta_0$. In step ${\mathrm{II}}$, a unitary correlates the pointer with the unknown qubit system.}
\label{fig:setup}
\end{center}
\end{figure}

In principle, there exist many models of refrigeration, e.g.,~\cite{ClivazSilvaHaackBohrBraskBrunnerHuber2019a}. To achieve ground state cooling, however, some form of resource has to diverge. For instance, infinite time is required in adiabatic Landauer erasure~\cite{VinjanampathyAnders2016}, infinite energy in finite size fridges~\cite{ClivazSilvaHaackBohrBraskBrunnerHuber2019a}, or infinite complexity (or time) in fridges of infinite size~\cite{AllahverdyanHovhannisyanJanzingMahler2011}. In short, quantum measurements inherit the limitations imposed by the third law of thermodynamics~\cite{MasanesOppenheim2017, WilmingGallego2017}. Since our example also aims to quantify the energy cost of correlations with increasing pointer size $N$, including fridges of arbitrary size may compromise statements about correlations costs at fixed $N$. We therefore consider a fridge model of the same type and size as the pointer, i.e., for each pointer qubit we add one fridge qubit, see Fig.~\ref{fig:setup}. For larger refrigeration systems the cost of cooling could be decreased by a constant factor (see~\cite{ClivazSilvaHaackBohrBraskBrunnerHuber2019a}), but would still diverge as one approaches zero temperature unless the fridge size is itself infinite.

%\vspace*{-3mm}
%%\noindent

Within this framework, we describe the measurement by two consecutive unitary operations, which we call \textit{cooling} and \textit{correlating}. The total transformation on the system, pointer, and fridge is
%\begin{align}
    $U_{\text{tot}} \,(\rho\Sys\otimes\tau\Poi(\beta)^{\otimes N}
    \otimes \tau\Fr(\beta)^{\otimes N})\,U_{\text{tot}}^\dagger$,
%\end{align}
%%
%\newpage
%%
where $\tau\Fr(\beta) = {1}/{\mathcal{Z}\Fr}(\ket{0}\!\!\bra{0} +e^{-\beta E\Fr} \ket{1}\!\!\bra{1})$ and $U_{\text{tot}} = (U_{corr}\otimes \mathbb{I}\Fr)\cdot (\mathbb{I}\Sys \otimes U_{\text{cool}})$, see Fig.~\ref{fig:setup}.
%%
%%
%It is clear that
Both unitaries drive the respective systems out of equilibrium and come at a thermodynamic cost. Neglecting the price for perfect control over these operations, the work cost of implementing them is lower-bounded by the total energy change of the system, pointer, and fridge, $W\ge \Delta E$.  The total cost in energy can be split into the sum of the two parts: cooling and correlating, which we write $\Delta E = \Delta E\subtiny{0}{0}{\mathrm{I}} +  \Delta E\subtiny{0}{0}{\mathrm{I\hspace*{-0.5pt}I}}$.
%\begin{align}\label{eq:ecost}
%    \Delta E &=\, \Delta E_{\text{cool }} +  \Delta E_{\text{corr }}\,.
%\end{align}
Our objective is to minimise  $\Delta E$ when performing a non-ideal measurement for a fixed value of the correlation function $C(\tilde{\rho}_{\SP}) = C_{\text{max}} < 1$.\\

%%%%%%%%%%%%%%%%%%%%%%%%%%%%%%%%%%%%%%%%%%%%%%%%%%%%%%%%%%%%%%%%%%%%%%%%%%%%%%%%%%%%%%%%%%%%%%%%%%%%%%%%%%%%%%%%%%%%%

\section{Minimal energy cost}

%We proceed to minimise each term in Eq.~\eqref{eq:ecost}.
%In order
To minimise $\Delta E\subtiny{0}{0}{\mathrm{I}}$ we use Ref.~\cite{ClivazSilvaHaackBohrBraskBrunnerHuber2019a}, which details the optimal cost for the single-qubit fridge. Cooling the pointer from $T=1/\beta$ to $E\Poi/(\beta E\Fr)$ such that $\tau\Poi(\beta)^{\otimes N}\mapsto \tau\Poi(\beta \tfrac{E\Fr}{E\Poi})^{\otimes N}$ requires at least
\begin{align}
    \Delta E\subtiny{0}{0}{\mathrm{I}}  &=N(E\Fr-1)  \left(\frac{1}{e^{-\beta E\Fr}+1}-\frac{1}{e^{-\beta E\Poi} +1}\right)\,.
\end{align}
To minimise $\Delta E\subtiny{0}{0}{\mathrm{I\hspace*{-0.5pt}I}}$ we are interested in determining $\min_{U_{\text{corr}}}\Delta E\subtiny{0}{0}{\mathrm{I\hspace*{-0.5pt}I}}$ such that $C(\tilde{\rho}_{\SP}) = C_{\text{max}} (\beta)$.
%\begin{align}\label{eq:optimise}
%%\begin{split}
%    \min_{U_{\text{corr}}}\Delta E_{\text{corr}} \ \ \text{s.t.} \ \ C(\tilde{\rho}_{\SP}) = C_{\text{max}} (\beta)
%\end{align}
For the case of a single-qubit system and $N$-qubit pointer (with $N$ odd), we have
%\footnote{Here we take $N$ to be odd. For $N$ even the formula is slightly different with the same qualitative behaviour.}
\begin{align}
    C_{\text{max}}(\beta) %&=
    %\max_{\substack{U_{\text{corr}}}} \;\;
   	%C\bigl(U_{\text{corr}}[
   	%\rho\Sys\otimes \tau(\beta)^{\otimes N}] U^\dagger_{\text{corr}}\bigr)
  	%\nonumber\\
    = \;\frac{1}{\mathcal{Z}^N}\sum\limits_{k=0}^{N/2} {{N}\choose{k}} e^{-kE\Poi\beta}\,.
    \label{eq:maxcor}
\end{align}
For even $N$, the formula is slightly different with the same qualitative behaviour.
As expected, in the limit of infinite pointer size ($N\rightarrow\infty$) for fixed $\beta$, or in the limit of zero temperature ($\beta\rightarrow\infty$) for any $N$, the correlations become perfect,
%\footnote{Note that this function is independent of the initial state of the system, a fact that is due to restricting to unbiased measurements.},
$\lim_{N\rightarrow\infty}C_{\text{max}}(\beta)=\lim_{\beta\rightarrow\infty}C_{\text{max}}(\beta)=1$.
%Since we assume unbiased measurements and we recover faithfulness in this limit, the measurement is also non-invasive, see Fig.~\ref{fig:triangle}.
In Appendix~\ref{app:qubits}, we construct the optimal unitary $U_{\text{opt}}$ that solves the optimisation problem for $\Delta E\subtiny{0}{0}{\mathrm{I\hspace*{-0.5pt}I}}$ for arbitrary $N$ and $\beta$,
%in~\eqref{eq:optimise}
i.e., the unitary that achieves the algebraic maximum correlation for minimal energy cost.
In particular, this construction allows us to specify an analytic expression for $\Delta E\subtiny{0}{0}{\mathrm{I\hspace*{-0.5pt}I}}$ in terms of $\beta$, $N$, and
%\footnote{$\Delta E_{\text{corr}}$ also depends on $E\Sys$, the gap of the system, which plays a role in the case that the initial state $\rho\Sys$ is known.}
$\rho\Sys$, which implies an achievable lower bound on the minimal energy cost of non-ideal measurements that approximate ideal ones as well as possible.
%Appendix~\ref{app:optimal}.

\begin{figure}[t!]
\begin{center}
\includegraphics[width=0.47\textwidth]{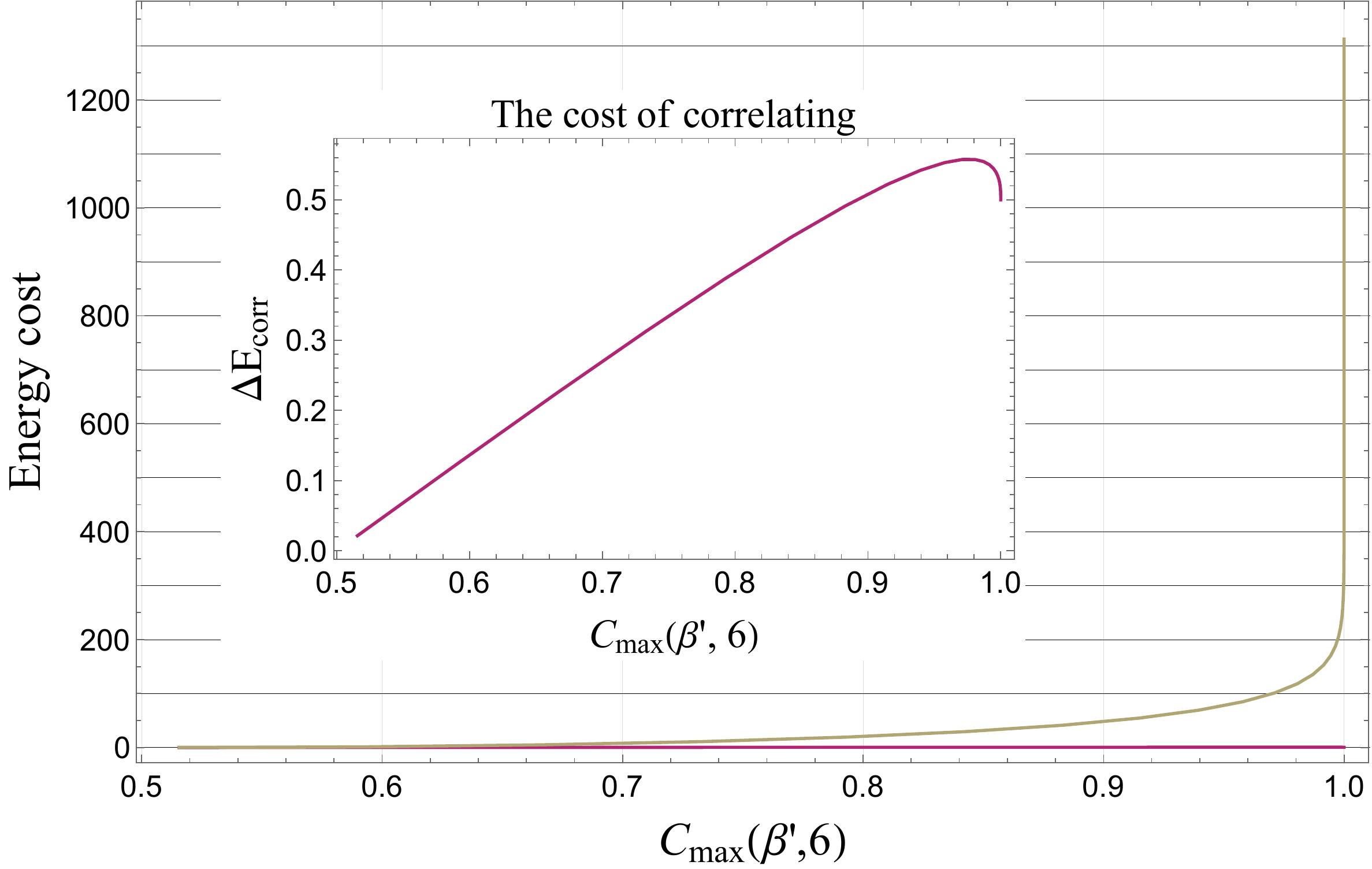}
\caption{Cost of a non-ideal projective measurement of a qubit system using a $6$-qubit pointer. Each pointer qubit is in the state $\tau\Poi(\beta)$.
%has energy gap $E\Poi$ and starts from inverse temperature $\beta$.
We start from room temperature ($\approx300$~K) and choose an energy gap in the microwave regime such that $\beta E\Poi=1/30$. Each point on the horizontal indicates the maximal algebraic correlation $C_{\text{max}}(\beta\pr, N=6)$ achievable for a fixed value $\beta\pr=\frac{\beta E\Fr}{E\Poi}$ (or equivalently, fixed $E\Fr$), which is the result of cooling the $6$-qubit pointer from $\beta$ to $\beta\pr$ using refrigeration qubits with gaps $E\Fr$. For each correlation value, the refrigeration cost $\Delta E\subtiny{0}{0}{\mathrm{I}}$ and the cost $\Delta E\subtiny{0}{0}{\mathrm{I\hspace*{-0.5pt}I}}$ of maximally correlating the thermal state at inverse temperature $\beta\pr$ are shown. The inset shows the relevant energy scale for correlating the system and pointer since the cooling cost significantly dominates the correlating cost.}
\label{fig:costplot}
\end{center}
\end{figure}

Note that in the limit $N\rightarrow\infty$ the energy cost of achieving $C_{\text{max}}(\beta)$ is finite but infinite time (or full control over $N$-body interactions with $N\rightarrow\infty$) is required (see Appendix~\ref{app:qubits}). For any finite $N$, the only way to achieve correlations higher than $C_{\text{max}}(\beta)$ is to cool the pointer. Thus, we consider the scenario where starting at some finite $\beta$, we cool the pointer ($\beta \rightarrow \beta_0>\beta$) and then correlate it with the system to the algebraic maximum for the new temperature $C_{\text{max}} (\beta_0)$. Results for exemplary temperatures are shown in Fig.~\ref{fig:costplot} for $N=6$. Within our cooling paradigm, the energy cost for reaching the ground state in finite time is infinite. Other paradigms allow cooling to the ground state using finite energy, but require infinite time~\cite{AllahverdyanHovhannisyanJanzingMahler2011}.
%For fixed $N$ and finite temperature, achieving perfect correlations is only possible if infinite resources are invested in cooling the pointer to the ground state.
Thus, without access to pure states, a measurement satisfying properties \eqref{def:unbiased} -- \eqref{def:noninv} has an infinite resource cost. The  cost for the maximally correlating unitary $U_{\text{corr}}$ is always finite and given by $\Delta E\subtiny{0}{0}{\mathrm{I\hspace*{-0.5pt}I}}^{\,(C = 1)} = \tfrac{1}{2} E\Poi$.

%\noindent

%%%%%%%%%%%%%%%%%%%%%%%%%%%%%%%%%%%%%%%%%%%%%%%%%%%%%%%%%%%%%%%%%%%%%%%%%%%%%%%%%%%%%%%%%%%%%%%%%%%%%%%%%%%%%%%%%%%%%

\section{Discussion}

The projection postulate is a central concept within the foundations of quantum mechanics, asserting that ideal projective measurements leave the system in a pure state corresponding to the observed outcome. All interpretations of quantum mechanics must be compatible with this statement together with the Born rule assigning the probabilities. However, the existence of such `true' projections is usually taken for granted. Here, we have discussed a self-contained description of measurements from a thermodynamic point of view. We have shown that, when their existence is not assumed, ideal finite-time projective measurements have an infinite cost.

We argued that a necessary condition for ideal measurements
%besides being unbiased and non-invasive,
is to be faithful, i.e., have perfect correlation between the system and pointer. However, this requirement incurs infinite costs unless pure states are freely available. Nonetheless, ideal measurements can be approximated by non-ideal, unbiased ones to arbitrary precision at finite energy cost.  We find that the correlation achieved by the best unbiased measurement is universally bounded by the largest $d\Poi/d\Sys$ eigenvalues of the pointer. To gain quantitative insight into this cost, we considered the measurement of a single qubit by an $N$-qubit pointer. We provided analytic expressions for the minimal energy cost for unitarily achieving maximal correlation for any initial temperature and any $N$. We find that correlations can be increased by increasing the pointer size and by cooling the pointer.

While the three mentioned properties capture the basic features of ideal measurements, they are not sufficient to characterise the `quantum to classical' transition. Classical outcomes additionally feature `robustness', where small perturbations of the pointer do not significantly alter the observed outcomes. This is an important consideration for broadcasting measurement outcomes. In our qubit model, this is taken into account by the size of the subspaces the system is correlated with, i.e., the number of pointer particles,~$N$.

%Apart from the relevance to quantum foundations, t
The insight that ideal measurements carry a diverging cost also sheds light on thermodynamics because it implies that, in practice, all measurements that can be performed are intrinsically non-ideal. Consequently, a central question is how well one can approximate ideal measurements and which consequences these approximate realizations have on tasks that arise in quantum thermodynamics. For instance, to interpret work as a random variable in the quantum regime, two projective measurements are commonly assumed to characterize work~\cite{Crooks1999, Tasaki2000, TalknerLutzHaenggi2007}. The impossibility of these measurements with finite work, prompts two questions (recently studied in~\cite{DebarbaEtAl2019}): (a) what is the impact of measurement imperfection on the observed fluctuations and (b) what is the total work cost of observing work fluctuations imperfectly. Furthermore, in quantum information-based engines~\cite{Vidrighin2016, ElouardJordan2018, AydinSismanKosloff2019}, it would be highly relevant to incorporate measurement imperfection and work cost into the efficiency in addition to other constraints~\cite{MohammadyAnders2017}.
% of the engine.
These insights could also be useful
for quantum information processing, e.g., bounding the minimal power consumption of quantum computers employing syndrome measurements for error correction.
% or in measurement-based architectures.
%fault-tolerant quantum computer using measurment-based architectures.

%%%%%%%%%%%%%%%%%%%%%%%%%%%%%%%%%%%%%%%%%%%%%%%%%%%%%%%%%%%%%%%%%%%%%%%%%%%%%%%%%%%%%%%%%%%%%%%%%%%%%%%%%%%%%%%%%%%%%

%\noindent \textit{Acknowledgments}.
\vspace*{-1.5mm}
\begin{acknowledgments}
\vspace*{-1.5mm}
We are grateful to Tiago Debarba and Karen Hovhannisyan for valuable comments and insights.
%We thank the editors of Science Advances for alerting us to the ``epty squares" at the end of the proofs.
We acknowledge support by the EU COST Action MP1209 ``Thermodynamics in the quantum regime"
and from the Austrian Science Fund (FWF) through the START project Y879-N27, the project P 31339-N27, and the joint Czech-Austrian project MultiQUEST (I 3053-N27 and GF17-33780L). 
YG acknowledges support by the Austrian Science Fund (FWF) through the Zukunftskolleg ZK03.
\end{acknowledgments}

%%%%%%%%%%%%%%%%%%%%%%%%%%%%%%%%%%%%%%%%%%%%%%%%%%%%%%%%%%%%%%%%%%%%%%%%%%%%%%%%%%%%%%%%%%%%%%%%%%%%%%%%%%%%%%%%%%%%%

\bibliographystyle{apsrev4-1fixed_with_article_titles_full_names}
\bibliography{bibfile}

%%%%%%%%%%%%%%%%%%%%%%%%%%%%%%%%%%%%%%%%%%%%%%%%%%%%%%%%%%%%%%%%%%%%%%%%%%%%%%%%%%%%%%%%%%%%%%%%%%%%

%\clearpage

%%%%%%%%%%%%%%%%%%%%%%%%%%%%%%%%%%%%%%%%%%%%%%%%%%%%%%%%%%%%%%%%%%%%%%%%%%%%%%%%%%%%%%%%%%%%%%%%%%%%

%\newpage
\hypertarget{sec:appendix}
\appendix

\section*{Appendix: Mathematical Model for Measurement Procedures}

\renewcommand{\thesubsubsection}{A.\Roman{subsection}.\arabic{subsubsection}}
\renewcommand{\thesubsection}{A.\Roman{subsection}}
\renewcommand{\thesection}{A}
\setcounter{equation}{0}
\numberwithin{equation}{section}
\setcounter{figure}{0}
\renewcommand{\thefigure}{A.\arabic{figure}}

In this Appendix, we give a detailed description of \emph{unbiased} measurement procedures introduced in the main text. As we have argued, ideal measurements (unbiased, faithful, and non-invasive) are not generally implementable in finite time or with finite energy. In practice, real measurements may nonetheless approximate ideal measurements by investing energy; loosely speaking, the approximation becomes better, the more energy that is invested. To make this more precise, we will explicitly determine the fundamental energy cost of projective\footnote{Arbitrary quantum measurements represented by positive-operator-valued measures (POVMs) can be realized by projective measurements on a Hilbert space obtained by appending an auxiliary system of, at most~\protect\cite{OszmaniecGueriniWittekAcin2017}, the same dimension as the original system. We therefore concentrate on projective measurements.} measurements.
%
%This requires a specific microscopic model for the measurement procedure, during which a quantum system $S$ to be measured interacts with a pointer system $P$ (the measurement device). In Sec.~\ref{sec:framework}, we formally define the mathematical models used for the system, pointer, and measurement procedure.
%
%\vspace*{-2mm}
\subsection{Framework}\label{sec:framework}
%\vspace*{-2mm}

\textbf{System.}\ We consider a quantum system $S$ with Hilbert space $\mathcal{H}\Sys$ of dimension $d\Sys=\dim(\mathcal{H}\Sys)$ initially in an arbitrary unknown quantum state represented by a density operator $\rho\Sys\in\mathcal{L}(\mathcal{H}\Sys)$, i.e., a Hermitian operator with $\tr(\rho\Sys)=1$ in the space of linear operators $\mathcal{L}(\mathcal{H}\Sys)$ over the system Hilbert space $\mathcal{H}\Sys$. %\equiv\mathbb{C}^{2}$.
We are then interested in describing (projective) measurements of the system w.r.t. a basis $\{\ket{i}\Sys\}_{i}$ of $\mathcal{H}\Sys$, which we take to be the eigenbasis of the system Hamiltonian $H\Sys$, i.e., we can write $H\Sys=\hbar\sum_{i}\Omega_{i}\ket{i}\!\!\bra{i}\Sys$, where $\hbar(\Omega_{j}-\Omega_{i})$ is the energy gap between the $i$-th and $j$-th levels. For instance, an example that we will focus on later is that of the simplest quantum system \textemdash\ a qubit \textemdash\ with vanishing ground state energy and energy gap $E\Sys=\hbar\Omega$. That is, $\mathcal{H}\Sys=\mathbb{C}^{2}$, and the system Hamiltonian $H\Sys$ has eigenstates $\ket{0}\Sys$ and $\ket{1}\Sys$ and spectral decomposition $H\Sys=E\Sys\ket{1}\!\!\bra{1}\Sys$.\\

\vspace*{-1mm}
\textbf{Pointer.}\ Similarly, we consider a pointer system $P$ with Hilbert space $\mathcal{H}\Poi$ of dimension $d\Poi=\dim(\mathcal{H}\Poi)$ and Hamiltonian $H\Poi$. We then take the resource-theoretic point of view that the pointer is initially in a state
%$\rho\Poi\in\mathcal{L}(\mathcal{H}\Poi)$
that is freely available, i.e., a thermal state $\tau\Poi(\beta)\in\mathcal{L}(\mathcal{H}\Poi)$ at ambient temperature $T=1/\beta$.
We order the spectrum of the pointer Hamiltonian in terms of its excitations into $d\Sys$ sectors of size $d\Poi/d\Sys$, i.e., $H\Poi=\sum_{k=0}^{d\Sys-1}\sum_{i=0}^{d\Poi/d\Sys -1}E\suptiny{0}{0}{(k)}_{i}\ket{E\suptiny{0}{0}{(k)}_{i}}\!\!\bra{E\suptiny{0}{0}{(k)}_{i}}$ with $E\suptiny{0}{0}{(k)}_{i}\leq E\suptiny{0}{0}{(k\pr)}_{j}\ \forall i,j$ for $k\pr>k$.
%
%
%where for arbitrary inverse temperatures $\beta$, and Hamiltonians $H\Poi$ with eigenbasis $\{\ket{\psi_{n}}\}_{n}$
%and corresponding eigenvalues $E_{n}$ such that $H\Poi\ket{\psi_{n}}\Poi=E_{n}\ket{\psi_{n}}\Poi$, the thermal (Gibbs) state is given by
%\begin{align}
%    \tau(\beta) &=\,\mathcal{Z}\Poi^{-1}e^{-\beta H\Poi}\,=\,\mathcal{Z}\Poi^{-1}\sum\limits_{n}e^{-E_{n}\beta}\ket{\psi_{n}}\!\!\bra{\psi_{n}}\Poi\,.
%\end{align}
The thermal (Gibbs) state is given by
\begin{align}\label{eq:diagonalise}
%\begin{split}
\tau\Poi(\beta) &= \sum_{k=0}^{d\Sys-1}\sum_{i=0}^{d\Poi/d\Sys -1} p\suptiny{0}{0}{(k)}_{i} \ket{E\suptiny{0}{0}{(k)}_{i}}\!\!\bra{E\suptiny{0}{0}{(k)}_{i}}
%\\
%\rho\Sys &= \sum_{i}^{} \rho_{ii} \ket{i}\bra{i}
%\end{split}
\end{align}
where $p\suptiny{0}{0}{(k)}_{i} = \exp\bigl(-\beta E\suptiny{0}{0}{(k)}_{i}\bigr)/\mathcal{Z}$ and $\mathcal{Z}$ is the pointer's partition function $\mathcal{Z} =\tr(e^{-\beta H\Poi})=\sum_{i,k}e^{-\beta E\suptiny{0}{0}{(k)}_{i}}$. \\
%%As an example that will also be of relevance later on, one could take the pointer system to be a collection of $N$ qubits with Hamiltonian $H\Poi=\sum_{i=1}^{N}E\Poi\ket{1}\!\!\bra{1}_{i}\in\mathcal{L}(\mathcal{H}\Poi)$, where $\mathcal{H}\Poi\equiv(\mathbb{C}^{2})^{\otimes N}$, and the subscript $i$ indicates an operator acting nontrivially only on the $i$-th (pointer) qubit, i.e., $\mathcal{O}_{i}\equiv\mathds{11}_{1}\otimes\ldots\otimes\mathds{1}_{i-1}\otimes \mathcal{O}\otimes\mathds{1}_{i+1}\otimes\ldots\otimes\mathds{1}_{N}$, while $\ket{0}_{i}$ and $\ket{1}_{i}$ denote the ground and excited states of the $i$-th qubit, respectively. Here, we have further assumed that all of these pointer qubits have the same energy gap $\omega$ for simplicity.\\

%\vspace*{-1mm}
\textbf{Measurement procedure.}\ We now wish to consider a measurement of the system's energy, i.e., of the observable $H\Sys$, or, in other words, a projective measurement of the system in the energy eigenbasis\footnote{Note that a projective measurement of the system in any other (orthonormal) basis can be subsumed into this discussion by including an additional unitary transformation (and its energy cost) on the initial state $\rho\Sys$ to switch between the energy eigenbasis and the desired measurement basis.}. The corresponding \emph{measurement procedure} may be defined via a completely positive and trace-preserving (CPTP) map $\mathcal{E}:\mathcal{L}(\mathcal{H}\SP)\rightarrow\mathcal{L}(\mathcal{H}\SP)$, where $\mathcal{H}\SP=\mathcal{H}\Sys\otimes\mathcal{H}\Poi$, that maps $\rho\SP=\rho\Sys\otimes\tau\Poi$ to a post-measurement state $\tilde{\rho}\SP\in\mathcal{L}(\mathcal{H}\SP)$. This may be understood as a generalized interaction between the system, the pointer, and some auxiliary system. Here, we do not wish to address the question of which measurement outcome is ultimately realized (which pure state the system is left in), or how and why this may be the case. That is, we do not attempt to make statements about what is often perceived as the ``measurement problem", but rather take the point of view that system and pointer are left in a joint state in which the internal states of the system are correlated with the internal states of the pointer. Each of the latter is designated to correspond to one of the system states $\ket{i}\Sys$, such that, upon finding the pointer in a state chosen to reflect $\ket{i}\Sys$, one concludes that the measurement outcome is ``$i$". More precisely, each eigenstate $\ket{i}\Sys$ of the measured observable of the system is assigned a set $\{\ket{\tilde{\psi}_{n}\suptiny{0}{0}{(i)}}\Poi\}_{n}$ of orthogonal states of the pointer corresponding to a projector $\Pi_{i}=\sum_{n} \ket{\tilde{\psi}_{n}\suptiny{0}{0}{(i)}}\!\!\bra{\tilde{\psi}_{n}\suptiny{0}{0}{(i)}}$. The projectors are chosen to be orthogonal and to form a complete set, i.e., $\Pi_{i}\Pi_{j}=\delta_{ij}\Pi_{i}$ and $\sum_{i}\Pi_{i}=\mathds{1}\Poi$. In an ideal measurement, upon obtaining the outcome ``$i$", one may further conclude that the post-measurement \emph{system} is left in the state $\ket{i}\Sys$. This is one of three features that can be identified for ideal projective measurements. As explained in the main text, ideal measurements are\\

\begin{tcolorbox}[colback=mycolor2!5,colframe=mycolor2,fonttitle=\bfseries,
left=2pt,
right=2pt,
top=0pt,
bottom=3pt]
\vspace*{1.5mm}
\textbf{unbiased}:
\begin{align}
    \tr\bigl[\mathds{1}\Sys\otimes\Pi_{i}\,\tilde{\rho}\SP\bigr]
    &=\tr\bigl[\ket{i}\!\!\bra{i}\Sys\rho\Sys]=\rho_{ii}\,\forall\,i,
    \label{def:unbiased appendix}
\end{align}
\textbf{faithful}:
\begin{align}
    C(\tilde{\rho}\SP)  &:=\sum_{i} \tr\bigl[\ket{i}\!\!\bra{i}\Sys\otimes \Pi_{i} \;\tilde{\rho}\SP\bigr]=1,
    \label{def:faithfulness appendix}
\end{align}
\textbf{non-invasive}:
\begin{align}
    \tr\bigl[\ket{i}\!\!\bra{i}\Sys\otimes \mathds{1}\Poi\ \tilde{\rho}\SP\bigr]
    =\tr\bigl[\ket{i}\!\!\bra{i}\Sys\rho\Sys\bigr]=\rho_{ii}\,\forall\,i.
    \label{def:noninv appendix}
\end{align}
\end{tcolorbox}
%where $\tilde{\rho}\Poi=\tr\Sys(\tilde{\rho}\SP)$ is the post-measurement state of the pointer.

\subsection{Example: 2-Qubit Measurements}\label{sec:Examples Two-Qubit Measurement Procedures}
%\vspace*{-2mm}

To illustrate the properties above and to understand why these conditions are not met by general non-ideal measurement procedures with finite energy input, we consider a simple example. Consider a measurement procedure using a single pointer qubit and assume that by some means it has been prepared in the ground state, i.e., $\rho\Poi=\ket{0}\!\!\bra{0}\Poi$. We model the interaction with the pointer by applying a controlled NOT operation $U\subtiny{0}{0}{\mathrm{CNOT}}=\ket{0}\!\!\bra{0}\Sys\otimes\mathds{1}\Poi+\ket{1}\!\!\bra{1}\Sys\otimes X\Poi$, with the usual Pauli operator $X=\ket{0}\!\!\bra{1}+\ket{1}\!\!\bra{0}$. Denoting the matrix elements of the initial state as $\rho_{ij}=\bra{i}\rho\Sys\ket{j}$, we can then write the post-measurement state $\tilde{\rho}\SP$ as
\begin{align}
    \tilde{\rho}\SP &=\,U\subtiny{0}{-1}{\mathrm{CNOT}}\,\rho\SP U\subtiny{0}{0}{\mathrm{CNOT}}^{\dagger}\,=\,\sum_{i,j=0,1}\rho_{ij}\ket{ii}\!\!\bra{jj}\,.
\end{align}
The system and pointer are now perfectly (classically) correlated in the sense that whenever the pointer is found in the state $\ket{0}\Poi$ ($\ket{1}\Poi$), the system is left in the corresponding state $\ket{0}\Sys$ ($\ket{1}\Sys$). In other words, for the choice $\Pi_{i}=\ket{i}\!\!\bra{i}\Poi$, we find that the measurement is faithful,
\begin{align}
    C(\tilde{\rho}\SP)
    	 &=\,\sum\limits_{i=0,1}
    	 	\tr\bigl[
    	 		\ket{ii}\!\!\bra{ii}\tilde{\rho}\SP\bigr]\,=
    	\,\sum\limits_{i=0,1}\rho_{ii}\,
    	=\,\tr[
    	\rho\Sys]\,=\,1\,.
    \label{eq:correlation function example}
\end{align}
The post-measurement system state $\tilde{\rho}\Sys=
	\tr\Poi[
		\tilde{\rho}\SP]
			=\sum_{i}\rho_{ii}\ket{i}\!\!\bra{i}\Sys$ is in general different from the initial system state since it not longer has any off-diagonal elements w.r.t. the measurement basis, but the measurement is nonetheless non-invasive since the diagonal elements match those of the initial system state $\rho\Sys$. At the same time, the chosen unitary $U\subtiny{0}{0}{\mathrm{CNOT}}$ guarantees that the probabilities for finding the pointer in the states $\ket{0}\Poi$ and $\ket{1}\Poi$, match those of the original system state, i.e., for $i=0,1$ we have
\begin{align}
    \tr\bigl(\ket{i}\!\!\bra{i}\Poi\tr\Sys(\tilde{\rho}\SP)\bigr)   &=\,\tr(\ket{i}\!\!\bra{i}\Sys\rho\Sys)\,=\,\rho_{ii}\,.
    \label{eq:example unbiasedenss condition}
    \vspace*{-2mm}
\end{align}
Consequently, the measurement is not biased towards one of the outcomes and reproduces the statistics of the original system state, while being perfectly correlated (i.e., faithful).

However, in general strong correlation and unbiasedness of the measurement do not imply one another. For instance, one can construct an unbiased but also generally uncorrelated measurement by replacing $U\subtiny{0}{0}{\mathrm{CNOT}}$ with $U\subtiny{0}{0}{\mathrm{SWA\hspace*{-0.7pt}P}}=\ket{00}\!\!\bra{00}+\ket{01}\!\!\bra{10}+\ket{10}\!\!\bra{01}+\ket{11}\!\!\bra{11}$, leaving the system in the state $\ket{0}\Sys$ no matter which state the pointer is in. Although all available information about the pre-measurement system is thus stored in the pointer, measuring the latter reveals no (additional) information about the post-measurement system. %In some sense this case is hence pathological, since the question of the energy cost of the projective measurement is just relayed from the original system to the pointer.
Alternatively, consider the unitary $\ket{1}\!\!\bra{1}\Sys\otimes\mathds{1}\Poi+\ket{0}\!\!\bra{0}\Sys\otimes X\Poi$ instead of $U\subtiny{0}{0}{\mathrm{CNOT}}$, both of which lead to the same correlation $C(\tilde{\rho}\SP)$, but the probabilities for observing the two outcomes are now exchanged w.r.t. to $\rho\Sys$, i.e., the pointer is found in the state $\ket{0}\Poi$ ($\ket{1}\Poi$) with probability $\rho_{11}$ ($\rho_{00}$) after the interaction.

For the purpose of examining real measurements, these examples are of course pathological due to the assumption of reliably preparing the pointer in a pure state (without having to have performed a projective measurement in order to model a projective measurement or having to cool to the ground state using finite resources~\cite{MasanesOppenheim2017}). Let us therefore relax this assumption and assume instead that the pointer has been prepared at some finite non-vanishing temperature $T=1/\beta$, such that $\rho\Poi=p\ket{0}\!\!\bra{0}\Poi+(1-p)\ket{1}\!\!\bra{1}\Poi$ for some $p=(1+e^{-\omega\beta})^{-1}=\mathcal{Z}^{-1}$ with $0<p<1$. A quick calculation then reveals that the previously perfect correlations are reduced to $C(\tilde{\rho}\SP)=p=\mathcal{Z}^{-1}<1$ and that the measurement procedure using $U\subtiny{0}{0}{\mathrm{CNOT}}$ is in general biased, i.e.,
\begin{align}
    \tr\bigl[
    	\ket{i}\!\!\bra{i}\Poi\tr\Sys[
    		\tilde{\rho}\SP]
    		\bigr]   &=\,\rho_{ii}(2p-1)+1-p\,.
    \label{eq:example}
\end{align}
However, while we generally have to give up the notion of a perfect projective measurement in the sense that the outcomes are perfectly correlated with the post-measurement states (as we have shown in the main text), one may retain the unbiasedness of the measurement. That is, if we replace  $U\subtiny{0}{0}{\mathrm{CNOT}}$ by $U\subtiny{0}{0}{\mathrm{unb.}}=\ket{00}\!\!\bra{00}+\ket{01}\!\!\bra{11}+\ket{11}\!\!\bra{10}+\ket{10}\!\!\bra{01}$, we obtain the same imperfect correlation value $C(\tilde{\rho}\SP)=p=\mathcal{Z}^{-1}$, but the unbiasedness condition of Eq.~(\ref{eq:example unbiasedenss condition}) is satisfied. To reiterate, measurement procedures using finite resources (finite time, finite energy, and finite complexity, i.e., operations with finite interaction range), cannot be ideal, since finite resources are not sufficient to prepare pointers in the required pure states. Realistic measurement procedures hence are non-ideal.

%%%%%%%%%%%%%%%%%%%%%%%%%%%%%%%%%%%%%%%%%%%%%%%%%%%%%%%%%%%%%%%%%%%%%%%%%%%%%%%%%%%%%%%%%%%%

\subsection{Non-Ideal Measurement Procedures}\label{sec:nonid meas procedures}
%\vspace*{-2mm}

When any one of the three properties (\ref{def:unbiased appendix}), (\ref{def:faithfulness appendix}) or (\ref{def:noninv appendix}) fails to hold, we call the corresponding measurement procedure \emph{non-ideal}. For non-ideal measurements, the relation between the remaining properties is more complicated. In particular, none of the three properties alone implies any of the other two. For instance, consider an ideal post-interaction state
\begin{align}
    \tilde{\rho}\SP &=\,
    \sum_{i}\rho_{ii}\,\ket{i}\!\!\bra{i}\otimes\rho\suptiny{0}{0}{(i)}
    +\,\text{off-diag.}\,,
    \label{eq:postin appendix}
    \vspace*{-2mm}
\end{align}
where we have not explicitly written the off-diagonal elements w.r.t. the basis $\{\ket{i}\otimes\ket{\tilde{\psi}_{n}\suptiny{0}{0}{(j)}}\}_{i,j,n}$ and $\rho\suptiny{0}{0}{(i)}$ is a pointer state that is associated to one and only one of the outcomes $i$, that is, $\Pi_{i}\rho\suptiny{0}{0}{(j)}=\delta_{ij}\rho\suptiny{0}{0}{(j)}$. Any measurement procedure based on a map $\mathcal{E}$ for which the values $\rho_{ii}$ in~(\ref{eq:postin appendix}) are replaced with arbitrary probabilities $p_{i}\neq\rho_{ii}$, results in a joint post-interaction state $\tilde{\rho}\SP$ satisfying (\ref{def:faithfulness appendix}), but not (\ref{def:unbiased appendix}) or (\ref{def:noninv appendix}), resulting in a non-ideal measurement that is faithful, but neither unbiased or non-invasive. Similarly, the state $\tilde{\rho}\SP=\rho\Sys\otimes \rho\Poi$ obtained from a trivial interaction $\mathcal{E}[\rho\SP]=\rho\SP$ complies with (\ref{def:noninv appendix}), but not with (\ref{def:faithfulness appendix}) or (\ref{def:unbiased appendix}). Finally, the map $\mathcal{E}$ realizing a complete exchange of the initial system and pointer states (assuming, for the purpose of this example that $d\Sys=d\Poi$), results in an unbiased~(\ref{def:unbiased appendix}) measurement procedure that does not satisfy either (\ref{def:faithfulness appendix}) or (\ref{def:noninv appendix}).\\

% and, as we will discuss in more detail later, unbiased measurements are generally neither faithful nor non-invasive.

Satisfying any single one of the three properties is hence not sufficient for distinguishing ideal from non-ideal measurements. When two out of the three properties hold, things become more subtle. In two cases, a joint final state $\tilde{\rho}\SP$ satisfying a \textit{pair} of properties implies the third property, and hence that the measurement is ideal for the particular given initial system state $\rho\Sys$. First, a measurement that is both faithful (\ref{def:faithfulness appendix}) \textit{and} unbiased (\ref{def:unbiased appendix}) implies that it is also non-invasive (\ref{def:noninv appendix}). To show this, we start with the property of unbiasedness and, summing the right-hand side of Eq.~(\ref{def:unbiased appendix}) over all $i$, we have $\sum_{i}\rho_{ii}=1$. The left-hand side of Eq.~(\ref{def:unbiased appendix}) thus gives
\begin{align}
    \sum_{i}\tr\bigl[\mathds{1}\Sys\otimes\Pi_{i}\,\tilde{\rho}\SP\bigr] &=\,
    \sum_{i,j}\tr\bigl[\ket{j}\!\!\bra{j}\Sys\otimes\Pi_{i}\,\tilde{\rho}\SP\bigr]
    \,=\,1.
    \label{eq:properites proof 1}
\end{align}
At the same time, property (\ref{def:faithfulness appendix}) demands that the sum in the second step of~(\ref{eq:properites proof 1}) yields $1$ already just for the terms where $i=j$, implying
\begin{align}
    \sum_{i\neq j} \tr\bigl[\ket{j}\!\!\bra{j}\Sys\otimes \Pi_{i} \;\tilde{\rho}\SP\bigr] &=\,0.
\end{align}
Since all diagonal matrix elements of a density operator are non-negative, this further implies $\tr\bigl[\ket{j}\!\!\bra{j}\Sys\otimes \Pi_{i} \;\tilde{\rho}\SP\bigr]=0$\ $\forall i\neq j$, which we can insert back into~(\ref{def:unbiased appendix}) to see that $\tr\bigl[\ket{i}\!\!\bra{i}\Sys\otimes \Pi_{i}\, \tilde{\rho}\SP\bigr]\,=\,\rho_{ii}$. Inserting all of this into the left-hand side of Eq.~(\ref{def:noninv appendix}) together with $\mathds{1}\Poi=\sum_{j}\Pi_{j}$, we obtain
\begin{align}
    \tr\bigl[\ket{i}\!\!\bra{i}\Sys\otimes \sum_{j}\Pi_{j}\, \tilde{\rho}\SP\bigr]
    &=\,\tr\bigl[\ket{i}\!\!\bra{i}\Sys\otimes \Pi_{i}\, \tilde{\rho}\SP\bigr]\,=\,\rho_{ii},
\end{align}
which concludes the proof that unbiased and faithful measurements are also non-invasive.

%\begin{align}
%\begin{split}
%\sum_i\tr[\sum_j &\ket{j}\bra{j}\otimes \Pi_i \tilde{\rho}\SP] = \sum_i\rho_{ii} \\
%&= 1\\
%&=\sum_i \tr[\ket{i}\bra{i}\otimes \Pi_i
%] \\
%\Longrightarrow \quad&
%\sum_{i\neq j\,i,j}\tr[\ket{j}\bra{j}\otimes \Pi_i \tilde{\rho}\SP]  = 0\\
%\Longrightarrow \quad &
%\tr[\ket{j}\bra{j}\otimes \Pi_i \tilde{\rho}\SP] = 0 \qquad \forall \; i \neq \,j\\
%\Longrightarrow \quad &
%\tr[\ket{i}\bra{i}\otimes \sum_j\Pi_j \tilde{\rho}\SP]  = \tr[\ket{i}\bra{i}\otimes \Pi_i \tilde{\rho}\SP] \\
%&\qquad\qquad\qquad\qquad\quad =\rho_{ii} \quad \forall \,i
%\end{split}
%\end{align}

Second, a measurement that is both faithful (\ref{def:faithfulness appendix}) \textit{and} non-invasive (\ref{def:noninv appendix}) is also unbiased (\ref{def:unbiased appendix}). Now starting with (\ref{def:noninv appendix}), we again sum the left-hand side over all $i$ and resolve the identity $\mathds{1}\Poi=\sum_{j}\Pi_{j}$, to obtain
\begin{align}
    \sum_{i,j}\tr\bigl[\ket{i}\!\!\bra{i}\Sys\otimes\Pi_{j}\,\tilde{\rho}\SP\bigr] &=\,
    \sum_{i}\rho_{ii}
    \,=\,1.
    \label{eq:properites proof 2}
\end{align}
This time, unbiasedness~(\ref{def:unbiased appendix}) implies
\begin{align}
    \sum_{i\neq j} \tr\bigl[\ket{i}\!\!\bra{i}\Sys\otimes \Pi_{j} \;\tilde{\rho}\SP\bigr] &=\,0,
\end{align}
and in turn $\tr\bigl[\ket{i}\!\!\bra{i}\Sys\otimes \Pi_{j} \;\tilde{\rho}\SP\bigr]=0$\ $\forall i\neq j$, as before. Inserting this into~(\ref{def:noninv appendix}) then implies $\tr\bigl[\ket{i}\!\!\bra{i}\Sys\otimes \Pi_{i}\, \tilde{\rho}\SP\bigr]\,=\,\rho_{ii}$. Finally inserting into the left-hand side of~(\ref{def:unbiased appendix}) yields
\begin{align}
    \tr\bigl[\mathds{1}\Sys\otimes\Pi_{i}\,\tilde{\rho}\SP\bigr]
    &=\sum_{j}\tr\bigl[\ket{j}\!\!\bra{j}\Sys\otimes\Pi_{i}\,\tilde{\rho}\SP\bigr]\nonumber\\
    &=\tr\bigl[\ket{i}\!\!\bra{i}\Sys\otimes \Pi_{i}\, \tilde{\rho}\SP\bigr]\,=\,\rho_{ii},
\end{align}
confirming the unbiasedness condition.

%\begin{align}
%\begin{split}
%\sum_i \tr[\ket{i}&\bra{i}\otimes \Pi_i
%]  = 1\\
%\Longrightarrow \quad&
%\sum_{i\neq j\,i,j}\tr[\ket{j}\bra{j}\otimes \Pi_i \tilde{\rho}\SP]  = 0\\
%\Longrightarrow \quad &
%\tr[\ket{j}\bra{j}\otimes \Pi_i \tilde{\rho}\SP] = 0 \qquad \forall \; i \neq \,j\\
%\Longrightarrow \quad & \tr[\sum_j\ket{j}\bra{j}\otimes \Pi_i\tilde{\rho}\SP] =  \tr[\sum_i \ket{j}\bra{j}\otimes \Pi_i\tilde{\rho}\SP]  \\
%& \qquad\qquad\qquad \qquad \quad =    \tr[\ket{i}\bra{i}\otimes \Pi_i \tilde{\rho}\SP]   \\
%& \qquad\qquad\qquad \qquad \quad =   \rho_{ii} \quad \forall \,i
%\,.
%\end{split}
%\end{align}

%Here we note that it is sufficient to check that the measurement procedure is faithful (perfectly correlating) and either one of the two remaining properties (unbiased or non-invasive) for any given \emph{fixed initial system state} $\rho\Sys$ to conclude that the measurement procedure is ideal for any state $\rho\Sys$.

For the remaining combination this is not the case. A measurement procedure that is unbiased (\ref{def:unbiased appendix}) and non-invasive (\ref{def:noninv appendix}) for a fixed system state $\rho\Sys$ is not necessarily faithful (\ref{def:faithfulness appendix}). Consider, e.g., the initial single-qubit system state $\rho\Sys=\tfrac{3}{4}\ket{0}\!\!\bra{0}+\tfrac{1}{4}\ket{1}\!\!\bra{1}$, i.e., where $\rho_{00}=\tfrac{3}{4}$ and $\rho_{11}=\tfrac{1}{4}$, and the two-qubit final state $\tilde{\rho}\SP=\sum_{m,n=0,1}\ket{m,n}\!\!\bra{m,n}$ with $p_{01}=p_{10}=p_{11}=\tfrac{1}{8}$ and $p_{00}=\tfrac{5}{8}$. For $\Pi_{i}=\ket{i}\!\!\bra{i}\Poi$, one has the reduced states $\tr\Sys(\tilde{\rho}\SP)=\tr\Poi(\tilde{\rho}\SP)=\rho\Sys$, so we have unbiasedness and non-invasiveness, but $C(\tilde{\rho}\SP)=\tfrac{3}{4}<1$.

%\begin{figure}[ht]
%\centering
%     \includegraphics[scale=0.42]{Fig_A1_triRelation.pdf}
%     \vspace*{-2mm}
%     \caption{The properties attributed to an ideal measurement. In a non-ideal measurement these three properties do not hold simultaneously, and satisfying one of them does not imply any of the other two. When $\rho\Sys$ is fixed, in two out of three cases, satisfying a pair of properties implies that the third property also holds. A measurement which is faithful and unbiased implies that it is also non-invasive, and a measurement that is faithful and non-invasive implies that it is unbiased. When $\rho\Sys$ is relaxed to \textit{all} initial system states, then satisfying any pair of properties implies the third, and one recovers the last relation, namely that a measurement which is unbiased and non-invasive (\textit{for all} $\rho\Sys$) implies that it is also faithful.}
%\label{fig:triangle}
%\end{figure}

Moreover, the measurement procedure (corresponding to the transformation above (whose details are not given in the example) may not be unbiased or non-invasive for other initial system states $\rho\Sys$. %As illustrated in Fig.~\ref{fig:tri},
This singles out the property of faithfulness when one is interested in checking the properties of a measurement procedure for \textit{any} given initial system state. Nonetheless, caution is advisable here. For a particular initial state $\rho\Sys$ of the system, all three properties may be satisfied, yet, this may not be so for other initial states. Simply consider the example in Eq.~(\ref{eq:example}). For an initial state with $\rho_{00}=\rho_{11}=\tfrac{1}{2}$, the measurement satisfies all three properties, but for $p<1$ the measurement procedure is biased and has non-perfect correlation for almost all (other) states $\rho\Sys$.

%%
%\begin{figure}[ht!]
%\begin{center}
%%%%trim={<left> <lower> <right> <upper>}
%\includegraphics[scale=0.55,trim={0cm 0.2cm 0cm 0cm},clip]{Fig_1_NewTriangle}
%\label{fig:tri}
%\caption{The properties attributed to an ideal measurement. In a non-ideal measurement these three properties do not hold simultaneously, and satisfying one of them does not imply any of the other two. However, in two cases, satisfying a pair of properties for a fixed initial system state $\rho\Sys$ implies that the third property also holds (for fixed $\rho\Sys$). A measurement which is faithful and unbiased implies that it is also non-invasive, and a measurement that is faithful and non-invasive implies that it is unbiased. In general, a measurement which is unbiased and non-invasive does not imply it is faithful, unless the measurement procedure is unbiased and non-invasive for all initial states $\rho\Sys$.}
%\end{center}
%\end{figure}
%%
%\noindent

Indeed, demanding that any of the properties hold only for particular initial system states $\rho\Sys$ is somewhat contradictory to the notion of performing a measurement that reveals previously unknown information about a system. In other words, measurements should not require detailed knowledge about $\rho\Sys$ to ensure that one may trust the measurement outcomes, or inferences made from them. The definitions of the properties (\ref{def:unbiased appendix}), (\ref{def:faithfulness appendix}) and (\ref{def:noninv appendix}) must hence be extended to demand that measurement procedures are only called unbiased, faithful, or non-invasive, if the respective properties (\ref{def:unbiased appendix}), (\ref{def:faithfulness appendix}) or (\ref{def:noninv appendix}) hold \emph{for all initial systems states} $\rho\Sys$.

With such an extended definition, one then indeed finds that any two properties imply the third. In particular, it is then the case that measurement procedures that are unbiased \emph{and} non-invasive, are also faithful, and thus ideal. The proof of this statement, relies on insights into the general structure of all maps representing unbiased measurement procedures, and as such appears later in Appendix~\ref{sec:missing proof}. In purely qualitative terms, maps that are either unbiased or non-invasive need to transfer the diagonal elements of the system state $\rho\Sys$ to particular (different) subspaces of the joint Hilbert space of system and pointer. The only way to simultaneously satisfy both the requirements for unbiasedness and non-invasiveness for arbitrary $\rho\Sys$ forces all information to be concentrated in the subspaces corresponding to the images of the projectors $\ket{i}\!\!\bra{i}\otimes\Pi_{i}$, such that the resulting state $\tilde{\rho}\SP$ satisfies (\ref{def:faithfulness appendix}) independently of the details of $\rho\Sys$.

Ultimately, this means that only one of the three properties can be satisfied exactly for all initial system states in any realistic measurement procedure. Given that the constraint of finite resources rules out that realistic non-ideal measurements are faithful, we have a choice between the measurement being unbiased or non-invasive. Arguably, biased measurements that are not even faithful are of little use, since the outcomes would not provide any level of certainty about either the pre- or post-measurement system state. In the following, we are therefore interested in unbiased measurement procedures for which the correlations between the system and the pointer are as large as possible. Given such non-ideal measurement, we then wish to minimize the associated energy costs.

\subsection{General Unbiased Measurements}\label{sec:Unbiased Measurement Procedures}
%\vspace*{-2mm}

Here, we will identify the basic structure and important properties of a general model of non-ideal measurement procedures. To do so, we separate what we believe to accurately model such a measurement procedure into two steps. These are:\\

\begin{tcolorbox}[colback=mycolor2!5,colframe=mycolor2,fonttitle=\bfseries,
left=1pt,
right=1pt,
top=1pt,
bottom=1pt]
\begin{enumerate}[\hspace*{-2mm}I]
    \item{\label{step i: preparation}\textbf{Preparation}:\ Some energy is invested to prepare the pointer system in a suitable quantum state.}
    \item{\label{step ii: correlating}\textbf{Correlating}:\ The pointer interacts with the system to be measured, creating correlations between them.}
\end{enumerate}
\end{tcolorbox}
%
%\vspace*{2mm}
%In Sections~\ref{sec:preparation} and~\ref{sec:correlation} we will motivate and describe these steps in more detail.
%\vspace{-3mm}

\subsubsection{Step I: Preparation}\label{sec:preparation}

Before interacting with the system, the pointer can be prepared in a suitable quantum state $\rho\Poi$ at the expense of some initial energy investment $\Delta E\subtiny{0}{0}{\mathrm{I}}$ accounting for the (CPTP) transformation $\mathcal{E}\subtiny{0}{0}{\mathrm{I}}:\mathcal{L}(\mathcal{H}\Poi)\rightarrow\mathcal{L}(\mathcal{H}\Poi)$ mapping $\tau(\beta)$ to $\rho\Poi = \mathcal{E}\subtiny{0}{0}{\mathrm{I}}[\tau(\beta)]$. In particular it may be desirable to lower the entropy of the initial pointer state. In principle, one may use any given amount of energy to prepare an arbitrary pointer state that is compatible with the specified energy and whose entropy is lower than that of $\tau(\beta)$. The energy cost for reaching a particular state $\rho\Poi$ is bounded from below by the free energy difference, i.e.,
\begin{align}
    \Delta E\subtiny{0}{0}{\mathrm{I}} &\geq\,\Delta F\bigl(\tau(\beta)\rightarrow\rho\Poi\bigr)  \,=\,\Delta E\Poi-T\Delta S\Poi\,,
\end{align}
with $\Delta E\Poi=\tr\bigl[H\Poi(\rho\Poi-\tau(\beta))\bigr]$ and $\Delta S\Poi=S\bigl(\rho\Poi\bigr)-S\bigl(\tau(\beta)\bigr)$, and where $S(\rho)=-\tr\bigl[
	\rho\log(\rho)\bigr]$ is the von~Neumann entropy. However, the exact work cost of the preparation depends on the control over the system and the available auxiliary degrees of freedom, and may exceed this bound. In particular, the free energy difference to the ground state is finite although this state cannot be reached with a finite work investment in finite time~\cite{MasanesOppenheim2017}. The precise resource requirements in terms of energy, control, and time for preparing arbitrary quantum states are hence difficult to capture\footnote{Certainly, any such preparation can be described by a CPTP map, which in turn can be seen as a unitary acting on $\rho\Poi\otimes\ket{\Phi}\!\!\bra{\Phi}$ in a Hilbert space enlarged by an auxiliary system with Hilbert space $\mathcal{H}\Aux\ni\ket{\Phi}$. However, this brings one back to the question of quantifying the cost for preparing the pure state $\ket{\Phi}$ of the auxiliary system.}, whereas the refrigeration of quantum systems is a well-understood task, whose energy cost %$\Delta E_{\mathrm{R}}$
has been quantified for various levels of control one assumes about the quantum systems involved in the cooling procedure~\cite{ClivazSilvaHaackBohrBraskBrunnerHuber2019a}.

It is therefore practically useful (and reasonable) to assume that the preparation only involves refrigeration. That is, we assume in the following that the temperature of the pointer is lowered from $T$ to $T_{0}\leq T$, reaching a thermal state $\tau(\beta_{0})$ with $\beta_{0}=1/T_{0}$. On the one hand, step~\ref{step i: preparation} thus becomes less general than it could potentially be since one does not explore the entire Hilbert space $\mathcal{H}\Poi$. On the other hand, the thermal state can be considered to be energetically optimal, since it minimizes the energy at fixed entropy. Moreover, at fixed energy the thermal state also maximizes the entropy and hence minimizes the free energy, which in turn bounds the work cost from below.

%%%%%%%%%%%%%%%%%%%%%%%%%%%%%%%%%%%%%%%%%%%%%%%%%%%%%%%%%%%%%%%%%%%%%%%%%%%%%%%%%%%%%%%%%%%%%%%%%%%%%%%%%%%

\subsubsection{Step II: Correlating}\label{sec:correlation}

During the second step of the measurement procedure, the system interacts with the pointer in such a way that correlations between the two are established via a CPTP map $\mathcal{E}\subtiny{0}{0}{\mathrm{I\hspace*{-0.5pt}I}}:\,\mathcal{L}(\mathcal{H}\SP)\rightarrow\mathcal{L}(\mathcal{H}\SP)$ that maps $\rho\SP=\rho\Sys\otimes\rho\Poi$ to $\tilde{\rho}\SP=\mathcal{E}\subtiny{0}{0}{\mathrm{I\hspace*{-0.5pt}I}}\bigl[\rho\SP\bigr]$.
A particularly important special case is the case of unitary correlating maps $U$, i.e., such that $\tilde{\rho}\SP=U\rho\SP U^{\dagger}$, representing measurement procedures where the joint system of $S$ and $P$ can be considered to be closed for the purpose of the correlating step. Then, the energy cost for the second step can be calculated via
\begin{align}\label{eq:deltaE}
    \Delta E\subtiny{0}{0}{\mathrm{I\hspace*{-0.5pt}I}}  &=\,\tr
    	\bigl[
    		(H\Sys+H\Poi)
    		(\tilde{\rho}\SP-\rho\SP)
    			\bigr].
\end{align}

In any case the generated correlations can in principle be (but need not be) genuine quantum correlations. For (non-ideal) projective measurements as defined here, it nonetheless suffices that classical correlations are established with respect to the measurement basis (here the eigenbasis of $H\Sys$) and a chosen basis of the pointer system. More specifically, we assign a set of orthogonal projectors
\begin{align}
    \Pi_{i} &:=\,\sum\limits_{n}\ket{\tilde{\psi}_{n}\suptiny{0}{0}{(i)}}\!\!\bra{\tilde{\psi}_{n}\suptiny{0}{0}{(i)}},
\end{align}
with $\Pi_{i}\Pi_{j}=\delta_{ij}\Pi_{i}$ (in particular, $\scpr{\tilde{\psi}_{m}\suptiny{0}{0}{(i)}}{\tilde{\psi}_{n}\suptiny{0}{0}{(j)}}=\delta_{ij}\delta_{mn}$) and $\sum_{i}\Pi_{i}=\mathds{1}\Poi$. The orthogonality and completeness of the projectors ensure that every pointer state is associated with a state of the measured system, i.e., every outcome provides a definitive measurement result ``i". We further amend the correlations function defined in Eq.~(\ref{eq:correlation function example}) for a single-qubit pointer to reflect the use of the more general projectors, i.e., we redefine the quantifier $C(\tilde{\rho}\SP)$ as
\begin{align}\label{eq:corfuncref}
    C(\tilde{\rho}\SP)    &:=\,\sum\limits_{i}\tr\bigl[
    	\ket{i}\!\!\bra{i}\Sys\otimes\Pi_{i}\,\tilde{\rho}\SP\bigr]\,.
\end{align}
\vspace*{-4mm}
%%%%%%%%%%%%%%%%%%%%%%%%%%%%%%%%%%%%%%%%%%%%%%%%%%%%%%%%%%%%%%%%%%%%%%%%%%%%%%%%%%%%%%%%%%%%%%%%%%%%%%%%%%%

\subsubsection{Unbiased measurements}\label{sec:unbiased measurements}

We are now in a position to give a formal definition of what we consider as an abstract measurement procedure.\\

\vspace*{-0.5mm}
\begin{Definitions}{\rm Measurement procedure}{measprocedure}
A measurement procedure $\mathcal{M}(\beta)$ that realizes a (non-ideal) projective measurement at ambient temperature $T=1/\beta$ of an (unknown) quantum state $\rho\Sys\in\mathcal{L}(\mathcal{H}\Sys)$ w.r.t. to an orthonormal basis $\{\ket{n}\}_{n}$ of $\mathcal{H}\Sys$ is given by the tuple $(\mathcal{H}\Poi,H\Poi,\Pi,\mathcal{E})$, consisting of a pointer Hilbert space $\mathcal{H}\Poi$, a pointer Hamiltonian $H\Poi\in\mathcal{L}(\mathcal{H}\Poi)$, a complete set $\Pi=\{\Pi_{i}\}_{i}$ of orthogonal projectors on $\mathcal{H}\Poi$, and a CPTP map $\mathcal{E}:\mathcal{L}(\mathcal{H}\SP)\rightarrow\mathcal{L}(\mathcal{H}\SP)$ with $\mathcal{H}\SP=\mathcal{H}\Sys\otimes\mathcal{H}\Poi$, together with the induced CPTP map $\mathcal{E}_{\mathcal{M}}:\mathcal{L}(\mathcal{H}\Sys)\rightarrow\mathcal{L}(\mathcal{H}\SP)$ given by
\begin{align}\label{eq:map}
    \mathcal{E}_{\mathcal{M}}:\ \rho\Sys\mapsto\tilde{\rho}\SP=\mathcal{E}[\rho\Sys\otimes\tau(\beta)].
\end{align}
\end{Definitions}

\vspace*{1.5mm}
Note that any definition of a thermal state $\tau(\beta)$ implies that the state has full rank. This definition includes, in particular, the case that we consider here, where the map $\mathcal{E}=\mathcal{E}\subtiny{0}{0}{\mathrm{I\hspace*{-0.5pt}I}} \circ(\mathds{1}\Sys\otimes\mathcal{E}\subtiny{0}{0}{\mathrm{I}})$ is split into two separate steps. As we have already motivated in our earlier example, we are interested in considering measurement procedures that represent the measured quantum state without bias. While perfect correlations cannot be guaranteed in this way, one may however ask that averages of the measured quantity match for the pointer and the system. Moreover, it is desirable that this is so independently of the specific initial states of the system and the pointer. All of this is captured by the following definition.\\

\begin{Definitions}{\rm Unbiased measurement}{unbiasedness}
A measurement procedure $\mathcal{M}(\beta)$ is called \emph{unbiased}, iff
$\tr\bigl[
	\Pi_{i}\tr\Sys(\tilde{\rho}\SP)\bigr] = \tr[
		\ket{i}\!\!\bra{i}\Sys\rho\Sys]$\ $\forall i$ and $\forall \rho\Sys$.
\end{Definitions}

\vspace*{2mm}
Since we wish to restrict our further considerations to unbiased measurements, it will be useful to know more about the structure of these measurement procedures, in particular, about the involved CPTP map $\mathcal{E}$ and projectors $\Pi_{i}$, given that one has selected a suitable pointer system with Hilbert space $\mathcal{H}\Poi$ and Hamiltonian $H\Poi$. To this end, note that our previous example using $U\subtiny{0}{0}{\mathrm{CNOT}}$ was unbiased only for pointers that can be prepared in the ground state (or any pure state for that matter). This can only be the case if the initial temperature vanishes or if infinite resources are available in step~\ref{step i: preparation}, whereas we are interested in describing more realistic conditions. To capture this, we formalise the following:\\

\begin{Definitions}{\rm Finite-resource measurement}{finresource}
A measurement procedure $\mathcal{M}(\beta)$ at a nonzero ambient temperature $T = 1/\beta$ uses \textit{finite resources} if the map $\mathcal{E}_{\mathcal{M}}$ is rank non-increasing.
\end{Definitions}

\newpage
On the other hand, measurement procedures which \textit{reduce} the rank use either infinite energy $E$, take infinite time $t$ (an infinite sequence of finite interaction range operations) or are infinitely complex (infinite interaction range operations) \cite{MasanesOppenheim2017, SchulmanMorWeinstein2005,WilmingGallego2017, ClivazSilvaHaackBohrBraskBrunnerHuber2019a, ClivazSilvaHaackBohrBraskBrunnerHuber2019b, AllahverdyanHovhannisyanJanzingMahler2011}. Now, in order to take a first step towards unraveling the structure of unbiased measurements we formulate the following lemma.\\

\begin{Lemmas}{}{structure of unbiased measurements}
All unbiased finite-resource measurement procedures $\mathcal{M}(\beta)$ with (thermal, full-rank) pointer system with Hilbert space $\mathcal{H}\Poi$, Hamiltonian $H\Poi$, and orthogonal projectors $\Pi_{i}$ can be realized by CPTP maps $\mathcal{E}$ of the form $\mathcal{E}=\mathcal{E}\subtiny{0}{0}{\mathrm{I\hspace*{-0.5pt}I}} \circ(\mathds{1}\Sys\otimes\mathcal{E}\subtiny{0}{0}{\mathrm{I}})$, where $\mathcal{E}\subtiny{0}{0}{\mathrm{I}}$ is a CPTP map from $\mathcal{L}(\mathcal{H}\Poi)$ to itself (achievable in finite time $t$ and satisfying $\Delta E\subtiny{0}{0}{\mathrm{I}}<\infty$), and the CPTP map $\mathcal{E}\subtiny{0}{0}{\mathrm{I\hspace*{-0.5pt}I}}$ from $\mathcal{L}(\mathcal{H}\SP)$ to itself has Kraus operators $K_{l}=\sum\limits_{i}K_{l}\suptiny{0}{0}{(i)}$ with\\
\vspace*{-5mm}
\begin{align}
    K_{l}\suptiny{0}{0}{(i)}    &=\,\sum\limits_{j=0}^{d\Sys-1}\sum\limits_{n=0}^{d\Poi-1}\sum\limits_{m=0}^{d_{i}-1}k_{jmn}\suptiny{0}{0}{(i,l)}
    \ket{j}\!\!\bra{i}\Sys\otimes\ket{\tilde{\psi}_{m}\suptiny{0}{0}{(i)}}\!\!\bra{\psi_{n}}\Poi,
    \label{eq:Kraus ops of ith map}
\end{align}
with $d\Sys=\dim(\mathcal{H}\Sys)$, $d\Poi=\dim(\mathcal{H}\Poi)\geq d\Sys$, and coefficients $k_{jmn}\suptiny{0}{0}{(i,l)}$ such that
\begin{align}
    \sum_{l}(K_{l}\suptiny{0}{0}{(i)})^{\dagger}K_{l}\suptiny{0}{0}{(i)}    &=\,\ket{i}\!\!\bra{i}\Sys\otimes\mathds{1}\Poi.
\end{align}
\end{Lemmas}

%\newpage
%\vspace*{3mm}
\textbf{Proof of Lemma}~\ref{lemma:structure of unbiased measurements}.\ Before we get into the technical details of the proof, let us phrase the Lemma~\ref{lemma:structure of unbiased measurements} more informally. It states that the map $\mathcal{E}$ consists of an arbitrary (finite energy, $\Delta E\subtiny{0}{0}{\mathrm{I}}<\infty$, finite time $t<\infty$) preparation of the pointer ($\mathcal{E}\subtiny{0}{0}{\mathrm{I}}$), followed by a map $\mathcal{E}\subtiny{0}{0}{\mathrm{I\hspace*{-0.5pt}I}}$ that maps the subspaces $\ket{i}\Sys$ to those corresponding to $\Pi_{i}$, respectively. Moreover, note that unbiasedness of course implies that the pointer system must be large enough ($d\Poi\geq d\Sys$) to accommodate all the possible measurement outcomes. Let us then prove the lemma. As mentioned before, the CPTP map $\mathcal{E}$ may be separated into a map $\mathcal{E}\subtiny{0}{0}{\mathrm{I}}$ acting nontrivially only on the pointer Hilbert space, and a CPTP map $\mathcal{E}\subtiny{0}{0}{\mathrm{I\hspace*{-0.5pt}I}}$ acting on the resulting state $\rho\SP=\rho\Sys\otimes\rho\Poi$, which we can write with respect to the basis $\ket{i}\Sys$ as
\begin{align}
    \rho\SP &=\,\left(\begin{array}{c|c|c|c}
        \tikzmarkin[ver=style green, line width=0mm,]{a1}\raisebox{2pt}{\protect\footnotesize{$\rho_{00}\rho\Poi$}}\tikzmarkend{a1}  & \cdot & \cdot & \cdots \\
        \hline
        \cdot & \tikzmarkin[ver=style orange, line width=0mm,]{a2}\raisebox{2pt}{\protect\footnotesize{$\rho_{11}\rho\Poi$}}\tikzmarkend{a2}  & \cdot & \cdots \\
        \hline
        \cdot  & \cdot & \tikzmarkin[ver=style cyan, line width=0mm,]{a33}\raisebox{2pt}{\protect\footnotesize{$\rho_{22}\rho\Poi$}}\tikzmarkend{a33} & \cdots \\
        \hline
        \vdots & \vdots & \vdots & \ddots
    \end{array}\right),\nonumber\\[-5mm]
    &\phantom{=\,(}\begin{array}{cccc}
        \hspace*{11.5mm} & \hspace*{8.0mm} & \hspace*{8.0mm} & \\[-2mm]
        \hspace*{1mm}$\upbracefill$ & \hspace*{-2.8mm}$\upbracefill$ & \hspace*{-2.6mm}$\upbracefill$  & \\[-1mm]
        \hspace*{2.4mm}\text{\protect\scriptsize{$\ket{0}\Sys$}} & \hspace*{0.1mm}\text{\protect\scriptsize{$\ket{1}\Sys$}} & \hspace*{0.1mm}\text{\protect\scriptsize{$\ket{2}\Sys$}} & \hspace*{0mm}\text{\protect\scriptsize{$\cdots$}}
    \end{array}
    \label{eq:instate}
\end{align}

Without loss of generality, we can then write the final state
$\tilde{\rho}\SP=\mathcal{E}\subtiny{0}{0}{\mathrm{I\hspace*{-0.5pt}I}}[\rho\SP]$ with respect to the product basis $\{\ket{i}\Sys\otimes\ket{\tilde{\psi}_{m}\suptiny{0}{0}{(j)}}\Poi\}_{i,j,m}$ in the form
\begin{center}
\includegraphics[scale=0.9]{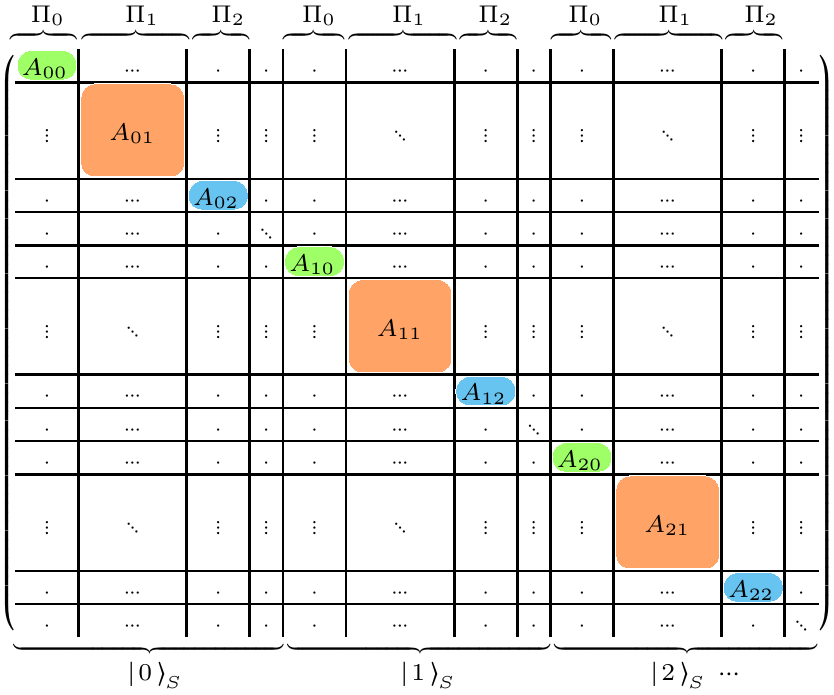}
\end{center}
\vspace*{-8mm}
\begin{align}
\label{eq:appendrow}
\end{align}
where we have indicated the columns corresponding to the subspaces of fixed vectors $\ket{i}\Sys$ and projectors $\Pi_{i}$ for $i=1,2,3$, and dots indicate matrix elements that may be nonzero but are not explicitly shown. In particular, the
%\newpage
%\noindent
latter can include subspaces for $i>3$, and the case for $d\Sys\leq3$ can be obtained by truncating the shown matrix by removing the corresponding rows and columns. The colored sub-blocks $A_{0i}$, $A_{1i}$, $A_{2i}$ and so forth are $d_{i}\times d_{i}$ block matrices in terms of which the unbiasedness condition of Def.~\ref{defi:unbiasedness} can be written as
\begin{align}
    \sum\limits_{j=0}^{d\Sys-1}\tr[A_{ji}]
    	 %A_{0i} + A_{1i} + A_{2i} +\cdots ]
    &=\,\rho_{ii}\ \ \forall i.
    \label{eq:unbiasedness cond proof}
\end{align}
Crucially, the unbiasedness condition in Eq.~(\ref{eq:unbiasedness cond proof}) is to hold for all possible system states $\rho\Sys$, and hence for all possible values of $\rho_{ii}$. This, in turn, implies that all sub-blocks with second subscript $i$ must be proportional to $\rho_{ii}$. That is $A_{ji}=\rho_{ii}\tilde{A}_{ji}$ $\;\forall \, j$ with
$\sum_{j=0}^{d\Sys-1}\tr[\tilde{A}_{ji}]=1$\
%$\tr[\tilde{A}_{0i} + \tilde{A}_{1i}  + \tilde{A}_{2i}  +\cdots ]=1$
$\forall i$. Since the unbiasedness condition is not sensitive to terms appearing in the off-diagonal blocks, a convenient representation of the relevant terms of the post-interaction state $\tilde{\rho}\SP$ under the map $\mathcal{E}\subtiny{0}{0}{\mathrm{I\hspace*{-0.5pt}I}}$ is :
\begin{align}
\Gamma_{\mathcal{E}\subtiny{0}{0}{\mathrm{I\hspace*{-0.5pt}I}}} =
\,\left[\begin{array}{c|c|c|c}
        \tikzmarkin[ver=style green, line width=0mm,]{d1}\raisebox{2pt}{\protect\footnotesize{$\rho_{00}\tilde{A}_{00}$}}\tikzmarkend{d1}  &  \tikzmarkin[ver=style orange, line width=0mm,]{dd2}\raisebox{2pt}{\protect\footnotesize{$\rho_{11}\tilde{A}_{01}$}}\tikzmarkend{dd2}  &    \tikzmarkin[ver=style cyan, line width=0mm,]{w1}\raisebox{2pt}{\protect\footnotesize{$\rho_{22}\tilde{A}_{02}$}}\tikzmarkend{w1}  &\cdots  \\
        \hline
        \tikzmarkin[ver=style green, line width=0mm,]{dd1}\raisebox{2pt}{\protect\footnotesize{$\rho_{00}\tilde{A}_{10}$}}\tikzmarkend{dd1} & \tikzmarkin[ver=style orange, line width=0mm,]{d2}\raisebox{2pt}{\protect\footnotesize{$\rho_{11}\tilde{A}_{11}$}}\tikzmarkend{d2}  & \tikzmarkin[ver=style cyan, line width=0mm,]{w3}\raisebox{2pt}{\protect\footnotesize{$\rho_{22}\tilde{A}_{12}$}}\tikzmarkend{w3} & \cdots \\
        \hline
        \tikzmarkin[ver=style green, line width=0mm,]{d33}\raisebox{2pt}{\protect\footnotesize{$\rho_{00}\tilde{A}_{20}$}}\tikzmarkend{d33}  &  \tikzmarkin[ver=style orange, line width=0mm,]{d44}\raisebox{2pt}{\protect\footnotesize{$\rho_{11}\tilde{A}_{21}$}}\tikzmarkend{d44} &  \tikzmarkin[ver=style cyan, line width=0mm,]{d5}\raisebox{2pt}{\protect\footnotesize{$\rho_{22}\tilde{A}_{22}$}}\tikzmarkend{d5} & \hdots\\
        \hline
        \vdots & \vdots & \vdots & \ddots
    \end{array}\right]
    \,,
    \label{eq:newrep}
\end{align}
which we call the \textit{correlation matrix}. Here it can immediately be seen that the unbiasedness condition, which implies $\sum_{j=0}^{d\Sys-1}\tr[\tilde{A}_{ji}]=1$ says that the sum of the trace of the blocks in column $i$ of $\Gamma_{\mathcal{E}}$ must be equal to $\rho_{ii}$ for unbiasedness to hold\footnote{Note that this representation is not square, since in principle the dimension of the matrix-valued entries of each column are different. In the case that the map $\mathcal{E}$ representing the measurement is unitary, the representation becomes square and the dimension of all blocks across all columns is equal.}.

%% If one compares this structure with the original state $\rho\SP$, which we can write with respect to the basis $\ket{i}\Sys$ as
%%\begin{align}
%%    \rho\SP &=\,\left(\begin{array}{c|c|c|c}
%%        \tikzmarkin[ver=style green, line width=0mm,]{a1}\raisebox{2pt}{\protect\footnotesize{$\rho_{00}\rho\Poi$}}\tikzmarkend{a1}  & \cdot & \cdot & \cdots \\
%%        \hline
%%        \cdot & \tikzmarkin[ver=style orange, line width=0mm,]{a2}\raisebox{2pt}{\protect\footnotesize{$\rho_{11}\rho\Poi$}}\tikzmarkend{a2}  & \cdot & \cdots \\
%%        \hline
%%        \cdot  & \cdot & \tikzmarkin[ver=style cyan, line width=0mm,]{a33}\raisebox{2pt}{\protect\footnotesize{$\rho_{22}\rho\Poi$}}\tikzmarkend{a33} & \cdots \\
%%        \hline
%%        \vdots & \vdots & \vdots & \ddots
%%    \end{array}\right),\nonumber\\[-5mm]
%%    &\phantom{=\,(}\begin{array}{cccc}
%%        \hspace*{10mm} & \hspace*{7.6mm} & \hspace*{7.6mm} & \\[-2mm]
%%        $\upbracefill$ & \hspace*{-1.5mm}$\upbracefill$ & \hspace*{-1.5mm}$\upbracefill$  & \\[-2mm]
%%        \hspace*{1.7mm}\text{\protect\scriptsize{$\ket{0}\Sys$}} & \hspace*{1.7mm}\text{\protect\scriptsize{$\ket{1}\Sys$}} & \hspace*{1.7mm}\text{\protect\scriptsize{$\ket{2}\Sys$}} & \hspace*{0mm}\text{\protect\scriptsize{$\cdots$}}
%%    \end{array}
%%    \label{eq:instate}
%%\end{align}
From the initial state \eqref{eq:instate} and the final state \eqref{eq:appendrow} it becomes apparent that unbiasedness can be guaranteed for maps that connect the subspaces corresponding to $\ket{i}\Sys$ only with those corresponding to $\Pi_{i}$. More precisely, each of these connecting maps can be viewed as an arbitrary CPTP map $\mathcal{E}\subtiny{0}{0}{\mathrm{I\hspace*{-0.5pt}I}}\suptiny{0}{0}{(i)}$ from the $d\Poi$-dimensional space spanned by the vectors in the set $\{\ket{i}\Sys\otimes\ket{\psi_{n}}\Poi\}_{n=0,\ldots,d\Poi-1}$, where $\{\ket{\psi_{n}}\Poi\}_{n}$ is an arbitrary basis of $\mathcal{H}\Poi$, to the $(d\Sys\times d_{i})$-dimensional space spanned by the vectors in the set $\{\ket{j}\Sys\otimes\ket{\tilde{\psi}_{m}\suptiny{0}{0}{(i)}}\Poi\}\subtiny{0}{0}{\substack{j=0,\ldots,d\Sys-1\\ m=0,\ldots,d_{i}-1}}$. The Kraus operators for these CPTP maps are precisely the $\{K_{l}\suptiny{0}{0}{(i)}\}_{l}$ of Eq.~(\ref{eq:Kraus ops of ith map}) and in matrix notation we may denote these maps as
\begin{scriptsize}
\begin{align}
    \rho_{ii}\rho_{P}  & \,\substack{\mathcal{E}\subtiny{0}{0}{\mathrm{I\hspace*{-0.5pt}I}}\suptiny{0}{0}{(i)}\\ \longmapsto}\,
    \left(\begin{array}{c|c|c|c}
        \!\!A_{0i}\!\!  & \!\!\cdot\!\! & \!\!\cdot\!\! & \!\!\cdots\!\! \\
        \hline
        \!\!\cdot\!\! & \!\!A_{1i}\!\!  & \!\!\cdot\!\! & \!\!\cdots\!\! \\
        \hline
        \!\!\cdot\!\!  & \!\!\cdot\!\! & \!\!A_{2i}\!\! & \!\!\cdots\!\! \\
        \hline
        \!\!\vdots\!\! & \!\!\vdots\!\! & \!\!\vdots\!\! & \!\!\ddots\!\!
    \end{array}\right)
    %\\ &\ \
    =\,\rho_{ii}
    \left(\begin{array}{c|c|c|c}
        \!\!\tilde{A}_{0i}\!\!  & \!\!\cdot\!\! & \!\!\cdot\!\! & \!\!\cdots\!\! \\
        \hline
        \!\!\cdot\!\! & \!\!\tilde{A}_{1i}\!\!  & \!\!\cdot\!\! & \!\!\cdots\!\! \\
        \hline
        \!\!\cdot\!\!  & \!\!\cdot\!\! & \!\!\tilde{A}_{2i}\!\! & \!\!\cdots\!\! \\
        \hline
        \!\!\vdots\!\! & \!\!\vdots\!\! & \!\!\vdots\!\! & \!\!\ddots\!\!
    \end{array}\right).
    \label{eq:subspace maps}
\end{align}
\end{scriptsize}
Since the domains as well as the images for different $i$ lie in orthogonal subspaces of $\mathcal{H}\SP$, the maps $\mathcal{E}\subtiny{0}{0}{\mathrm{I\hspace*{-0.5pt}I}}\suptiny{0}{0}{(i)}$ can be combined to the map $\mathcal{E}\subtiny{0}{0}{\mathrm{I\hspace*{-0.5pt}I}}$ with Kraus operators\footnote{Note that the number of nonzero Kraus operators may be different for each $i$, but one may always add trivial (vanishing) Kraus operators to each set $\{K_{l}\suptiny{0}{0}{(i)}\}_{l}$ with fixed $i$.} $\{K_{l}=\sum_{i}K_{l}\suptiny{0}{0}{(i)}\}_{l}$. Once can check that the unbiasedness condition is satisfied for these Kraus operators by a simple calculation, which we will not repeat here. If the initial ambient temperature is nonzero and the measurement procedure uses finite resources (time and energy), the pointer state $\rho\Poi$ has full rank and unbiasedness can only be achieved via maps of the form mentioned, as claimed in Lemma~\ref{lemma:structure of unbiased measurements}, which concludes the proof. \qed\\

Inspecting again the example from Appendix~\ref{sec:Examples Two-Qubit Measurement Procedures}, one notes that the controlled NOT operation $U\subtiny{0}{0}{\mathrm{CNOT}}$ is not of the form required for a finite-resource unbiased measurement, as expected. However, when the pointer can be prepared in a pure state (w.l.o.g. the ground state $\ket{0}\Poi$) one observes that the measurement procedure using $U\subtiny{0}{0}{\mathrm{CNOT}}$ in the correlating step becomes unbiased because some of the sub-blocks $A_{ji}$  are only trivially proportional to $\rho_{ii}$. In particular, $A_{01}=\tilde{A}_{01}=A_{10}=\tilde{A}_{10}=0$.

Having understood the general structure of all unbiased measurements, we now want to turn to some specific instances of such measurement procedures.

%%%%%%%%%%%%%%%%%%%%%%%%%%%%%%%%%%%%%%%%%%%%%%%%%%%%%%%%%%%%%%%%%%%%%%%%%%%%%%%%%%%%%%%%%%%%%%%%%%%%

\subsection{Extremal Measurements}\label{sec:extremal meas procedures}
%\vspace*{-2mm}

With the help of Lemma~\ref{lemma:structure of unbiased measurements} we can now describe the set of \emph{all} unbiased measurements for a given quantum system $\rho\Sys$ and pointer. The measurement within this set may further be categorized according to their specific properties, in particular, their energy cost, the amount of correlations created between the system and the pointer (how faithful the measurement is), and the level of control required for their implementation (e.g., what type of auxiliary systems are available and which operations can be performed on the system, pointer, and auxiliaries). Given an (unknown) quantum system $S$ it would ideally be desirable to answer the question: What is the maximal correlation achievable between the system and \emph{any suitable} pointer given a fixed work input $\Delta E$? A more restricted version of this question is: What is the maximal correlation achievable between the system and \emph{a particular} pointer given a fixed work input $\Delta E$?

Since we assume that the system state $\rho\Sys$ is unknown before the measurement, the correlation measure $\bar{C}$ that we are interested in optimizing is obtained from averaging $C(\tilde{\rho}\SP)$ from Eq.~(\ref{def:faithfulness appendix}) over all system states. We observe that for any particular systems state $\rho\Sys$, the correlation measure $C(\tilde{\rho}\SP)$ does not depend on any of the matrix elements of $\rho\Sys$ except for those on the diagonal, i.e.,
\begin{align}
    C(\tilde{\rho}\SP)  &=\,\rho_{00}\tr[
    	\tilde{A}_{00}
    		]+\rho_{11}\tr[
    		\tilde{A}_{11}]+\rho_{22}\tr[
    		\tilde{A}_{22}
    			]+\dots\,.
\end{align}
Averaging over all states $\rho\Sys$ is  hence equivalent to an average over all probability distributions corresponding to the diagonal of $\rho\Sys$. Moreover, for each of these values $\rho_{ii}$ $(i=0,\ldots,d\Sys-1)$, the average over all probability distributions results in the value $1/d\Sys$, such that the average of $C(\tilde{\rho}\SP)$ is given by
\begin{align}
    \bar{C} &=\,\tfrac{1}{d\Sys}\tr\bigl[
    	\tilde{A}_{00}+\tilde{A}_{11}+\tilde{A}_{22}+\dots\bigr],
    \label{eq:average corr}
\end{align}
which, in terms of the representation presented in Eq.~\eqref{eq:newrep}, corresponds to taking the trace of the blocks appearing along the diagonal. \\
Despite this simple form of $\bar{C}$, the optimization over all pointer systems and operations thereon is a daunting task. Indeed, even for a fixed pointer at initial temperature $T=1/\beta$, identifying the optimal measurement procedure in terms of the best ratio of (average) correlation increase per unit energy cost (averaged over the input system states) is highly nontrivial. To illustrate the difficulty, first note that an (attainable) bound exists for correlating (quantified by the mutual information) two arbitrary systems that are initially thermal at the same temperature at optimal energy expenditure~\cite{BruschiPerarnauLlobetFriisHovhannisyanHuber2015, HuberPerarnauHovhannisyanSkrzypczykKloecklBrunnerAcin2015}. While the known protocol for attaining this bound is in general not unitary (it involves lowering the temperature), in some cases the bound is tight already when one correlates the systems unitarily. However, it was recently shown~\cite{VitaglianoKloecklHuberFriis2019, BakhshinezhadEtAl2019} that the optimal (in the sense of the mentioned bound being tight) trade-off between correlations and energy cost cannot always be achieved unitarily.

Of course, in our case, the initial state of the system is not known, and cannot be expected to be thermal in general. Moreover, the mutual information is not a suitable figure of merit for quantifying the desired correlations between system and pointer because the latter don't distinguish between classical and genuinely quantum correlations. For instance, for a single-qubit system and pointer, the states $\ket{\Phi^{+}}\SP=\bigl(\ket{00}+\ket{11}\bigr)/\sqrt{2}$ and $\rho\SP=\tfrac{1}{2}\bigl(\ket{00}\!\!\bra{00}+\ket{11}\!\!\bra{11}\bigr)$ have different values of mutual information but are equally well (i.e., perfectly) correlated w.r.t. to the desired measurement basis. The above arguments on optimally correlating protocols hence do not apply directly, but with the added complication of the unknown system state and the unbiasedness condition, we cannot rule out the possibility that the optimal unbiased measurement procedures are not realized by a unitary correlating step.

Nonetheless, it can be argued that any nonunitary realization of the second part $\mathcal{E}\subtiny{0}{0}{\mathrm{I\hspace*{-0.5pt}I}}$ of the CPTP map $\mathcal{E}$ must require higher levels of control than a corresponding unitary realization due to the requirement of realizing nonunitary maps $\mathcal{E}\subtiny{0}{0}{\mathrm{I\hspace*{-0.5pt}I}}$ as unitaries on a larger Hilbert space. Specifically, any CPTP map $\mathcal{E}\subtiny{0}{0}{\mathrm{I\hspace*{-0.5pt}I}}$ can be thought of as a unitary on a larger Hilbert space $\mathcal{H}\SP\otimes\mathcal{H}\subtiny{0}{0}{E}$ (with a factoring initial condition)~\cite{Paris2012}, that is, one may write any $\mathcal{E}\subtiny{0}{0}{\mathrm{I\hspace*{-0.5pt}I}}$ as
\begin{align}
    \mathcal{E}\subtiny{0}{0}{\mathrm{I\hspace*{-0.5pt}I}}[\rho\SP] &=\,\tr\subtiny{0}{0}{E}
    	\bigl[
    		U\SPE \bigl[\rho\SP\otimes\ket{\chi}\!\!\bra{\chi}\bigr] U\SPE^{\dagger} \bigr]
\end{align}
for some unitary $U\SPE$ on $\mathcal{H}\SP\otimes\mathcal{H}\subtiny{0}{0}{E}$ and for some pure state $\ket{\chi}\in\mathcal{H}\subtiny{0}{0}{E}$. At the same time, employing a unitary to correlate pointer and system enables us to unambiguously quantify the work cost of the correlating step without assumptions about the Hamiltonian of potential auxiliary systems.

We are therefore particularly interested in describing all unbiased measurement procedures, where $\mathcal{E}\subtiny{0}{0}{\mathrm{I\hspace*{-0.5pt}I}}$ is realized unitarily, such that
\begin{align}
    \tilde{\rho}\SP &=\,\mathcal{E}\subtiny{0}{0}{\mathrm{I\hspace*{-0.5pt}I}}[\rho\SP]\,=\,U\rho\SP U^{\dagger}
\end{align}\label{eq:map2}
with $UU^{\dagger}=U^{\dagger}U=\mathds{1}\SP$. In this sense, our focus lies on unbiased measurement procedures where all control that one may have over external systems (beyond $S$ and $P$) is used in the initial step represented by $\mathcal{E}\subtiny{0}{0}{\mathrm{I}}$ to prepare the pointer in a suitable state, e.g., by lowering its temperature. Here we make use of the fact that the work cost of refrigeration with various levels of control has been extensively studied~\cite{ClivazSilvaHaackBohrBraskBrunnerHuber2019a}. This leaves us with the task of analyzing the structure of the representations $U$ of the unitary maps $\mathcal{E}\subtiny{0}{0}{\mathrm{I\hspace*{-0.5pt}I}}$.

A first step towards the completion of this task is to note that the unbiasedness condition for measurement procedures with a \emph{unitary} correlating step $\mathcal{E}\subtiny{0}{0}{\mathrm{I\hspace*{-0.5pt}I}}$ means that it is inefficient (in terms of energy cost) to use a pointer Hilbert space $\mathcal{H}\Poi$ whose dimension is not an integer multiple of the system dimension. This can be explained in the following way. By inspection of the maps in Eq.~(\ref{eq:subspace maps}), one notes that $\mathcal{E}\subtiny{0}{0}{\mathrm{I\hspace*{-0.5pt}I}}\suptiny{0}{0}{(i)}$ maps the $d\Poi\times d\Poi$ density matrix $\rho\Poi$ to a $d\Sys\operatorname{rank}(\Pi_{i})\times d\Sys\operatorname{rank}(\Pi_{i})$ density matrix. That is, the size of each of the $d\Sys$ blocks $\tilde{A}_{ji}$ $\; \forall \, j$ is determined by the rank of $\Pi_{i}$. If the map $\mathcal{E}\subtiny{0}{0}{\mathrm{I\hspace*{-0.5pt}I}}$ is unitary, this implies that all $\mathcal{E}\subtiny{0}{0}{\mathrm{I\hspace*{-0.5pt}I}}\suptiny{0}{0}{(i)}$ are unitary, and hence
\begin{align}
    d\Poi  &=\,d\Sys\,\operatorname{rank}(\Pi_{i})\quad \forall\,i.
    \label{eq:dimensions for unitary unbiased}
\end{align}
Conversely, this implies that all projectors $\Pi_{i}$ have the same rank $d\Poi/d\Sys$, which must be an integer larger or equal to 1, $d\Poi=\lambda d\Sys$ for $\lambda\in\mathbb{N}$. The implication of this for the correlation matrix in~\eqref{eq:newrep} is that it is now square
\begin{align}
\Gamma_{U_{\text{unb}}} =
\,\left[\begin{array}{c|c|c|c}
        \tikzmarkin[ver=style green, line width=0mm,]{D1}\raisebox{2pt}{\protect\footnotesize{$\rho_{00}\tilde{A}_{00}$}}\tikzmarkend{D1}  &  \tikzmarkin[ver=style orange, line width=0mm,]{DD2}\raisebox{2pt}{\protect\footnotesize{$\rho_{11}\tilde{A}_{01}$}}\tikzmarkend{DD2}  &    \tikzmarkin[ver=style cyan, line width=0mm,]{W1}\raisebox{2pt}{\protect\footnotesize{$\rho_{22}\tilde{A}_{02}$}}\tikzmarkend{W1}  &\cdots  \\
        \hline
        \tikzmarkin[ver=style green, line width=0mm,]{DD1}\raisebox{2pt}{\protect\footnotesize{$\rho_{00}\tilde{A}_{10}$}}\tikzmarkend{DD1} & \tikzmarkin[ver=style orange, line width=0mm,]{D2}\raisebox{2pt}{\protect\footnotesize{$\rho_{11}\tilde{A}_{11}$}}\tikzmarkend{D2}  & \tikzmarkin[ver=style cyan, line width=0mm,]{W3}\raisebox{2pt}{\protect\footnotesize{$\rho_{22}\tilde{A}_{12}$}}\tikzmarkend{W3} & \cdots \\
        \hline
        \tikzmarkin[ver=style green, line width=0mm,]{D33}\raisebox{2pt}{\protect\footnotesize{$\rho_{00}\tilde{A}_{20}$}}\tikzmarkend{D33}  &  \tikzmarkin[ver=style orange, line width=0mm,]{D44}\raisebox{2pt}{\protect\footnotesize{$\rho_{11}\tilde{A}_{21}$}}\tikzmarkend{D44} &  \tikzmarkin[ver=style cyan, line width=0mm,]{D5}\raisebox{2pt}{\protect\footnotesize{$\rho_{22}\tilde{A}_{22}$}}\tikzmarkend{D5} & \hdots\\
        \hline
        \vdots & \vdots & \vdots & \ddots
    \end{array}\right]
    \,.
    \label{eq:newrepunit}
\end{align}
In principle, one could initially consider a pointer with a Hilbert space dimension larger than required for the desired $\lambda$. However, the energy levels exceeding $\lambda d\Sys$ would have to be truncated before the preparation step to avoid unnecessary additional energy costs. The general form of the unitaries realising such unbiased measurement procedures is summarised below.\\

\begin{Lemmas}{}{structure of unitary unbiased measurements}
Let $\mathcal{M}_{U}(\beta)$ be an unbiased finite-resource ($\Delta E<\infty, T=1/\beta>0$) measurement procedure with \emph{unitary} correlating step $\mathcal{E}\subtiny{0}{0}{\mathrm{I\hspace*{-0.5pt}I}}$ using a pointer system Hilbert space $\mathcal{H}\Poi$ and Hamiltonian $H\Poi$ with $d\Poi=\lambda d\Sys$ for $\lambda\in\mathbb{N}$. The unitary map $U$ realizing the correlating step, i.e.,
\begin{align}
    \mathcal{E}\subtiny{0}{0}{\mathrm{I\hspace*{-0.5pt}I}}[\rho\Sys\otimes\rho\Poi] &=\,U\rho\Sys\otimes\rho\Poi U^{\dagger},
\end{align}
can be split into two consecutive unitary operations, $U=V\tilde{U}$, where $\tilde{U}$ and $V$ are of the form
\begin{align}
    \tilde{U}    &=\,\sum\limits_{i=0}^{d\Sys-1}\ket{i}\!\!\bra{i}\Sys\otimes \tilde{U}\suptiny{0}{0}{(i)},\\
    V   &=\,\sum\limits_{i,j=0}^{d\Sys-1}\sum\limits_{m=1}^{\lambda}
    \ket{j}\!\!\bra{i}\Sys\otimes
    \ket{\tilde{\psi}_{m}\suptiny{0}{0}{(i)}}\!\!\bra{\tilde{\psi}_{m}\suptiny{0}{0}{(j)}}\Poi,
    \label{eq:lemma 2 V}
\end{align}
and $\tilde{U}\suptiny{0}{0}{(i)}$ are arbitrary unitaries on $\mathcal{H}\Poi$.
\end{Lemmas}
%\newpage
\textbf{Proof of Lemma}~\ref{lemma:structure of unitary unbiased measurements}.\ The structure of the unitaries in the correlating step can be understood by noting that unitaries have only a single non-trivial Kraus operator. The operators $\tilde{U}\suptiny{0}{0}{(i)}$ in the first unitary $\tilde{U}$ then simply correspond to the single Kraus operators of the maps $\mathcal{E}\subtiny{0}{0}{\mathrm{I\hspace*{-0.5pt}I}}\suptiny{0}{0}{(i)}$ from Eq.~(\ref{eq:subspace maps}), rearranging the joint density matrix only in the subspaces of fixed $\ket{i}\Sys$, creating the distinction between the sub-blocks $\tilde{A}_{ji}$ for different $j$. The second part, realized by the unitary $V$ then just swaps these sub-blocks, such that all $\tilde{A}_{0i}$ are left in the subspace corresponding to $\ket{0}\Sys$ and $\Pi_{i}$, all $\tilde{A}_{1i}$ are left in the subspace corresponding to $\ket{1}\Sys$ and $\Pi_{i}$, and so forth. The only freedom in choosing unitary correlation steps for unbiased measurements hence lies in the choice of the $\tilde{U}\suptiny{0}{0}{(i)}$. \qed

%%%%%%%%%%%%%%%%%%%%%%%%%%%%%%%%%%%%%%%%%%%%%%%%%%%%%%%%%%%%%%%%%%%%%%%%%%%%%%%%%%%%%%%%%%%%%%%%%%

\subsection{Unbiased and Non-invasive Measurements are Faithful}\label{sec:missing proof}

With the compact result of Lemmas~\ref{lemma:structure of unbiased measurements} and~\ref{lemma:structure of unitary unbiased measurements} at hand, let us now briefly return to the relationship between the properties unbiasedness and non-invasiveness. As we have now seen, measurement procedures that are unbiased for all initial system states $\rho\Sys$ are required to map the subspace of the joint system-pointer Hilbert space corresponding to the image of the operator $\ket{i}\!\!\bra{i}\Sys\otimes\mathds{1}\Poi$ to the subspace corresponding to the image of $\mathds{1}\Sys\otimes\Pi_{i}$ for all $i$. At this point it becomes clear that, formulating the analogous statements to Lemmas~\ref{lemma:structure of unbiased measurements} and~\ref{lemma:structure of unitary unbiased measurements} for measurement procedures that are non-invasive instead of unbiased for all $\rho\Sys$, results in maps from the subspace corresponding to the image of the operator $\ket{i}\!\!\bra{i}\Sys\otimes\mathds{1}\Poi$ to itself. Since the image of $\ket{i}\!\!\bra{i}\Sys\otimes\mathds{1}\Poi$ is spanned by the set of non-trivial joint eigenvectors of the set of projectors $\{\ket{i}\!\!\bra{i}\Sys\otimes\Pi_{j}\}_{j}$, and the image of $\mathds{1}\Sys\otimes\Pi_{i}$ is spanned by the set of non-trivial joint eigenvectors of the set of projectors $\{\ket{j}\!\!\bra{j}\Sys\otimes\Pi_{i}\}_{j}$, a map that is supposed to satisfy both unbiasedness and non-invasiveness for all $\rho\Sys$ must be a map from the image of $\ket{i}\!\!\bra{i}\Sys\otimes\mathds{1}\Poi$ to the span of the set of non-trivial eigenvectors of $\ket{i}\!\!\bra{i}\Sys\otimes\Pi_{i}$, such that $\tr(\ket{i}\!\!\bra{i}\Sys\otimes\Pi_{i}\tilde{\rho}\SP)=\rho_{ii}\ \forall i$. By construction, one thus has $\sum_{i}\tr(\ket{i}\!\!\bra{i}\Sys\otimes\Pi_{i}\tilde{\rho}\SP)=\sum_{i}\rho_{ii}=1$, and the measurement is faithful.

%%%%%%%%%%%%%%%%%%%%%%%%%%%%%%%%%%%%%%%%%%%%%%%%%%%%%%%%%%%%%%%%%%%%%%%%%%%%%%%%%%%%%%%%%%%%%%%%%%

\subsection{Maximally Correlating Unbiased Measurements}\label{app:max corr unbiased}

To gain further insight into the fundamental limitations of non-ideal measurements, we now wish to focus on a special case where Lemma~\ref{lemma:structure of unitary unbiased measurements} applies, that is, an unbiased measurement procedure with unitary correlating step, such that \textemdash\ at least for the purpose of controlling their interaction \textemdash\ the joint system of pointer and measured system can be considered closed. That is, procedures where $\mathcal{E}\subtiny{0}{0}{\mathrm{I\hspace*{-0.5pt}I}}[\rho\Sys\otimes\rho\Poi] =U\rho\Sys\otimes\rho\Poi U^{\dagger}$. Apart from this restriction, we will only consider preparation steps that modify the temperature of the initial pointer system, i.e. $\mathcal{E}\subtiny{0}{0}{\mathrm{I}}$ is a refrigeration step.
%such that $\rho\Poi=\tau\Poi(\beta)$.
In such a scenario, $T=1/\beta$ might be the initial temperature of the pointer, or, e.g., one below the ambient temperature, reached by investing energy for cooling the pointer. For such measurement procedures, we now wish to find the maximum attainable correlation between the system and pointer. As we show, there is an algebraic maximum $C_{\text{max}}$ for the correlations that can be unitarily created between the system and the thermal pointer, regardless of the initial system state $\rho\Sys$. Recall the definition of the correlation function in Eq.~\eqref{eq:corfuncref}, which we rewrite as
\begin{align}\label{eq:maxcorrel}
    C(\tilde{\rho}\SP)    &:=\,\sum\limits_{i}\tr\bigl[\tilde{\Pi}_{ii}\tilde{\rho}\SP\bigr]\,,
\end{align}
by making the association $\tilde{\Pi}_{ij} = \ket{i}\!\!\bra{i}\otimes \Pi_j \;\forall \,i,j$. By definition, the correlation function is only sensitive to terms appearing along the diagonal w.r.t. any chosen common eigenbasis (with nontrivial eigenvalues) of the set of operators $\tilde{\Pi}_{ii}$. We refer to the subspace of $\mathcal{H}\Sys\otimes\mathcal{H}\Poi$ spanned by these eigenvectors as $\mathcal{H}_{\mathrm{corr}}$, and to its complement as $\mathrm{H}_{\mathrm{nc}}$, such that $\mathcal{H}\Sys\otimes\mathcal{H}\Poi=\mathcal{H}_{\mathrm{corr}}\oplus\mathcal{H}_{\mathrm{nc}}$. In particular, this implies that the unitary transformation achieving the algebraic maximum (over all unitaries $U_{\mathrm{unb}}$ realizing unbiased measurement procedures in the sense of Lemma~\ref{lemma:structure of unitary unbiased measurements})
\begin{align}
 \max_{U_{\mathrm{unb}}} C(\tilde{\rho}\SP)  = C_{\text{max}}
\end{align}
is not unique since $C(\tilde{\rho}\SP)$ is invariant under operations of the form $U_{\mathrm{corr}}\oplus U_{\mathrm{nc}}$, where $U_{\mathrm{corr}}$ and $U_{\mathrm{nc}}$ act nontrivially only on the subspaces $\mathcal{H}_{\mathrm{corr}}$ and $\mathcal{H}_{\mathrm{nc}}$, respectively.

Within the orbit of all unitaries that one may perform (including those corresponding to biased measurements) on $\rho\Sys\otimes\tau\Poi$, the global maximum value of the function $C(\tilde{\rho}\SP)$ is achieved when the state $\tilde{\rho}\SP$ is block-diagonal w.r.t. to the subspace partition into $\mathcal{H}_{\mathrm{corr}}\oplus\mathcal{H}_{\mathrm{nc}}$, and the eigenvalues of the joint final state restricted to the $d\Poi$-dimensional correlated subspace $\mathcal{H}_{\mathrm{corr}}$, given by $\tilde{\rho}_{\mathrm{corr}}=\Pi_{\mathrm{corr}}\tilde{\rho}\SP\Pi_{\mathrm{corr}}$ with $\Pi_{\mathrm{corr}}=\sum_{i}\ket{i}\!\!\bra{i}\otimes \Pi_{i}$, are the $d\Poi$ largest eigenvalues of $\tilde{\rho}\SP$. These eigenvalues depend on $\rho\Sys$. However, we have to take into account unbiasedness and the fact that we are looking for a unitary. In particular, from Eq.~(\ref{eq:dimensions for unitary unbiased}) we know that the $\Pi_i$ must all have the same rank, namely $\operatorname{rank}(\Pi_{i})=d\Poi/d\Sys=2^{N-1}\ \forall\,i$. It then becomes apparent that one is restricted to selecting the $d\Poi/d\Sys$ largest eigenvalues of $\rho\Poi=\tau\Poi(\beta)$ for each of the $d\Sys$ subspaces corresponding to the image of a projector $\ket{i}\!\!\bra{i}\otimes \Pi_{i}$. Since the assignment of eigenvalues to each subspace labelled by $i$ is the same, and the corresponding matrix elements of the initial system state sum to 1, the maximal achievable correlation is independent of $\rho\Sys$.
%
%
%will be one that is a product of swaps along the diagonal in the subspaces $\tilde{\Pi_i}$.
%%Thus, all possible populations of $\tilde{\rho}\SP$ as a result of such swaps can be seen as sets of the pointer probabilities $p_i^{{\suptiny{0}{0}{(k)}}}$ modulated by the system populations $\rho_{ii}$. Written in terms of the sectors, we have
%%\begin{align}\label{eq:partition}
%%\begin{split}
%%\rho_{00}&\{
%%p_0^{{\suptiny{0}{0}{(0)}}}, p_1^{{\suptiny{0}{0}{(0)}}}, \cdots ,
%%			p_{0}^{{\suptiny{0}{0}{(i)}}}, p_1^{{\suptiny{0}{0}{(i)}}} \cdots , p_{0}^{{\suptiny{0}{0}{(d\Sys -1)}}},  p_{1}^{{\suptiny{0}{0}{(d\Sys -1)}}}, \cdots
%%				\},\\
%%\rho_{11}  & \{  p_0^{{\suptiny{0}{0}{(0)}}}, p_1^{{\suptiny{0}{0}{(0)}}}, \cdots ,
%%p_{0}^{{\suptiny{0}{0}{(i)}}}, p_1^{{\suptiny{0}{0}{(i)}}} \cdots ,
%%p_{0}^{{\suptiny{0}{0}{(d\Sys -1)}}},  p_{1}^{{\suptiny{0}{0}{(d\Sys -1)}}}, \cdots
%% \},\\\
%%&\vdots
%%\end{split}
%%\end{align}
%When the initial state of the system is unknown, i.e. $\rho\Sys = \tfrac{1}{d\Sys}\mathds{1}_{d\Sys}$,
%
%Irrespective of the initial state of the system, the algebraic maximum of the correlation function is achieved by assigning the largest probability populations to the correlated subspaces, i.e., to the subspaces corresponding to the images of $\tilde{\Pi}_{ii}$, each of dimension $\tfrac{d\Poi}{d\Sys}$.
In the notation of Eq.~\eqref{eq:diagonalise}, this corresponds to all probability populations that belong to the sector where $(k=0)$, i.e.,
\begin{align}
    C_{\text{max}}(\beta) &= \sum_{i=0}^{d\Sys -1}\sum_{j=0}^{d\Poi/d\Sys-1}
    \rho_{ii}\, p\suptiny{0}{0}{(0)}_{j}
    \,=\,  \tfrac{1}{\mathcal{Z}}\sum_{i=0}^{d\Poi/d\Sys-1} e^{-\beta E\suptiny{0}{0}{(0)}_{i}}\;.
    \label{eq:cmax}
\end{align}
As mentioned in Eq.~\eqref{eq:average corr}, this function is independent of the system. The remaining probability weights (i.e., the $p\suptiny{0}{0}{(k)}_{j}$ for $k\neq0$ and $j=0,\ldots, d\Poi/d\Sys-1$) are distributed in the non-correlated subspace.

The corresponding general form of the joint final state $\tilde{\rho}\SP$ of any unitarily maximally correlating unbiased measurement procedure (starting from an initially thermal pointer state) can then be compactly specified in terms of its correlation matrix $\Gamma_{U_{C_{\mathrm{max}}}}$ as defined in Eq.~(\ref{eq:newrep}). To write $\Gamma_{U_{C_{\mathrm{max}}}}$ in a simple form, let $a\suptiny{0}{0}{(0)}_{i}$ for $i=0,1,\ldots,d\Sys$ be $d\Poi/d\Sys\times d\Poi/d\Sys$ Hermitian matrices whose eigenvalues are the $d\Poi/d\Sys$ largest eigenvalues of the initial pointer state $\rho\Poi=\tau\Poi(\beta)$, i.e.,
\begin{align}
    a\suptiny{0}{0}{(0)}_{i}  = M\suptiny{0}{0}{(0)}_{i}  \bigl(
	\operatorname{diag}(p\suptiny{0}{0}{(0)}_{0}, p\suptiny{0}{0}{(0)}_{1}, \ldots,
    p\suptiny{0}{0}{(0)}_{d\Poi/d\Sys-1}
	\bigr) {M\suptiny{0}{0}{(0)}_{i}}^{\dagger},
    \label{eq:adiag}
%\quad
%	\in \mathbb{R}^{\tfrac{d\Poi}{d\Sys} \times \tfrac{d\Poi}{d\Sys} }
%\, .
\end{align}
where the $M\suptiny{0}{0}{(0)}_{i}$ are $d\Poi/d\Sys\times d\Poi/d\Sys$ unitary matrices. The correlation matrix $\Gamma_{U_{C_{\mathrm{max}}}}$ is then of the form
\begin{align}
\Gamma_{U_{C_{\mathrm{max}}}} &=
\,\left[\begin{array}{c|c|c|c}
        \tikzmarkin[ver=style cyan, line width=0mm,]{n1}\raisebox{2pt}{\protect\footnotesize{$\rho_{00}a\suptiny{0}{0}{(0)}_{0}$}}\tikzmarkend{n1}  &  \tikzmarkin[ver=style green, line width=0mm,]{n2}\raisebox{2pt}{\protect\footnotesize{$\rho_{11}\tilde{A}_{01}$}}\tikzmarkend{n2}  &    \tikzmarkin[ver=style green, line width=0mm,]{n3}\raisebox{2pt}{\protect\footnotesize{$\rho_{22}\tilde{A}_{02}$}}\tikzmarkend{n3}  &\cdots  \\
        \hline
        \tikzmarkin[ver=style green, line width=0mm,]{n4}\raisebox{2pt}{\protect\footnotesize{$\rho_{00}\tilde{A}_{10}$}}\tikzmarkend{n4} & \tikzmarkin[ver=style cyan, line width=0mm,]{n5}\raisebox{2pt}{\protect\footnotesize{$\rho_{11}a\suptiny{0}{0}{(0)}_{1}$}}\tikzmarkend{n5}  & \tikzmarkin[ver=style green, line width=0mm,]{n6}\raisebox{2pt}{\protect\footnotesize{$\rho_{22}\tilde{A}_{12}$}}\tikzmarkend{n6} & \cdots \\
        \hline
        \tikzmarkin[ver=style green, line width=0mm,]{n7}\raisebox{2pt}{\protect\footnotesize{$\rho_{00}\tilde{A}_{20}$}}\tikzmarkend{n7}  &  \tikzmarkin[ver=style green, line width=0mm,]{n8}\raisebox{2pt}{\protect\footnotesize{$\rho_{11}\tilde{A}_{21}$}}\tikzmarkend{n8} &  \tikzmarkin[ver=style cyan, line width=0mm,]{n9}\raisebox{2pt}{\protect\footnotesize{$\rho_{22}a\suptiny{0}{0}{(0)}_{2}$}}\tikzmarkend{n9} & \hdots\\
        \hline
        \vdots & \vdots & \vdots & \ddots
    \end{array}\right]
    \,,
    \label{eq:gammaU}
\end{align}
where the block matrices on the diagonal (blue) are the corresponding entries of $\tilde{\rho}\SP$ restricted to the correlated subspace $\mathcal{H}_{\mathrm{corr}}$, whereas the block matrices on the off-diagonal of $\Gamma_{U_{C_{\mathrm{max}}}}$ (here shown in green) correspond to the diagonal blocks of $\tilde{\rho}\SP$ restricted to the non-correlated subspace $\mathcal{H}_{\mathrm{nc}}$. Additional off-diagonal entries may appear in the projection of $\tilde{\rho}\SP$ onto $\mathcal{H}_{\mathrm{nc}}$ between blocks $\tilde{A}_{ij}$ and $\tilde{A}_{i\pr j}$ with the same column index $j$ but different row indices $i\neq i\pr$ with $i,i\pr\neq j$, while maintaining an unbiased measurement with maximal correlation. The additional constraint of Eq.~\eqref{eq:unbiasedness cond proof} ensuring an unbiased measurement procedure can here be written as
$\tr[a\suptiny{0}{0}{(0)}_{j}] + \sum_{i\,,  \\i\neq j} \tr[\tilde{A}_{ij} ]= 1,  \, \forall \, j$. The remaining freedom of applying unitaries that leave the subspaces $\mathcal{H}_{\mathrm{corr}}$ and $\mathcal{H}_{\mathrm{nc}}$ invariant and are compatible with unbiasedness can be used for minimization of the corresponding energy cost. Before we discuss this procedure for arbitrary system and pointer dimensions, it will be instructive to consider the special case where the system is a single qubit and the pointer consists of $N$ identical two-level systems.

\subsection{Optimally Correlating Unitary for a Single-Qubit System and $N$-Qubit Pointer}\label{app:qubits}

In the previous appendix, we have identified the structure of all unitarily correlating unbiased measurements that create maximal correlations $C_{\mathrm{max}}$ (a subclass of the maps $\mathcal{E}\subtiny{0}{0}{\mathrm{I\hspace*{-0.5pt}I}}$). We are now interested in further restricting this set of measurements to identify those unitaries that achieve $C_{\mathrm{max}}$ for the least energy.
%In general, the energy
That is, we wish to determine the optimal $U_{\mathrm{opt}}$ which solves the optimisation problem
\begin{align}
    \min_{U_{\text{corr}}} \, \Delta E\subtiny{0}{0}{\mathrm{I\hspace*{-0.5pt}I}}
 \; \;  \text{s.t.} \;\; C(\tilde{\rho}_{\SP}) &= C_{\text{max}}
 %\quad \forall \;\beta, N
 \,.
 \label{eq:opt problem supplemental}
\end{align}
For arbitrary system dimensions and Hamiltonians, the explicit form of the solutions $U_{\mathrm{opt}}$ is rather involved and the proofs of optimality become very technical in nature. Before we move on to such general cases in Appendix~\ref{app:optimal}, let us therefore here illustrate the general method by focusing on an example of interest.

Here, we consider a two-dimensional system, i.e., a qubit with Hilbert space $\mathcal{H}\Sys=\mathbb{C}^{2}$, dimensions $d\Sys=2$, and a Hamiltonian $H\Sys$ with eigenstates $\ket{0}\Sys$ and $\ket{1}\Sys$ and spectral decomposition $H\Sys=E\Sys\ket{1}\!\!\bra{1}\Sys$. In addition, we assume that the system state is initially unknown such that the corresponding density operator is maximally mixed, $\rho\Sys = \tfrac{1}{2}\mathds{1}_{2}$. Meanwhile, we consider a measurement apparatus modelled as an $N$-qubit pointer, $d\Poi=2^{N}$, where each qubit has the same local Hamiltonian with vanishing ground state energy and energy gap matching the system energy gap, $E\Poi=E\Sys$. The total pointer Hamiltonian is thus $H\Poi= \sum_{k=0}^1\sum_{i=0}^{2^{N-1}-1} E\suptiny{0}{0}{(k)}_{i}\ket{E\suptiny{0}{0}{(k)}_{i}}\!\!\bra{E\suptiny{0}{0}{(k)}_{i}}$, where we have adopted the sector notation introduced in Eq.~\eqref{eq:diagonalise}. Note that the pointer spectrum is highly degenerate since there are $2^{N}$ eigenvalues but only $N+1$ different energy levels, $E\suptiny{0}{0}{(k)}_{i}/E\Sys\in\{0,1,\ldots,N\}$. With respect to the energy eigenbasis the initial pointer state before the correlating step is
\begin{align}
    \tau\Poi(\beta)^{\otimes N} &= \sum_{k=0}^{1}\sum_{i=0}^{2^{N-1} -1} p\suptiny{0}{0}{(k)}_{i}
	 \ket{E\suptiny{0}{0}{(k)}_{i}}\!\!\bra{E\suptiny{0}{0}{(k)}_{i}},
    \label{eq:thermal}
\end{align}
with $p\suptiny{0}{0}{(k)}_{i} = e^{-\beta E\suptiny{0}{0}{(k)}_{i}}/\mathcal{Z}$ and $\mathcal{Z} =\tr(e^{-\beta H\Poi})=\sum_{i,k}e^{-\beta E\suptiny{0}{0}{(k)}_{i}}$. For this setting, we will now solve the optimization problem of Eq.~(\ref{eq:opt problem supplemental}) for $\tilde{\rho}\SP = U_{\mathrm{corr}}\,(\rho\Sys\otimes \tau\Poi(\beta)^{\otimes N}\,)U_{\mathrm{corr}}^{\dagger}$.

%Since the details of the proof are very technical for arbitrary system dimensions and arbitrary Hamiltonians, in this section we focus on an example to illustrate the general method, which follows in Section~\ref{app:optimal}. Here, we examine the particularly interesting case when the measured system is a single qubit and the measurement apparatus is modelled as an $N$-qubit pointer. We assume the system Hamiltonian to have vanishing ground state and energy gap $E\Sys$, such that $H\Sys=E\Sys\ket{1}\!\!\bra{1}\Sys$ and that the state is initially unknown $\rho_S = \frac{1}{2}\mathds{1}_2$.

%Similarly, for each of the pointer qubits we have a vanishing ground state $E_0 =0$ and gap $E\Poi =1$; each qubit has the same local Hamiltonian $H\Poin = E\Poi\ket{E_1}\!\!\bra{E_1}\Poi$, such that

From \eqref{eq:cmax}, the maximum algebraic correlation achievable between an $N-$qubit pointer and a qubit system is
\begin{align}
C_{\text{max}}( \beta)
 &= \;\tfrac{1}{\mathcal{Z}} \sum_{i=0}^{2^{N-1} -1}
%\sum_{k=0}^{\frac{N}{2}} {{N}\choose{k}} e^{-kE\Poi\beta}
e^{-\beta E\suptiny{0}{0}{(0)}_{i}}
\,,
\end{align}
and the post-interaction correlation matrix associated with this scenario is given by
\begin{align}
\Gamma_{U_{C_{\text{max}}}} =
\,\left[\begin{array}{c|c}
        \tikzmarkin[ver=style cyan, line width=0mm,]{Q1}\raisebox{2pt}{\protect\footnotesize{$\rho_{00}a\suptiny{0}{0}{(0)}_{0}$}}\tikzmarkend{Q1}  & \tikzmarkin[ver=style green, line width=0mm,]{Q2}\raisebox{2pt}{\protect\footnotesize{$\rho_{11}\tilde{A}_{01}$}}\tikzmarkend{Q2}   \\
        \tikzmarkin[ver=style green, line width=0mm,]{Q3}\raisebox{2pt}{\protect\footnotesize{$\rho_{00}\tilde{A}_{10}$}}\tikzmarkend{Q3} & \tikzmarkin[ver=style cyan, line width=0mm,]{Q4}\raisebox{2pt}{\protect\footnotesize{$\rho_{11}a\suptiny{0}{0}{(0)}_{1}$}}\tikzmarkend{Q4}
    \end{array}\right]
    \, .
\end{align}
Here, $a\suptiny{0}{0}{(0)}_{0}$ and $a\suptiny{0}{0}{(0)}_{1}$ are $2^{N-1}\times 2^{N-1}$ dimensional Hermitian matrices whose eigenvalues are the $2^{N-1}$ largest eigenvalues of $\tau\Poi^{\otimes N}$, that is, there are a unitary matrices $M\suptiny{0}{0}{(0)}_{0}$ and $M\suptiny{0}{0}{(0)}_{1}$  such that
\begin{align}
a\suptiny{0}{0}{(0)}_{i} = M\suptiny{0}{0}{(0)}_{i}  \bigl(\text{diag}(p\suptiny{0}{0}{(0)}_{0}, \cdots, p\suptiny{0}{0}{(0)}_{2^{N-1} -1}) \bigl) {M\suptiny{0}{0}{(0)}_{i}}^{\dagger}\ \ \text{for}\ i=0,1.
\end{align}
In our example, we further have $\rho_{00}=\rho_{11}=\tfrac{1}{2}$ but we leave the symbols $\rho_{00}$ and $\rho_{11}$ for clarity where necessary. For the interaction to be unbiased according to Eq.~\eqref{eq:gammaU}, there is now no choice but to set $\tilde{A}_{01} = a\suptiny{0}{0}{(1)}_{1}$ and $\tilde{A}_{10} = a\suptiny{0}{0}{(1)}_{0}$, where
\begin{align}
    a\suptiny{0}{0}{(1)}_{i}  = M\suptiny{0}{0}{(1)}_{i}  \bigl(
	\operatorname{diag}(p\suptiny{0}{0}{(1)}_{0}, p\suptiny{0}{0}{(1)}_{1}, \ldots,
    p\suptiny{0}{0}{(1)}_{2^{N-1} -1}
	\bigr) {M\suptiny{0}{0}{(1)}_{i}}^{\dagger},
    \label{eq:adiag example 1}
\end{align}
and $M\suptiny{0}{0}{(1)}_{i}$ for $i=0,1$ are unitaries not yet fixed by the requirements of unbiasedness or maximal correlation. The eigenvalues $\{p\suptiny{0}{0}{(1)}_{0}, p\suptiny{0}{0}{(1)}_{1}, \ldots, p\suptiny{0}{0}{(1)}_{2^{N-1} -1}\}$ of $a\suptiny{0}{0}{(1)}_{0}$ and $a\suptiny{0}{0}{(1)}_{1}$ correspond to the second (smaller) half of the eigenvalues of $\tau\Poi^{\otimes N}$, and hence we have $\tr(a\suptiny{0}{0}{(0)}_{i})+\tr(a\suptiny{0}{0}{(1)}_{i})=1\ \forall\,i$. The correlation matrix becomes
\begin{align}
\Gamma_{U_{C_\text{max}}} =
\,\left[\begin{array}{c|c}
        \tikzmarkin[ver=style cyan, line width=0mm,]{QQ1}\raisebox{2pt}{\protect\footnotesize{$\rho_{00}a\suptiny{0}{0}{(0)}_{0}$}}\tikzmarkend{QQ1}  & \tikzmarkin[ver=style green, line width=0mm,]{QQ2}\raisebox{2pt}{\protect\footnotesize{$\rho_{11}a\suptiny{0}{0}{(1)}_{1}$}}\tikzmarkend{QQ2}   \\
        \tikzmarkin[ver=style green, line width=0mm,]{QQ3}\raisebox{2pt}{\protect\footnotesize{$\rho_{00}a\suptiny{0}{0}{(1)}_{0}$}}\tikzmarkend{QQ3} & \tikzmarkin[ver=style cyan, line width=0mm,]{QQ4}\raisebox{2pt}{\protect\footnotesize{$\rho_{11}a\suptiny{0}{0}{(0)}_{1}$}}\tikzmarkend{QQ4}
    \end{array}\right]
    \,.
    \label{eq:gammaopt}
\end{align}
In the present case where $d\Sys=2$ and a maximally correlated, unbiased measurement, the correlation matrix $\Gamma_{U_{C_\text{max}}}$ indeed catches all nonzero elements of $\tilde{\rho}\SP$, which is block diagonal,
\begin{align}
    \tilde{\rho}\SP &=\diag(\rho_{00}a\suptiny{0}{0}{(0)}_{0},\rho_{11}a\suptiny{0}{0}{(1)}_{1},\rho_{00}a\suptiny{0}{0}{(1)}_{0},\rho_{11}a\suptiny{0}{0}{(0)}_{1}).
\end{align}
In general, for $d\Sys>2$, additional nonzero off-diagonal elements may appear in $\tilde{\rho}\SP$ that are not explicitly captured by $\Gamma_{U_{C_\text{max}}}$.

The cost of correlating, and the function we wish to minimise, is given by the energy difference of the initial and final states,
\begin{align}\label{eq:ecorr}
\Delta E\subtiny{0}{0}{\mathrm{I\hspace*{-0.5pt}I}}
	 = E(\tilde{\rho}\SP) -  E({\rho}\SP).
\end{align}
Since the initial energy is fixed by the initial temperature (and any preparation one wishes to include), we focus on minimising $E(\tilde{\rho}\SP)$.
In order to facilitate the computation, it will be useful to decompose $E(\tilde{\rho}\SP)$ in terms of the correlated and non-correlated subspaces
\begin{align}
E(\tilde{\rho}\SP) &=
E_{\mathrm{corr}}
(\tilde{\rho}\SP) +
 E_{\mathrm{nc}}
 (\tilde{\rho}\SP) \\[1.5mm]
&=
 \tr
[
\sum_{i}
\tilde{\Pi}_{ii}
H\SP
 \tilde{\rho}\SP]%\nonumber
 %\\
 %&\qquad\qquad
 \,+\,\tr[
\sum
_{\substack{i,j\\ i\ne j}}
\tilde{\Pi}_{ij}
 H\SP
\;\tilde{\rho}\SP
],
\nonumber
\end{align}
where, again we have $\tilde{\Pi}_{ij} = \ket{i}\!\!\bra{i}\otimes \Pi_j$ and the combined system-pointer Hamiltonian is $H\SP = H\Sys\otimes \mathds{1}\Poi + \mathds{1}\Sys \otimes H\Poi$.
For our qubit example, measured by an $N$-qubit pointer, this becomes
\begin{align}\label{eq:energycorr}
E(\tilde{\rho}\SP)
&=
 \tr
[(\tilde{\Pi}_{00} + \tilde{\Pi}_{11})
H\SP
\tilde{\rho}\SP]%\nonumber
 \\[1.5mm]
 &\ \
 +\tr[
(\tilde{\Pi}_{01} + \tilde{\Pi}_{10})
 H\SP
\;\tilde{\rho}\SP
].
\nonumber
\end{align}
The class of unitaries that achieve $C_{\text{max}}$ rearranges the elements of $\rho\SP$ to place the `heaviest' populations of probabilities (eigenvalues of $\tau\Poi$) into the correlated subspaces of $\tilde{\rho}\SP$. From this constraint we already know which elements (eigenvalues) of the post-interaction state $\tilde{\rho}\SP$ are assigned to which subspaces. In order to minimise the energy, one must therefore find the optimal assignment of the energy eigenbasis to these subspaces, which amounts to determining the $\Pi_i$.

We now proceed as follows: First, we will minimise the energy in the correlated subspaces, after which we will minimise the energy in the non-correlated subspaces, a strategy that presents a global energy minimum for the entire state.\\

Noting that $\tilde{\Pi}_{ij}  \perp \tilde{\Pi}_{i\pr j\pr}$ whenever  $i\neq i\pr$ or $j\neq j\pr$ we observe that also $(\tilde{\Pi}_{00} + \tilde{\Pi}_{11})$ is a projector. We can hence rewrite the first term on right-hand side of Eq.~\eqref{eq:energycorr} as
\begin{align}
    &\tr\bigl[\sum\limits_{i=0,1}\tilde{\Pi}_{ii}\,H\SP\,\tilde{\rho}\SP\bigr]  \,=\,\tr\bigl[\bigl(\sum\limits_{i=0,1}\tilde{\Pi}_{ii}\bigr)^{2}\,H\SP\,\tilde{\rho}\SP\bigr]\nonumber\\
    &\ \ \ \ =\,\tr\bigl[\bigl(\sum\limits_{i=0,1}\tilde{\Pi}_{ii}\bigr)\,H\SP\,\tilde{\rho}\SP\,\bigl(\sum\limits_{j=0,1}\tilde{\Pi}_{jj}\bigr)\bigr]\nonumber\\
    &\ \ \ \ =\,\sum\limits_{i=0,1}\tr\bigl[\tilde{\Pi}_{ii}\,H\SP\,\tilde{\rho}\SP\,\tilde{\Pi}_{ii}\bigr],
    \label{eq:energy minimisation step 1}
\end{align}
where we have used the orthogonality of the projectors again in the last step. Once the trace has been restricted to the subspace corresponding to the space spanned by the nontrivial eigenvectors of $\tilde{\Pi}_{ii}$, we can further rewrite Eq.~(\ref{eq:energy minimisation step 1}) as
\begin{align}
    &\sum\limits_{i=0,1}\tr\bigl[\tilde{\Pi}_{ii}\,H\SP\,\tilde{\rho}\SP\,\tilde{\Pi}_{ii}\bigr]  \label{eq:energy minimisation step 2}\\
    &\ \ =\,\sum\limits_{i=0,1}\tr\bigl[\tilde{\Pi}_{ii}\,(H\Sys+H\Poi)\,\tilde{\rho}\SP\,\tilde{\Pi}_{ii}\bigr]\nonumber\\
    &\ \ =\,\rho_{11}\,E\Sys\,\tr\bigl[a\suptiny{0}{0}{(0)}_{1}\bigr]\,+\,\sum\limits_{i=0,1}\rho_{ii}\,\tr\bigl[\Pi_{i}H\Poi\Pi_{i}\,a\suptiny{0}{0}{(0)}_{i}\bigr].\nonumber
    %\nonumber\\
    %&\ \ =\,\tr\bigl[\bigl(\rho_{00}\,{M\suptiny{0}{0}{(1)}_{0}}^{\dagger}H\Poi {M\suptiny{0}{0}{(1)}_{0}}\nonumber\\
    %&\ \ \ \ +\rho_{11}{M\suptiny{0}{0}{(1)}_{1}}^{\dagger}(E\Sys+\Pi_{1}H\Poi\Pi_{1}){M\suptiny{0}{0}{(1)}_{1}}\bigr)a\suptiny{0}{0}{(0)}\bigr],
\end{align}
Here, one should note that, strictly speaking, $\Pi_{i}H\Poi\Pi_{i}$ are $d\Poi\times d\Poi$ matrices, while $a\suptiny{0}{0}{(0)}_{i}$ are $d\Poi/d\Sys\times d\Poi/d\Sys$ matrices. However, the image of $\Pi_{i}H\Poi\Pi_{i}$ is also of dimension $d\Poi/d\Sys$, and one may hence think of $\Pi_{i}H\Poi\Pi_{i}$ as nonzero $d\Poi/d\Sys\times d\Poi/d\Sys$ matrices padded by rows and columns of zeros. In a slight abuse of notation we use the same symbol for the entire operator and its nontrivial block, since it is clear from the context, which object is referred to. In particular, $(\Pi_{i}H\Poi\Pi_{i})\,a\suptiny{0}{0}{(0)}_{i}$ refers to the product of two $d\Poi/d\Sys\times d\Poi/d\Sys$ matrices. To simplify the last line of Eq.~(\ref{eq:energy minimisation step 2}), let us first write $a\suptiny{0}{0}{(0)}:=\text{diag}(p\suptiny{0}{0}{(0)}_{0}, \cdots, p\suptiny{0}{0}{(0)}_{2^{N-1} -1})$, where we
may assume w.l.o.g. that $a\suptiny{0}{0}{(0)}$ is diagonal w.r.t. the same basis as $\Pi_{i}H\Poi\Pi_{i}$ and the populations are ordered in non-increasing order. Any mismatch can be absorbed into the choice of the ${M\suptiny{0}{0}{(0)}_{i}}$. Further note for the first term that
$\tr\bigl[a\suptiny{0}{0}{(0)}_{1}\bigr]=\tr\bigl[a\suptiny{0}{0}{(0)}\bigr]$, whereas the second term can be expressed as
\begin{align}
    \tr\bigl[\Pi_{i}H\Poi\Pi_{i}\,a\suptiny{0}{0}{(0)}_{i}\bigr]
    &=\,\tr\bigl[\Pi_{i}H\Poi\Pi_{i}\,
    {M\suptiny{0}{0}{(0)}_{i}}a\suptiny{0}{0}{(0)}{M\suptiny{0}{0}{(0)}_{i}}^{\dagger}\bigr]\nonumber\\
    &=\,\tr\bigl[{M\suptiny{0}{0}{(0)}_{i}}^{\dagger}\Pi_{i}H\Poi\Pi_{i}\,
    {M\suptiny{0}{0}{(0)}_{i}}a\suptiny{0}{0}{(0)}\bigr].
    \label{eq:energy minimisation step 3}
\end{align}
The quantity that we wish to minimize in this first step is thus of the form
\begin{align}
    E_{\mathrm{corr}}(\tilde{\rho}\SP) &=\,
    \tr\bigl[\,\bigl(\rho_{00}{M\suptiny{0}{0}{(0)}_{0}}^{\dagger}\Pi_{0}H\Poi\Pi_{0}{M\suptiny{0}{0}{(0)}_{0}}\\
    &\ +\rho_{11}{M\suptiny{0}{0}{(0)}_{1}}^{\dagger}\Pi_{1}(H\Poi+E\Sys)\Pi_{1}{M\suptiny{0}{0}{(0)}_{1}}\bigr)a\suptiny{0}{0}{(0)}\bigr].
    \nonumber
\end{align}
The minimization is to be carried out over all choices of projectors $\Pi_{i}$, i.e., choosing the basis $\{\ket{\tilde{\psi}\suptiny{0}{0}{(i)}_{n}}\}_{i,n}$ in relation to the eigenbasis of $H\SP$, as well as over all choices of the ${M\suptiny{0}{0}{(0)}_{i}}$, or, in general the ${M\suptiny{0}{0}{(j)}_{i}}$. While the minimisation over the $\Pi_{i}$ requires some more in-depth analysis (that we perform below), a simple observation can made right away. For a given initial state, the state in the unitary orbit of the initial state with minimal energy w.r.t. to a given Hamiltonian must be diagonal in the eigenbasis of this Hamiltonian, i.e., the corresponding passive state. Meanwhile, the conditions of unbiasedness and maximal algebraic correlations impose a certain block structure once a basis has been fixed and allow for but do not require off-diagonal elements. Therefore, it is clear that the unbiased, minimal energy solution with maximal algebraic correlations must be diagonal w.r.t. $H\SP$, restricting the unitaries connecting the bases $\{\ket{\tilde{\psi}\suptiny{0}{0}{(i)}_{n}}\}_{i,n}$ with the eigenbasis of $H\SP$ as well as the unitaries ${M\suptiny{0}{0}{(j)}_{i}}$ to be \emph{permutation} matrices. Moreover, suppose that for the fixed choice of ${M\suptiny{0}{0}{(0)}_{i}}=\mathds{1}\ \forall i$ one has found an optimal choice of $\Pi_{i}$. Then any nontrivial modification of any of the ${M\suptiny{0}{0}{(0)}_{i}}$ can only increase the energy in the respective subspace. Without loss of generality we therefore set ${M\suptiny{0}{0}{(0)}_{i}}=\mathds{1}\ \forall i$.

%In the last line, the trace may be taken with respect to the eigenbases of $a\suptiny{0}{0}{(0)}_{0}$ and  $a\suptiny{0}{0}{(0)}_{1}$, respectively, meaning that the
%
%\begin{align}
%\tr[(\tilde{\Pi}_0 &+ \tilde{\Pi}_1 ) H\SP \tilde{\rho}\SP (\tilde{\Pi}_0 + \tilde{\Pi}_1 )] \\
%&= \tr[
%\tilde{\Pi}_0 H\SP \tilde{\rho}\SP \tilde{\Pi}_0] + \tilde{\Pi}_1 H\SP \tilde{\rho}\SP \tilde{\Pi}_1]
%] \\
%&=\tr[\tilde{\Pi}_0 H\SP \tilde{\Pi}_0\tilde{\Pi}_0\tilde{\rho}\SP \tilde{\Pi}_0] + \tr[\tilde{\Pi}_1 H\SP \tilde{\Pi}_1\tilde{\Pi}_1\tilde{\rho}\SP \tilde{\Pi}_1]\\
%&=\tr[\tilde{\Pi}_0 H\SP \tilde{\Pi}_0\rho_{00}a\suptiny{0}{0}{(0)} ] + \tr[\tilde{\Pi}_1 H\SP \tilde{\Pi}_1\rho_{11}a\suptiny{0}{0}{(0)}]
%\end{align}
With this, we can now rewrite the energy in the correlated subspace as
\begin{align}
E_{\mathrm{corr}}
	(\tilde{\rho}\SP)
%&= \tr[  \tilde{\Pi}_0) H\SP\tilde{\rho}\SP  ] \\
%% &=\tr[ \rho_{00}\tilde{\Pi}_{00} H\SP  \tilde{\Pi}_{00}\,a\suptiny{0}{0}{(0)} ] \\\nonumber
%%&\qquad\qquad + \tr[	\rho_{11}\tilde{\Pi}_{11} H\SP  \tilde{\Pi}_{11}
%%a\suptiny{0}{0}{(0)}
%% ]\\
%  &=\tr[
%( \rho_{00}\tilde{\Pi}_{00} H\SP  \tilde{\Pi}_{00} + \rho_{11}\tilde{\Pi}_{11} H\SP  \tilde{\Pi}_{11})
% \,a\suptiny{0}{0}{(0)}
%  ]\\
%%%%%%%%%%%%%
&=\,\tr\bigl[\,\bigl(\rho_{00}\Pi_{0}H\Poi\Pi_{0}\label{eq:energyCorr}\\
&\ \ \ \ \ \ +\rho_{11}\Pi_{1}(H\Poi+E\Sys)\Pi_{1}\bigr)a\suptiny{0}{0}{(0)}\bigr]%\nonumber\\
%%%%%%%%%%%%%
 %&
 =\mathbf{x}\cdot \mathbf{a}^{0},\nonumber
\end{align}
where $\mathbf{a}^{0}$ is the vector of diagonal entries of the matrix $a\suptiny{0}{0}{(0)}$
and
$\mathbf{x}\in\mathbb{R}^{2^{N-1}}$ is the vector of diagonal entries of the matrix $\bigl(\rho_{00}\Pi_{0}H\Poi\Pi_{0}+\rho_{11}\Pi_{1}(H\Poi+E\Sys)\Pi_{1}\bigr)$.
%Without loss of generality we take the permutation $M_{{\suptiny{0}{0}{(0)}}}=\mathds{1}$ to be the permutation that orders the populations in non-increasing order such that $a\suptiny{0}{0}{(0)} =\text{diag}(  p_0^{\suptiny{0}{0}{(0)}}, \cdots, p^{\suptiny{0}{0}{(0)}}_{2^{N-1} -1},) $.
Every component $x_{i}$ of $\mathbf{x}$ is seen to be sum of two energies from the $\Pi_{0}$ and $\Pi_{1}$ subspaces of $H\Poi$, respectively modulated by the respective system populations $\rho_{ii}$.  \\

We now switch to a slightly less cumbersome notation for the pointer. Let the set $\mathcal{S}_N$ with elements $s_i $ for $ i\in \{0, \cdots, d\Poi-1\}$ be the set of energies (in units of $E\Sys$) in the energy spectrum of the pointer, ordered in non-decreasing order, such that $H\Poi = \sum_i s_i \ket{s_i}\!\!\bra{s_i}$ and $s_i\leq s_j \;\forall i<j$. For example, a $3$-qubit pointer with gap $E\Poi = E\Sys$ and vanishing ground state would be associated with the set $\mathcal{S}_3 =\{0, 1, 1, 1, 2, 2, 2,3\}$.
The elements of the vector $\mathbf{x}$ can now be written as
%, respectively, i.e., from the matrices $\tilde{A}_0$ and $\tilde{B}_1$
\begin{align}\label{eq:xVec}
\begin{split}
x_i = \rho_{00}s_j +\rho_{11}(s_l + E\Sys) \quad &j\neq l,
\\
	& 0\le  j, l \le (2^{N}-1 ),\\
	&  s_{j,l} \in \mathcal{S}_N
		\;,
		\end{split}
\end{align}
such that the indices $j,l$ are used only once. Thus, the $x_i$ are composed by selecting pairs of elements, without replacement, from the set $\mathcal{S}_N$.
There are several statements we can make immediately about the set $\mathcal{S}_N$. First, it has $2^{N}$ elements which are distributed binomially such that the energy $k E\Poi $ appears ${{N}\choose{k}}$ times. Second, in the case that we are probing an unknown state, $\rho\Sys = \tfrac{1}{2}\mathds{1}_2$, the sum of the elements
%%%%%%%%%%%%%%%%%%%%%%%%%%%%%%%%%%%%%
\begin{figure}[ht!]
\begin{center}
\includegraphics[scale=0.4]{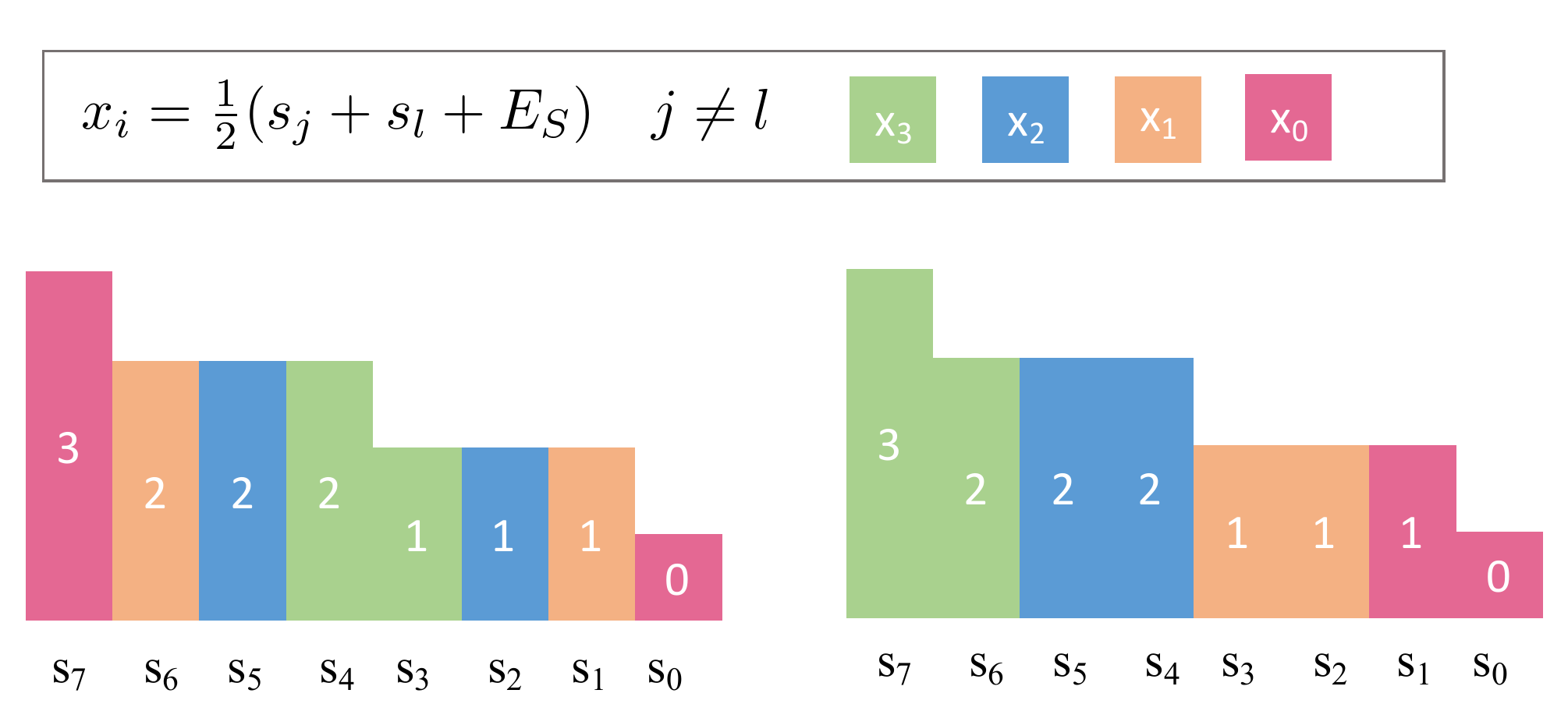}
\caption{The energies of a $3-$qubit pointer $ \mathcal{S}_3 =\{s_0, \cdots, s_7\}$ with  gap $E\Poi =1$ are arranged in non-increasing order. The schematic show two ways of choosing $x_i$ from the set $\mathcal{S}_3$. The right hand side selects nearest neighbour pairs and thus represents the optimal pairing that minimises $E_{\mathrm{corr}}(\tilde{\rho}\SP) $ in~\eqref{eq:minprob}.}
\label{fig:strategy}
\end{center}
\end{figure}
\noindent
%%%%%%%%%%%%%%%%%%%%%%%%%%%%%%%%%%%%
of $\mathbf{x}$ is constant. Namely
\begin{align}
\sum_{i = 0}^{2^{N-1} -1}x_i &=\sum_{i = 0}^{2^{N} -1} (s_i + \tfrac{1}{4}E\Sys)
%&\qquad  s_i \in \mathcal{S}_N
= c\,.
\end{align}
Since $x_i \ge 0 \;\forall i$, this means we can treat the set $\left\{{x_i}/{c}\right\}_i$ as a normalised probability distribution.
Let $X$ denote the set of all possible vectors $\mathbf{x}$, then, minimising the energy in the correlated subspace amounts to
\begin{align}\label{eq:minprob}
\min E_{\mathrm{corr}}
	(\tilde{\rho}\SP)   &= \min_{\mathbf{x}\in X}(\mathbf{x}\cdot\mathbf{a}^{0})\,=\,\mathbf{x}^{*}\cdot\mathbf{a}^{0}.
\end{align}
The set $X$ can be understood as the set of all possible ways of choosing pairs from $\mathcal{S}_N$ without replacement. The size (i.e., the cardinality) of $X$, denoted $|X|$,  grows factorially with $N$, so searching by brute-force for the optimal vector is not feasible.
The solution $\mathbf{x}^{*}$ for the minimization problem in Eq.~\eqref{eq:minprob} is given by the vector that pairs the smallest weights $p\suptiny{0}{0}{(0)}_{i}$ with the largest values $x_{i}$. Specifically, let $\mathbf{v}, \mathbf{w} \in \mathbb{R}^{2^{N-1}}$ be two normalised vectors with their components ordered in non-increasing order such that $v_0\ge v_1\cdots $ and $w_0\ge w_1\cdots $. We say that $\mathbf{v}$ majorises $\mathbf{w}$, written $\mathbf{v}\succ\mathbf{w}$, when $\sum_i^k v_i \ge \sum_i^k w_i\;\forall \;k$. In other words the cumulative sum of the components of the vector $\mathbf{v}$ grows faster than for $\mathbf{w}$. The vector $\mathbf{x}^{*}$ that presents the solution to the optimization problem is hence the vector that majorises all other vectors $\mathbf{x}\in X$, i.e.,
\begin{align}
    \mathbf{x}^{*}\succ\mathbf{x}\ \qquad\forall\,\mathbf{x}\in X.
    \label{eq:majorisation relation}
\end{align}
This vector is constructed by maximising each $x_i^*$ term by term from the bottom up, populating the components of $\mathbf{x}^*$ such that $x_{2^{N-1}-1} \ge \cdots \ge x_0$. This construction amounts to picking nearest neighbour pairs from the set $\mathcal{S}_N$, starting with the largest pair, as illustrated in Fig.~\ref{fig:strategy}. Thus, the components of the optimal solution take the form
\begin{align}
    x_i^*= \rho_{00}s_{2i} +\rho_{11}(s_{2i+1} +E\Sys)\qquad s_i \in \mathcal{S}_N
\,.
\end{align}
By construction, the majorisation of Eq.~(\ref{eq:majorisation relation}) is satisfied, and we have found the minimum energy solution in the correlated subspace.
%\noindent

By constructing $\mathbf{x}^*$ from nearest neighbour pairs in $\mathcal{S}_N$, we have fixed the energy eigenbasis in the pointer Hilbert space. The projectors on the pointer are then
\begin{align}\label{eq:proj}
\Pi_0 = \sum_{i=0}^{2^{N-1}-1} \ket{s_{2i}}\!\!\bra{s_{2i}}, \ \
\Pi_1 = \sum_{i=0}^{2^{N-1}-1}  \ket{s_{2i+1}}\!\!\bra{s_{2i+1}}
\,.
\end{align}

We now proceed to minimise the energy in the non-correlated subspaces.
Following a similar calculation and series of arguments as leading to Eq.~\eqref{eq:energyCorr}, we write the energy as
\begin{align}
    E_{\mathrm{nc}}(\tilde{\rho}\SP)
 %&=\tr[ \rho_{00}\tilde{\Pi}_{01} H\SP  \tilde{\Pi}_{01}\,a^{\suptiny{0}{0}{(1)}} ] \\\nonumber
%&\qquad\qquad + \tr[	\rho_{11}\tilde{\Pi}_{10} H\SP  \tilde{\Pi}_{10}
%a^{\suptiny{0}{0}{(1)}}
% ]\\
    &=\,\tr\bigl[\,\rho_{11}\Pi_{1}H\Poi\Pi_{1}\,{M\suptiny{0}{0}{(1)}_{1}}a\suptiny{0}{0}{(1)}{M\suptiny{0}{0}{(1)}_{1}}^{\dagger} \nonumber\\
    &\ +\rho_{00}\Pi_{0}(H\Poi+E\Sys)\Pi_{0}{M\suptiny{0}{0}{(1)}_{0}}a\suptiny{0}{0}{(1)}{M\suptiny{0}{0}{(1)}_{0}}^{\dagger}\bigr]\nonumber\\[1mm]
    &=\mathbf{y}_{1}\cdot \mathbf{a}^1_{1}+\mathbf{y}_{0}\cdot \mathbf{a}^1_{0}\nonumber\\[1mm]
    &=\mathbf{y}_{1}\cdot {M\suptiny{0}{0}{(1)}_{1}}\mathbf{a}^1+\mathbf{y}_{0}\cdot {M\suptiny{0}{0}{(1)}_{0}}\mathbf{a}^1,
\label{eq:energyunCorr}
\end{align}
where $\mathbf{{a}}^{1}_{1}$ and $\mathbf{{a}}^{1}_{0}$ are vectors whose components are the eigenvalues of the matrix $a\suptiny{0}{0}{(1)}$ in, as of yet, undetermined permutations (fixed by ${M\suptiny{0}{0}{(1)}_{1}}$ and ${M\suptiny{0}{0}{(1)}_{0}}$). The vector $\mathbf{a}^1$, in turn, collects exactly these eigenvalues in non-increasing order. However, since the energy basis for the pointer has been fixed by Eq.~\eqref{eq:proj}, the vectors $\mathbf{y}_{1}$ and $\mathbf{y}_{0}$ are completely determined.
Their components are given by
\begin{align}\label{eq:yVec}
%\begin{split}
%y_i = \rho_{11}s_{2i+1} +\rho_{00}(s_{2i} + E\Sys) \quad &j\neq l
(y_1)_i \,=\, \rho_{11}s_{2i+1},\quad (y_0)_i \,=\, \rho_{00}(s_{2i} + E\Sys)\,.
%\\
%	& 0\le  j, l \le (2^{N}-1 )\\
%	&  s_{j,l} \in \mathcal{S}_N
%		\;.
%		\end{split}
\end{align}
To minimise the energy in the non-correlated subspace, we are thus looking for the solution to the optimisation problem
\begin{align}\label{eq:minprobY}
\min E_{\mathrm{nc}}
(\tilde{\rho}\SP)   &= \min_{M\suptiny{0}{0}{(1)}_{0},M\suptiny{0}{0}{(1)}_{1}}(
\mathbf{y}_{0}\cdot M\suptiny{0}{0}{(1)}_{0}\mathbf{a}^1+\mathbf{y}_{1}\cdot M\suptiny{0}{0}{(1)}_{1}\mathbf{a}^1)%\\ \
%&=\mathbf{y}\cdot M_{*}\mathbf{a}^1
\,,
\end{align}
i.e., to find the optimal permutation matrices $M\suptiny{0}{0}{(1)}_{0}$ and $M\suptiny{0}{0}{(1)}_{1}$. %$M_*$ of $\mathbf{a}^1$.
Because of the freedom to choose these two permutations independently, the optimizations in the two subspaces decouple and it can be easily seen that the optimal solution for both is to pair up the smallest energies with the largest weights. In other words, to select $M\suptiny{0}{0}{(1)}_{0}=M\suptiny{0}{0}{(1)}_{1}=\mathds{1}$. For $\rho_{00}=\rho_{11}$, this in turn implies
\begin{align}\label{eq:minprobY}
\min E_{\mathrm{nc}}(\tilde{\rho}\SP)
&=\,(\mathbf{y}_{0}+\mathbf{y}_{1})\cdot \mathbf{a}^1
\,=\,\mathbf{x}^{*}\mathbf{a}^1
\,.
\end{align}
where we have noted that for the special case\footnote{This equality holds in the special case that $d\Sys =2$, for the more general case see Appendix~\protect\ref{app:optimal}.} of $d\Sys=2$ one may collect $\mathbf{y}_{0}$ and $\mathbf{y}_{1}$ into $\mathbf{y}_{0}+\mathbf{y}_{1} = \mathbf{x}^{*}$.

%By a similar majorisation argument to the correlated subspace we conclude that
$M^{*}\mathbf{a}^1$ must be ordered in non-increasing order to achieve the global minimum, which in turn implies that $M_* = \mathds{1}$.

Collecting the results for the correlated and non-correlated subspaces and substituting for the forms of $\mathbf{a}^0$ and $\mathbf{a}^1$, the total energy after the interaction is
\begin{align}\label{eq:optcorren}
\min E(\tilde{\rho}\SP)
    &= \min ( E_{\mathrm{corr}}(\tilde{\rho}\SP) + E_{\mathrm{nc}}(\tilde{\rho}\SP) )\nonumber\\[1mm]
    &=\,\mathbf{x}^{*}\cdot (\mathbf{{a}^0} +\mathbf{{a}^1}) \\
    &= \tfrac{1}{2}\sum_{i=0}^{2^{N-1} -1} \bigl(s_{2i} +s_{2i+1} +E\Sys\bigr) (p\suptiny{0}{0}{(0)}_i + p\suptiny{0}{0}{(1)}_i ).\nonumber
\end{align}
Since the initial state is diagonal w.r.t. the energy eigenbasis, the initial energy can also be easily computed to be
\begin{align}\label{eq:initialE}
E(\rho\SP)
%&= \sum_{i=0}^{2^{N-1} -1}
%( \rho_{00} s_i + \rho_{11} ( s_i  + E\Sys) )p_i^{\suptiny{0}{0}{(0)}}   +\\
%&\qquad( \rho_{00} s_{(2^{N-1}+i)} + \rho_{11} ( s_{(2^{N-1}+i)} + E\Sys) )p_i^{\suptiny{0}{0}{(1)}}
&= \tfrac{1}{2}\sum_{i=0}^{2^{N-1} -1}\bigl[
( 2 s_i  + E\Sys )p\suptiny{0}{0}{(0)}_i   \\
&\ \ \ \ \ \ \ \ \ \ \ \ \ \ + ( 2 s_{(2^{N-1}+i)} + E\Sys )p\suptiny{0}{0}{(1)}_{i}\bigr].\nonumber
\end{align}
Thus from the above and Eq.~\eqref{eq:initialE} we have
\begin{align}
\Delta E\subtiny{0}{0}{\mathrm{I\hspace*{-0.5pt}I}}
	&=
 E(\tilde{\rho}\SP) -  E(\rho\SP) \\
 &=\tfrac{1}{2}\sum_{i=0}^{2^{N-1} -1} ( s_{2i} +s_{2i+1}  -2s_i ) p\suptiny{0}{0}{(0)}_i \nonumber\\
&\qquad\quad + ( s_{2i}  + s_{2i+1} -  2 s_{(2^{N-1}+i)}) p\suptiny{0}{0}{(1)}_{i},\nonumber
\end{align}
where we note that, $E\Sys$ (the gap of the system) no longer plays any role. Finally, observe that the cost of correlating is always finite. If one substitutes for the $p\suptiny{0}{0}{(j)}_i$ from Eq.~\eqref{eq:thermal} and takes the limit in which the pointer is in a pure state, i.e., $\beta \rightarrow \infty$, then the maximal correlation indeed is perfect correlation, $C=1$, and the corresponding cost of correlating is given by
\begin{align}
\lim_{\beta \rightarrow \infty} \Delta E\subtiny{0}{0}{\mathrm{I\hspace*{-0.5pt}I}}
	 =\Delta E\subtiny{0}{0}{\mathrm{I\hspace*{-0.5pt}I}}
	 ^{\,(C = 1)} =  \tfrac{1}{2} E\Poi
\,.
\end{align}
Notably, this expression is independent of $N$ and hence true also when $N=1$. Therefore, regardless of how many qubits the pointer consists of, if these qubits are initially in the ground state, the cost of correlating the system and pointer is precisely the cost of exciting only a single qubit (modulated by $\rho_{00}=\rho_{11}=\tfrac{1}{2}$).
%In other words, regardless of the dimension of the pointer, if at least one particle is in the ground state state, the cost of correlating the system and pointer is precisely the cost of exciting the particle.\\

These results can equivalently be expressed in terms of the sector notation introduced in Appendix~\ref{sec:framework}. In this case, the projectors in Eq.~\eqref{eq:proj} become
\begin{subequations}
\begin{align}
\Pi_0 &= \sum_{k=0}^1 \sum_{i=0}^{2^{N-2}-1} \ket{E_{2 i}\suptiny{0}{0}{(k)}}\!\!\bra{E_{2i}\suptiny{0}{0}{(k)}}, \\
\Pi_1 &=\sum_{k=0}^1 \sum_{i=0}^{2^{N-2}-1} 	\ket{E_{2i+1}\suptiny{0}{0}{(k)}}\!\!\bra{E_{2i+1}\suptiny{0}{0}{(k)}}
\,.
\end{align}
\end{subequations}
Similarly, the energy after optimally correlating in Eq.~\eqref{eq:optcorren} is
\begin{align}
E(\tilde{\rho}\SP)=
&\tfrac{1}{2}\sum_{j=0}^1\sum_{i=0}^{2^{N-2} -1}
\bigl(E_{2 i }\suptiny{0}{0}{(j)}  + E_{2 i +1}\suptiny{0}{0}{(j)} + E\Sys \bigl)\\
&\ \ \ \ \ \ \ \times( p_{i+j2^{N-2}}\suptiny{0}{0}{(0)}+  p_{i+j2^{N-2}}\suptiny{0}{0}{(1)}),\nonumber
\end{align}
and the cost of correlating is
\begin{align}
\Delta E\subtiny{0}{0}{\mathrm{I\hspace*{-0.5pt}I}}
	 &=
\tfrac{1}{2}\sum_{j=0}^{1}\!\!\!\sum_{i=0}^{2^{N\!-\!2}\! -\!1}\! \!\!\bigl[\bigl(
E_{2 i }\suptiny{0}{0}{(j)}  \!+\! E_{2 i +1}\suptiny{0}{0}{(j)}  \!-\! 2E_{i+j2^{N-2}}\suptiny{0}{0}{(0)}
\bigl) p_{i+j2^{N-2}}^{\suptiny{0}{0}{(0)}}\nonumber \\[1mm]
&\qquad\ \ \ + \bigl(
E_{2 i }\suptiny{0}{0}{(j)}  + E_{2 i +1}\suptiny{0}{0}{(j)}  - 2E_{i+j2^{N-2}}\suptiny{0}{0}{(1)}
\bigl) p_{i+j2^{N-2}}\suptiny{0}{0}{(1)}\bigr].
\end{align}

%%%%%%%%%%%%%%%%%%%%%%%%%%%%%%%%%%%%%%%%%%%%%%%%%%%%%%%%%%%%%%%%%%%%%%%%%%%%%%%%%%%%%%%%%%%%%%%%%%%%%%%%%%%%%%%%%%%%%

\subsection{Construction of the optimal unitary for arbitrary systems}\label{app:optimal}

In the previous appendix we provided the construction for correlating a qubit system with an $N-$qubit pointer to $C_{\text{max}}$. We also proved that this construction was an energy minimum. This construction generalises to any unknown quantum systems $\rho\Sys = \tfrac{1}{d\Sys}\mathds{1}_{d\Sys}$ and  thermal pointers $\tau\Poi (\beta)$, with arbitrary Hamiltonians $H\Sys$ and $H\Poi$. Below we provide the recipe for constructing such a unitary.\\

Consider the thermal pointer $\tau\Poi(\beta)$ and order the spectrum of the pointer Hamiltonian in terms of its excitations into $d\Sys$ sectors of size $d\Poi/d\Sys$, i.e., $H\Poi=\sum_{k=0}^{d\Sys-1}\sum_{i=0}^{d\Poi/d\Sys -1}E\suptiny{0}{0}{(k)}_{i}\ket{E\suptiny{0}{0}{(k)}_{i}}\!\!\bra{E\suptiny{0}{0}{(k)}_{i}}$ with $E\suptiny{0}{0}{(k)}_{i}\leq E\suptiny{0}{0}{(k\pr)}_{j}\ \forall i,j$ for $k\pr>k$.
%\\
%
%$\{ E_0^{\Poi}, E_1^{\Poi}, \cdots, E_{d_p -1}^{\Poi}\}$ s.t. $E_i^{\Poi} \le E_j^{\Poi}$ and $\mathcal{P} = \{ p_0, p_1, \cdots, p_{d_p-1 }\}$ s.t. $p_i \le p_j$ and
%$
%\tau\Poi(\beta) = \sum_{i}^{d_p} p_i \ket{E_i^{\Poi}}\bra{E_i^{\Poi}}
%$.
%%The populations are given by $p_i = 1/\mathcal{Z} e^{-\beta E_i}$ and the Hamiltonian is $H\Poi = \sum_i E_i^{\Poi} \ket{E_i^{\Poi}}\bra{E_i^{\Poi}}$. Note that in this way all degeneracies are enumerated.
Diagonalise the pointer and the system in their ordered energy eigenbases,
\begin{align}
\begin{split}
\tau\Poi(\beta) &= \sum_{k=0}^{d\Sys-1}\sum_{i=0}^{d\Poi/d\Sys -1} p\suptiny{0}{0}{(k)}_i \ket{E\suptiny{0}{0}{(k)}_i}\!\!\bra{E\suptiny{0}{0}{(k)}_i}, \\
\rho\Sys &= \sum_{i}^{} \rho_{ii} \ket{i}\!\!\bra{i},
\end{split}
\end{align}
where $p\suptiny{0}{0}{(k)}_i = {1}/{\mathcal{Z}} e^{-\beta E\suptiny{0}{0}{(k)}_i}$.\\

Assign the largest $d\Poi/d\Sys$ eigenvalues of the pointer state $\tau\Poi(\beta)$ (captured in the matrix $a\suptiny{0}{0}{(0)}$) to the correlated subspace. The form of the correlation matrix is given in Eq.~\eqref{eq:gammaU}.
To minimise the energy contribution from the correlated subspace, given by $ E_{\mathrm{corr}}(\tilde{\rho}\SP)=\tr[\sum_{i}\tilde{\Pi}_{ii}H\SP \tilde{\rho}\SP]$,
%**
%
%The construction of the optimal unitary presents further constraints to the final state and also the form of the correlation matrix, since we are looking for a transformation that obeys the condition above, as well as being a minimum of global energy. Although the final state \eqref{eq:finalstate} is diagonal in the energy eigenbasis, we still have not specified the form of the projectors $\Pi_i$ in eq.~\eqref{eq:maxcorrel} onto the pointer Hilbert space. Specifying these determines the energy of the final state, up to degeneracies.
%The choice of projectors presenting the optimal solution is:
choose the pointer Hilbert space projectors to be
  \begin{align}\label{eq:projgen}
  \Pi_i &= \sum_{k=0}^{{d\Sys} -1}
   \sum_{i=0}^{(d\Poi/d\Sys^2)-1}\!\!\!
  	\ket{E_{d\Sys \cdot j +i}\suptiny{0}{0}{(k)}}\!\!\bra{E_{d\Sys \cdot j +i}\suptiny{0}{0}{(k)}}\\
  	&\ \ \ \ \ \ \ \ \ \ 	 \forall \,j \in \{0, \cdots, d\Sys -1\}\,.\nonumber
  \end{align}
This choice fixes the basis vectors for the pointer and thus it remains to distribute the remaining probability weights (the remaining eigenvalues of $\tau\Poi$) in the non-correlated subspace. This is achieved by pairing the largest weights with the smallest energies. The remaining weights are
\begin{align}
    a\suptiny{0}{0}{(i)} &=\,  \bigl(\text{diag}(p\suptiny{0}{0}{(0)}_0, \cdots, p\suptiny{0}{0}{(0)}_{2^{N-1} -1}) \bigl)  \quad i \in\{1, \cdots, d\Sys-1\},
\end{align}
and the resulting correlation matrix, arising from the optimal unitary $U_{\text{opt}}$ has the form
\begin{align}
\Gamma_{U_{\text{opt}}} =\!\!
    	\begin{array}{c@{\!\!\!}l}
\left[\begin{array}{c|c|c|c}
        \!\tikzmarkin[ver=style cyan, line width=0mm,]{A11}\raisebox{2pt}{\protect\footnotesize{$\rho_{00}{a\suptiny{0}{0}{(0)}}$}}\tikzmarkend{A11}\!  &
        \!\tikzmarkin[ver=style green, line width=0mm,]{A12}\raisebox{2pt}{\protect\footnotesize{$\rho_{11}{a\suptiny{0}{0}{(1)}}$}}\tikzmarkend{A12}\!  & \!\cdots\!  &
         \!\tikzmarkin[ver=style green, line width=0mm,]{A15}\raisebox{2pt}{\protect\footnotesize{$\rho_{d\Sys-1 \, d\Sys-1 }{a\suptiny{0}{0}{(1)}}$}}\tikzmarkend{A15}\!  \\
        \hline
         \!\tikzmarkin[ver=style green, line width=0mm,]{A21}\raisebox{2pt}{\protect\footnotesize{$\rho_{00}{a\suptiny{0}{0}{(1)}}$}}\tikzmarkend{A21}\!  & \!\tikzmarkin[ver=style cyan, line width=0mm,]{A22}\raisebox{2pt}{\protect\footnotesize{$\rho_{11}{a\suptiny{0}{0}{(0)}}$}}\tikzmarkend{A22}\!   & \!\cdots\! &          \!\tikzmarkin[ver=style green, line width=0mm,]{A25}\raisebox{2pt}{\protect\footnotesize{$\rho_{d\Sys-1 \, d\Sys-1 }{a\suptiny{0}{0}{(2)}}$}}\tikzmarkend{A25}\!   \\
        \hline
        \!\vdots\! & \!\vdots\!  & \!\ddots\! & \!\vdots\! \\
          \hline
        \!\tikzmarkin[ver=style green, line width=0mm,]{A55}\raisebox{2pt}{\protect\footnotesize{$\rho_{00}{a\suptiny{0}{0}{(d\Sys-1)}}$}}\tikzmarkend{A55}\! &         \!\tikzmarkin[ver=style green, line width=0mm,]{A65}\raisebox{2pt}{\protect\footnotesize{$\rho_{11}{a\suptiny{0}{0}{(d\Sys-1)}}$}}\tikzmarkend{A65}\!  & \!\cdots\!  &
        \!\tikzmarkin[ver=style cyan, line width=0mm,]{aa2}\raisebox{2pt}{\protect\footnotesize{$\rho_{d\Sys-1 \, d\Sys-1 }{a\suptiny{0}{0}{(0)}}$}}\tikzmarkend{aa2}\!
    \end{array}\right]
\end{array}.
%\\\nonumber
\end{align}
In turn, this fixes the the matrices $\tilde{A}_{ij} $ in Eq.~\eqref{eq:gammaU} to be
\begin{align}\label{eq:unbias}
\tilde{A}_{ij} = a^{(\pi [i, j])}
\,
\end{align}
where $\pi[m,k]$ denotes the $m, k-$th element of the  $d\Sys\times d\Sys$ matrix composed of permutations of the entries of the set $\{0,1,\ldots, d\Sys-1\}$ under the constraint that the diagonal entries $\pi[m, m]=0\ \,,\forall m$. For the optimal energy solution, this permutation matrix takes the form
\begin{align}
\pi = \left( \begin{array}{ccccc}
0\ & 1\ & 1\  & 1\ & \hdots\\
1\ & 0\ & 2\ & 2\ &\hdots \\
2\ & 2\ & 0\ & 3\ &\hdots \\
3\ & 3\ & 3\ &0\ &\hdots\\
\vdots & \vdots & \vdots  &\vdots & \ddots\\
 \end{array} \right),
\end{align}
which encodes how the correlations in the matrices $a\suptiny{0}{0}{(i)}$ are paired with the non-correlated subspaces.
The final state admits a simplified form, namely
\begin{align}\label{eq:finalstate}
   	&\tilde{\rho}\SP =  \sum_{k=0}^{d\Sys-1}\frac{\rho_{kk}}{\mathcal{Z}}\Bigl(\sum_{i=0}^{d\Poi/d\Sys -1 }
   		e^{-\beta E\suptiny{0}{0}{(0)}_{i}}\ket{k}\!\!\bra{k}\otimes
   	    \ket{E\suptiny{0}{0}{(i)}_k}\!\!
   	    \bra{E\suptiny{0}{0}{(i)}_k} \\
  & \ \ +%\sum_{k=1}^{d_S}\frac{\rho_{kk}}{\mathcal{Z}}
  \sum_{m\neq k}^{d\Sys - 1}
  	\sum_{i=0}^{d\Poi/d\Sys -1}
  			e^{-\beta E\suptiny{0}{0}{(\pi[m,k])}_{i}}
  				\ket{m}\!\!\bra{m} \otimes
  				 \ket{E\suptiny{0}{0}{(i)}_k}\!\!
   	    \bra{E\suptiny{0}{0}{(i)}_k}
  						\Bigr).\nonumber
\end{align}
It can be seen that this state is not unique due to the inherent description in terms of energy. Thus the final state $\tilde{\rho}\SP$ depends on one's choice of how to represent the basis and excitations and in general is degenerate. We leave it as an open investigation as to whether, within this class there is a preferred state with special and interesting properties.

\end{document}